\documentclass[aps,prc,reprint,groupedaddress,showpacs,preprintnumbers,nofootinbib,nobibnotes,amsmath,amssymb,floatfix,superscriptaddress]{revtex4-1}
\usepackage[colorlinks=false]{hyperref}
\usepackage{dcolumn}
\usepackage{bm}
\usepackage[normalem]{ulem}
\usepackage{graphicx,color,xcolor}
\usepackage{siunitx}
\usepackage{enumitem}
\usepackage[utf8]{inputenc}
\usepackage{mathrsfs}

\DeclareSIUnit{\mev}{\mega\electronvolt}

\newcommand{\mmax}[1]{M^\star_{#1}}
\newcommand{\fmax}[1]{f^\star_{#1}}
\newcommand{\cmax}[1]{C^\star_{#1}}
\newcommand{\rmax}[1]{R^\star_{#1}}
\newcommand{\xmax}[1]{x^\star_{#1}}
\usepackage[normalem]{ulem}  

\def\mnras{Mon. Notices Royal Astron. Soc.} 
\def\physrep{Phys. Rep.} 
\def\nar{New Astron. Rev.} 

\begin{document}

\title{Maximum mass of compact stars from gravitational wave events with finite-temperature equations of state}

\author{Sanika Khadkikar}\email{sanikakhadkikar@outlook.com}
\affiliation{Birla Institute of Technology Sciences, Pilani, Hyderabad Campus, India}%
\author{Adriana R. Raduta}\email{araduta@nipne.ro}
\affiliation{National Institute for Physics and Nuclear Engineering
(IFIN-HH),  RO-077125 Bucharest, Romania}%
\author{Micaela Oertel}\email{micaela.oertel@obspm.fr}
\affiliation{LUTH, Observatoire de Paris, Universit\'e PSL, CNRS,
  Universit\'e de Paris, 92190 Meudon, France}%
\author{Armen Sedrakian}\email{sedrakian@fias.uni-frankfurt.de}
\affiliation{Frankfurt Institute for Advanced Studies, D-60438 
Frankfurt-Main, Germany}
\affiliation{Institute of Theoretical Physics,  University of Wroc\l{}aw,
50-204 Wroc\l{}aw, Poland
}
\date{03/08/2021}  
\begin{abstract}
We conjecture and verify a set of relations between global parameters
of {\it hot and fast-rotating} compact stars which do not depend
  on the equation of state (EoS), including a relation connecting the
masses of the mass-shedding (Kepler) and static configurations.  We
apply these relations to the GW170817 event by adopting the scenario
in which a hypermassive compact star remnant formed in a merger
evolves into a supramassive compact star that collapses into a black
hole once the stability line for such stars is crossed.  We deduce an
upper limit on the maximum mass of static, cold neutron stars $
2.15^{+0.18}_{-0.17}\le \mmax{\mathrm{TOV}} \le
2.24^{+0.45}_{-0.44} $ for the typical range of entropy per
baryon $2 \le S/A \le 3$ and electron fraction $Y_e = 0.1$
characterizing the hot hypermassive star. Our result implies that
accounting for the finite temperature of the merger remnant relaxes
previously derived constraints on the value of the maximum mass of a
cold, static compact star.
\end{abstract}

\pacs{
}

\maketitle


\section{Introduction}\label{s:intro}

Neutron (or compact) stars, containing matter at densities exceeding
that at the centers of atomic nuclei, represent unique laboratories to
probe the matter under extreme conditions. Considerable effort is
underway to pin down the dense matter equation of state (EoS) as
present in neutron stars, which is pressed ahead by many recent
observations and the prospects opened by the dawn of multi-messenger
astrophysics. Among these are the precise pulsar mass determinations
from the pulsar timing
analysis~\cite{Demorest:2010bx,Antoniadis:2013pzd,Cromartie:2019kug,Ozel:2016oaf,Watts:2014tja},
measurements of compact star masses and radii through the x-ray
observations of their surface
emission~\cite{Watts:2016uzu,Watts:2018iom} in particular, the results
of the NICER experiment~\cite{Riley_2019,Miller_2019}, and the
gravitational wave detection of binary neutron star (BNS) mergers by
the LIGO-Virgo collaboration \cite{Abbott_2017,Abbott:2020uma}.  Among
the events in the last category, the GW170817 event is currently
outstanding, since it has been possible to measure not only the
neutron star tidal deformability during inspiral, but also
electromagnetic
counterparts~\cite{LIGO_Virgo2017b,LIGO_Virgo2018a}. As a result, the
GW170817 event has triggered a large number of works which are aimed
at constraining neutron star properties and the EoS, either from the
analysis of the tidal deformability alone (see for example
\cite{LIGO_Virgo2018b,Malik_18,Paschalidis2018,Dexheimer_18,Capano_19,Li:2019tjx,Guven_20,Li2020PhRvD,Marczenko2020AA}),
or from a combination of tidal deformability and the electromagnetic
signal~\cite{Radice_17,Margalit_17,Shibata_17,Bauswein_17,Coughlin_18,Rezzolla_2018,
  Shibata_2019a,Ruiz2018}. Including
the information from the electromagnetic signal requires as an input
numerical modeling of the merger process, which introduces additional
uncertainties but, at the same time, broadens the experimental base of
the analysis.

Another interesting event is GW190814, where the mass of the lighter object has been determined (at 90\% credible level) to be 2.50-2.67
$M_\odot$~\cite{Abbott_2020}. In the standard interpretation~\cite{Zhang2020,Sedrakian2020,Dexheimer2020,Tan2020,Tsokaros2020,Biswas2020,Fattoyev2020,Nathanail2021}
this is either the most massive neutron star observed to date or
is a black hole that is located in the so-called mass-gap. Other,
more exotic models include for example a strange
star~\cite{Horvath2020,Bombaci2020} or a compact star in an
alternative theory of gravity~\cite{Astashenok2020}.  The neutron star
interpretation of the light companion in the GW190814 challenges our
current understanding of the EoS, even if one assumes that this star
is rotating very
rapidly~\cite{Zhang2020,Sedrakian2020,Dexheimer2020,Tan2020,Tsokaros2020,Biswas2020,Fattoyev2020,Nathanail2021}.

An important aspect of the merger process is that before the merger
the two stars are well described by a one-parameter EoS of cold matter
in weak ($\beta$-)equilibrium, which typically relates pressure to
(energy) density. This means that the measured tidal deformabilities
and masses of the two merging stars essentially concern this cold EoS
of dense matter in $\beta$-equilibrium. In contrast, after the merger
the evolution of the post-merger remnant (if there is no prompt black
hole formation) requires as an input an EoS at non-zero temperature
and out of (weak) $\beta$-equilibrium, {\it i.e.,} the pressure
becomes a function of three thermodynamic
parameters~\cite{Shibata_11,Rosswog_15,Baiotti_2017,Chatziioannou2020GReGr}.
Most commonly, these are chosen to be baryon number density, $n_B$,
temperature $T$ and charge fraction $Y_Q = n_Q/n_B$, where $n_Q$ is
defined as the total hadronic charge
density~\citep{Oertel_RMP_2017}. The electron fraction $Y_e = Y_Q$ due
to electrical charge neutrality. In the following, when referring to
cold compact stars, we will assume that they are in
$\beta$-equilibrium. Small deviations from $\beta$-equilibrium, which
can lead to some kinematical effect (bulk viscosity, etc) will be
neglected. 

Alongside full-fledged hydrodynamics simulations of the post-merger
phase, different studies focused on stationary solutions for compact
star configurations, which give, among other things,  hints on the magnitude of
the maximum mass supported by a post-merger object and thus the
conditions for the formation of a black hole. As evidenced by
numerical simulations, post-merger objects are rapidly rotating and
support significant internal flows. Therefore, to assess the stability
of the post-merger object rapidly and differentially rotating
configurations of compact stars should be studied.

Universal relations, {\it i.e.}, relations between different global
quantities of the star found empirically to be independent of the EoS,
have attracted much attention in this context. Such relations have
been established for both uniformly
\cite{Cook1994,Doneva:2013rha,Maselli_PRD_2013,Breu_2016} and differentially
rotating stars~\cite{Bozzola_17,Bozzola_19,Weih2018MNRAS} in the case of cold stars,
described by zero-temperature EoS with the matter under
$\beta$-equilibrium.  However, for the merger remnant the thermal
effects cannot be ignored and can influence, among other observables,
the maximum mass of a static or rapidly rotating
star~\cite{Marques_PRC_2017,Nunna_ApJ_2020} as well as the applicability of
universal relations.  In
Refs.~\cite{Martinon_PRD_2014,Lenka_JPG_2019,Marques_PRC_2017}, it has
been shown that thermal effects induce deviations from the universal
relations obtained for $\beta$-equilibrated matter at zero
temperature. Subsequently, Ref.~\cite{Raduta_MNRAS_2020} demonstrated
that universality is restored if finite-temperature configurations
with the same entropy per baryon and electron fraction are
considered. Here we will extend the study of Ref.~\cite{Raduta_MNRAS_2020}
which has focussed on non-rotating or slowly rotating stars to rapid
rotation.

As a consequence of our findings on universality for hot stars, we
revisit the inference of the maximum mass of a compact star from the
analysis of the GW170817 event. This problem has been addressed by
several authors, see
Refs.~\cite{Margalit_17,Ruiz2018,Rezzolla_2018,Shibata_2019a} using
the scenario of the formation of a hypermassive compact star in the merger
event and its delayed collapse to a black hole close to the neutral
stability line for supramassive compact stars. Some of these authors
employed the universality of the linear relation between the {\it
  maximum gravitational mass} for uniformly rotating stars at the
Kepler limit, $\mmax{K}$, and the same quantity for a non-rotating
star $\mmax{\mathrm{TOV}}={\rm max}~(
M_{\mathrm{TOV}})$~\cite{Cook1994,Lasota_ApJ_1996,Breu_2016}
\begin{equation}
  \mmax{K} = \cmax{M} \mmax{\mathrm{TOV}}~.
  \label{eq:KTOV}
\end{equation}
Here and below the superscript $\star$ refers to quantities
characterizing the maximum mass objects. The employed value for
$\cmax{M} \approx 1.2$~\cite{Cook1994,Lasota_ApJ_1996,Breu_2016},
relating $\mmax{K}$ and $\mmax{\mathrm{TOV}}$ has, however, been
determined assuming that the star rotating at Kepler frequency is cold
and in $\beta$-equilibrium, which is not necessarily the case for the
merger remnant. Therefore we will revisit this question and will
determine the impact of nonzero temperature and matter out of
$\beta$-equilibrium on the value of $\cmax{M}$.

This paper is organized as follows.  In Sec.~\ref{sec:model} we
describe briefly the numerical setup for modeling fast-rotating hot
{compact} stars and our collection of EoS.  In
Section~\ref{sec:universal} we investigate different universal relations
for fast-rotating stars.  Section~\ref{sec:mkepler} is devoted to the
discussion of the maximum mass of fast-rotating compact stars. We
derive a new upper limit on $\mmax{\mathrm{TOV}}$ using the universal relations
in Sec.~\ref{sec:GW170817}, Our conclusions are collected in 
Sec.~\ref{sec:conclusions}. Throughout this paper we use natural units
with $c=\hbar=k_B=G=1$.


\section{Setup}
\label{sec:model}
This section is devoted to a description of our strategy to solve for
the structure of a hot rapidly and rigidly rotating relativistic
star. More details on the formalism can be found
in~\cite{Gouss1,Villain2004,Marques_PRC_2017}. Combined Einstein and
equilibrium equations are solved, assuming stationarity and
axisymmetry. Besides, we assume the absence of meridional currents
such that the energy-momentum tensor fulfills the circularity
condition, {\it i.e.} there is no convection. An EoS is needed to
close the system of equations.  In neutron stars older than several
minutes matter is cold, neutrino-transparent, and in (approximate)
$\beta$-equilibrium. Its EoS is barotropic, {\it i.e.}  it depends
only on one variable, which commonly is chosen as baryon number
density $n_B$. In contrast, the merger-remnant matter is hot and not
necessarily in $\beta$-equilibrium, such that the EoS depends in
addition to $n_B$ on temperature $T$ and electron fraction $Y_e =
n_e/n_B$ or thermodynamically equivalent variables.  Under the
above-mentioned assumptions, in particular stationarity, the most
general solution for the star's structure becomes again barotropic,
{\it i.e.}, the electron fraction and the temperature need to be
related to $n_B$~\cite{Gouss1,Gouss2,Villain2004}\footnote{If the
  assumption of rigid rotation is relaxed, then stationary solutions
  can be constructed with non-barotropic equations of state, see for
  example, Refs.~\cite{Camelio2019,Camelio2020}}.  To fulfill this
requirement, we consider below stars characterized by constant entropy
per baryon $S/A$ and some fixed value of the electron fraction or
constant electron lepton fraction $Y_L = (n_e + n_\nu)/n_B = n_L/n_B$
($n_\nu$ and $n_L$ being the neutrino and electron lepton number
densities, respectively). It should be stressed that this simplified
setup does not reflect realistic conditions in the merger remnant. A
variation of the values of $S/A$ and $Y_e$ or $Y_L$ should
nevertheless allow us to cover the relevant conditions and thus to
estimate the sensitivity of the universal relations and those
observables needed to place limits on $\mmax{\mathrm{TOV}}$ on the
thermal and out of $\beta$-equilibrium effects and to give an
uncertainty range.

\subsection{Numerical models of rapidly rotating hot stars}
For computing numerical models of hot rapidly rotating stars, we have
used the \textsc{LORENE}
library~\cite{Lorene}\footnote{\url{https://lorene.obspm.fr}}. \textsc{LORENE}
is a set of C++ classes developed for solving problems in numerical
relativity. It contains tools for computing equilibrium
configurations of relativistic rotating bodies~\cite{BGSM} for which
combined Einstein and equilibrium equations are solved assuming
stationarity, axisymmetry, asymptotic flatness, and circularity.

Using a quasi-isotropic gauge, the line element expressed in
spherical-like coordinates reads~\cite{BGSM}
\begin{eqnarray}
  \label{e:QI_metric}
  ds^2&=&-N^2dt^2 + A^2\left(dr^2 + r^2 d\theta^2\right) \nonumber\\
  && + B^2r^2 \sin^2
  \theta \left( d\varphi^2 + N^\varphi dt\right)^2,
\end{eqnarray}
with $N, N^\varphi,A$, and $B$ being functions of coordinates
$(r,\theta)$.  Under the present symmetry assumptions, Einstein
equations for the four metric potentials reduce to a set of four
elliptic (Poisson-like) partial differential equations, in which
source terms contain both contributions from the energy-momentum
tensor (matter) and non-linear terms with non-compact support,
involving the gravitational field itself. More details and explicit
expressions can be found in~\cite{BGSM}.

The matter is assumed to behave as a perfect fluid such that
the energy-momentum tensor can be written as
\begin{equation}
 T^{\alpha\beta}=(\varepsilon+p)\, u^{\alpha}u^{\beta}+p\,g^{\alpha\beta},
\end{equation}
where $\varepsilon$ is the total energy density (including rest
mass), $p$ the pressure, and $u^\alpha$ the fluid four-velocity.  The
angular velocity of the fluid then becomes $\Omega := u^\varphi/
u^t$. Equilibrium equations are derived from energy and momentum
conservation, $\nabla_\alpha T^{\alpha\beta} = 0$, and become within
the present setup~\cite{Gouss1,Gouss2,Villain2004,Marques_PRC_2017}
\begin{eqnarray}
\partial_i \left( H + \ln N - \ln \Gamma\right) &=& \nonumber \\   
                                                && \hspace{-2.2cm} \frac{e^{-H}}{m_B}\left[ T\, \partial_i (S/A)
                                                   + \mu_L \partial_i Y_L\right] - u_\varphi u^t\partial_i\Omega,
\label{finalDivT}
\end{eqnarray}
where $\Gamma = N u^t$ is  the Lorentz factor of the fluid with
respect to the Eulerian observer and $S/A$ the entropy
per baryon ($k_B =1$),
\begin{equation}
H = \ln\left({\frac{\varepsilon + p}{ m_B\,n_B}}\right)~,
\end{equation}
is the pseudo-log enthalpy with $m_B$ being a constant of the
dimension of a mass~\footnote{We chose the value $m_B= 939.565$ MeV.}.
Since in this work we consider only uniform rotations with
$\Omega = \textrm{const}$, constant $S/A$, and constant $Y_e$ with
$\mu_L = 0$ or constant $Y_L$, the right hand side of
Eq.~(\ref{finalDivT}) vanishes and the equilibrium equation takes the
same form as in the zero temperature and $\beta$-equilibrium case.

Upon computing models of rotating stars, at finite temperature, an additional difficulty arises from the fact that the surface of the
star is no longer well defined since an extended dilute atmosphere can
form, see for instance the discussion in
\cite{Raduta_MNRAS_2020,Stone2019}. For simplicity, we assume that the
surface corresponds to the density $n_B = 10^{-5}\, \mathrm{fm}^{-3}$
for all EoS models and any considered combinations of $S/A$ and
$Y_e/Y_L$. We have checked that our conclusions do not depend on the
choice of the definition of the surface, see Appendix~\ref{sec:app}.
\begin{table*}
    \begin{tabular}{l||ccccccccc}
    \hline
    Model & $\mmax{\mathrm{TOV}}$  & $\mmax{B} $ & $R_{1.4}$ & $\tilde\Lambda $ & $E_B$& $n_s$& $K$ & $E_S$ & $L$   \\
    & ($M_\odot$)& ($M_\odot$) & (km) & &  (MeV) & ($\mathrm{fm}^{-3}$) & (MeV) & (MeV)& (MeV) \\  
    \hline
    RG(SLy4) & 2.06 & 2.46 & 11.73 & 322-353 & -15.97 & 0.159 & 230.0 & 32.0 & 46.0 \\
    HS(DD2)  & 2.42 & 2.92 & 13.2 & 758-799   & -16.00 & 0.149 & 242.6 & 31.7 & 55.0 \\
    HS(IUF)  & 1.95 & 2.27 & 12.64 & 499-530 & -16.40 & 0.155 & 231.3 & 31.3 & 47.2 \\
    SFHo     & 2.06 & 2.45 & 11.9 & 366-401  & -16.19 & 0.158 & 245.4 & 31.6 & 47.1 \\
    NL3-$\omega \rho$ & 2.75 & 3.39 & 13.82 & 1042-1051  & -16.24 & 0.148 & 271.6 & 31.7 & 55.5 \\
    FSU2H    & 2.39 & 2.86 & 13.28 & 635-655  & -16.28 & 0.150 & 238.0 & 30.5 & 44.5 \\
    SRO(APR) & 2.17 & 2.66 & 11.33 & 271-295  & -16.00 & 0.160 & 266.0 & 32.6 & 57.6 \\
    BHB$\Lambda \phi$    & 2.10 & 2.45 & 13.22 & 754-789   & -16.00 & 0.149 & 242.6 & 31.7 & 55.0 \\
    SFHoY    & 1.99 & 2.34 & 11.9 & 366-401  & -16.19 & 0.158 & 245.4 & 31.6 & 47.1 \\
    NL3-$\omega \rho$ NY & 2.35 & 2.77 & 13.82 & 1042-1051   & -16.24 & 0.148 & 271.6 & 31.7 & 55.5 \\
    FSU2H NY & 1.99 & 2.37 & 13.28 & 637-653  & -16.28 & 0.150 & 238.0 & 30.5 & 44.5\\
    \hline 
  \end{tabular}
    \caption{ Global parameters of cold neutron stars (first
      four columns) for EoS considered in this work.  These columns
      list (from left to right) the EoS model acronym, maximum
      gravitational and baryonic masses, radius of a $1.4 M_{\odot}$
      star and the tidal deformability $\tilde\Lambda$ range for the 
      GW170817 event.  The latter quantity is calculated assuming for
      the merger stars the masses $m_1 \in (1.36,1.60)M_{\odot}$ and
      $m_2 \in(1.16,1.36)M_{\odot}$, which corresponds to the mass
      ratio range $0.73 \leq q=m_2/m_1 \leq 1$.  The remaining columns
      list properties of symmetric nuclear matter at saturation
      density according to the employed EoS model: the binding energy per
      nucleon $E_B$, saturation density $n_s$, compression modulus
      $K$, symmetry energy $E_S$ and its slope $L$.  Presently
      available observational and experimental constraints on listed
      quantities include a lower limit on the maximum gravitational
      mass $\mmax{\mathrm{TOV}}\ge 2.01 \pm 0.04 M_{\odot}$
      \cite{Antoniadis:2013pzd}, simultaneous constraint on the radius
      and mass of a compact star from the NICER experiment for PSR
      J0030+0451 $R(1.44^{+0.15}_{-0.14}
      M_{\odot})=13.02^{+1.24}_{-1.06}$ km \citep{Miller_2019} and
      $R(1.34^{+0.15}_{-0.16} M_{\odot})=2.71^{+1.14}_{-1.19}$ km
      \citep{Riley_2019}, and a range for the tidal deformability obtained
      from the GW170817 event $\tilde \Lambda =300^{+500}_{-190}$
      (90\% credible interval) or $\tilde \Lambda
      =300^{+420}_{-230}$ (90\% highest posterior density) for a low
      spin prior \cite{Abbott_2019}. The nuclear matter properties
      have been determined as $E_B =-15.8 \pm 0.3$ MeV
      \cite{Margueron_PRC_2018}, $n_s=0.155 \pm 0.005$ fm$^{-3}$
      \cite{Margueron_PRC_2018}, $K=230 \pm 40$ MeV
      \cite{Khan_PRL_2012}, $E_s=31.7 \pm 3.2$ MeV
      \cite{Oertel_RMP_2017}, $L=58.7 \pm 28.1$ MeV
      \cite{Oertel_RMP_2017}.  }
  \label{table:eos}
\end{table*}

Employing
\textsc{LORENE}, we find global stellar parameters such as gravitational, $M_G$, and baryon, $M_B$, mass, and equatorial circumferential radius $R$.  We additionally compute the angular
momentum, the moment of inertia, and the quadrupole moment. The
corresponding expressions for the quadrupole moment can be found in
\cite{Salgado1994} and \cite{Pappas2012}. For our setup with constant
$S/A$, the star's total entropy is simply given by $S/A\, M_B$.
\begin{figure}
    \includegraphics[angle=0, width=\columnwidth]{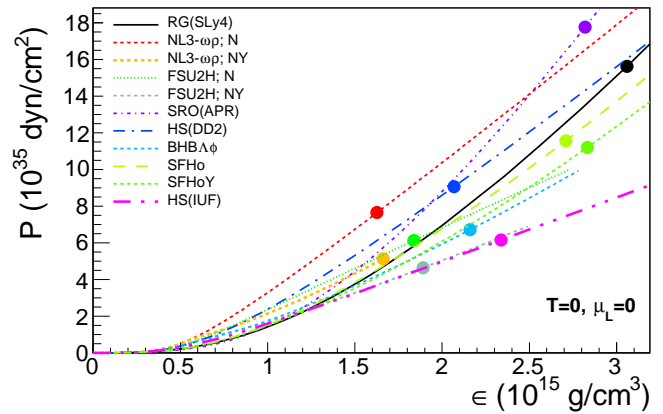}
    \caption{ Pressure of cold, $\beta$-equilibrated neutron star matter
      as function of its energy density according to the EoS models employed
      in this work. The symbols indicate the central energy density of the maximum mass configuration for cold, $\beta$-equilibrated matter.  }
  \label{fig:eos}
\end{figure}

\subsection{Equations of state}
\label{sec:eos}
The system of equations for solving for the star's structure discussed
in the preceding section is closed by an EoS.  To ensure that our
results are not an artifact of a particular choice of EoS model, we
have performed the same calculations for a set of different EoS
models.  There exists a large number of EoS models obtained for cold
matter in compact stars. The number of EoS covering the regimes of
finite temperature and varying electron fraction is however
small. These are mostly based on density functional theory. Here we
choose a set of EoS models that are based either on relativistic
density functional theory with various parameterizations or a
non-relativistic model based on Skyrme functional and an empirical
extension of a variational microscopic model. These models are
reasonably compatible with existing constraints from nuclear
experiments, theory, and astrophysics, in particular, they predict
maximum masses above 2
$M_\odot$~\cite{Demorest:2010bx,Antoniadis:2013pzd,Arzoumanian2018,Cromartie:2019kug}
or at least marginally consistent with this value.  To be specific, we
consider one non-relativistic density-functional (DFT) model,
RG(SLy4)~\cite{Raduta_PRC_2015,Raduta_NPA_2019}; five variants of
relativistic DFT, one with density-dependent couplings,
HS(DD2)~\cite{HS,Typel_PRC_2010}, and four with non-linear couplings,
HS(IUF)~\cite{Fischer2014,Fattoyev_PRC_2010},
SFHo~\cite{Steiner_2013}, NL3$\omega\rho$~\cite{Pais16,Horowitz01} and
FSU2H~\cite{Tolos17,Tolos17b}; as well as the SRO(APR)
model~\cite{Constantinou2014,Schneider2019}. The latter is based on
the APR EoS~\cite{APR}, which itself is partly adjusted to the
variational calculation of \cite{Akmal_1997}. If available, we compare
the above purely nucleonic EoS models with the corresponding EoS
allowing for the presence of hyperons. These are
BHB$\Lambda\Phi$~\cite{Banik_2014}\footnote{The EoS model
  BHB$\Lambda\phi$ contains only $\Lambda$-hyperons and not the full
  baryon octet. There exists a version, DD2Y~\cite{Marques_PRC_2017},
  based on the same nucleonic HS(DD2) EoS, which contains the full
  baryon octet. For the present purpose, both give very similar
  results.}, the extension of HS(DD2); SFHoY~\cite{Fortin_PASA_2018},
extension of SFHo; NL3$\omega\rho$Y, an extension of NL3$\omega\rho$;
and FSU2HY, an extension of FSU2H.  For NL3$\omega\rho$Y and FSU2HY we
adopt the parameterizations in \cite{Fortin_PRD_2020} but disregard
the $\sigma^*$-meson field.  Except for FSU2H(Y) and
NL3$\omega\rho$(Y), EoS data are publicly available on the
\textsc{Compose} data
base~\cite{Typel_2013}~\footnote{\url{https://compose.obspm.fr}}.  Key
properties of our collection of the EoS are summarized in
Table~\ref{table:eos} together with present constraints. The value for
the tidal deformability of NL3$\omega\rho$ lies above the 90\%
confidence interval given by the GW170817 event~\cite{Abbott_2019},
but in view of the large uncertainty we feel it premature to exclude a
certain EoS model and keep the NL3$\omega\rho$ model as representative
of a large deformability in our EoS sample. In Fig. \ref{fig:eos} we
show the pressure as a function of energy density for cold,
$\beta$-equilibrated matter.

\section{Universal relations for fast rotating stars at finite temperature}
\label{sec:universal}

Although the properties of static and rotating stars depend strongly
on the EoS, a series of ``universal'' relations have been found between
global parameters of static stars which are almost EoS
independent~(see for a review \cite{Yagi2017}). These were later
extended to slowly and maximally fast-rotating
stars~\cite{HZ_Nature_1989,Friedman_PRL_1989,Shapiro_Nature_1989,Haensel_AA_1995,Lasota_ApJ_1996,Haensel_AA_2009,Koliogiannis_PRC_2020}. The practical importance of such relations resides in their potential to
provide constraints on quantities that are difficult to access
experimentally.

It was previously shown that most of the universal relations for
slowly rotating stars remain valid at finite temperature if the same
thermodynamic conditions are maintained (for example by fixing $S/A$
and $Y_L$)~\cite{Raduta_MNRAS_2020}. Here we extend this investigation
to rapidly rotating hot stars.  In Sec.~\ref{ssec:KSmax} we first
address the universal relations between global properties of
non-rotating and Keplerian configurations for stars with constant
$S/A$ and $Y_e$. In the subsequent Sec.~\ref{ssec:KKmax} we address the universal relations among the normalized moment of inertia, quadrupole moment, and the compactness for the maximum mass configuration of a compact star at the Kepler limit.

\subsection{Relations between global properties of
  non-rotating and Keplerian configurations}

\label{ssec:KSmax}

\begin{figure}
    \includegraphics[angle=0, width=0.9\columnwidth]{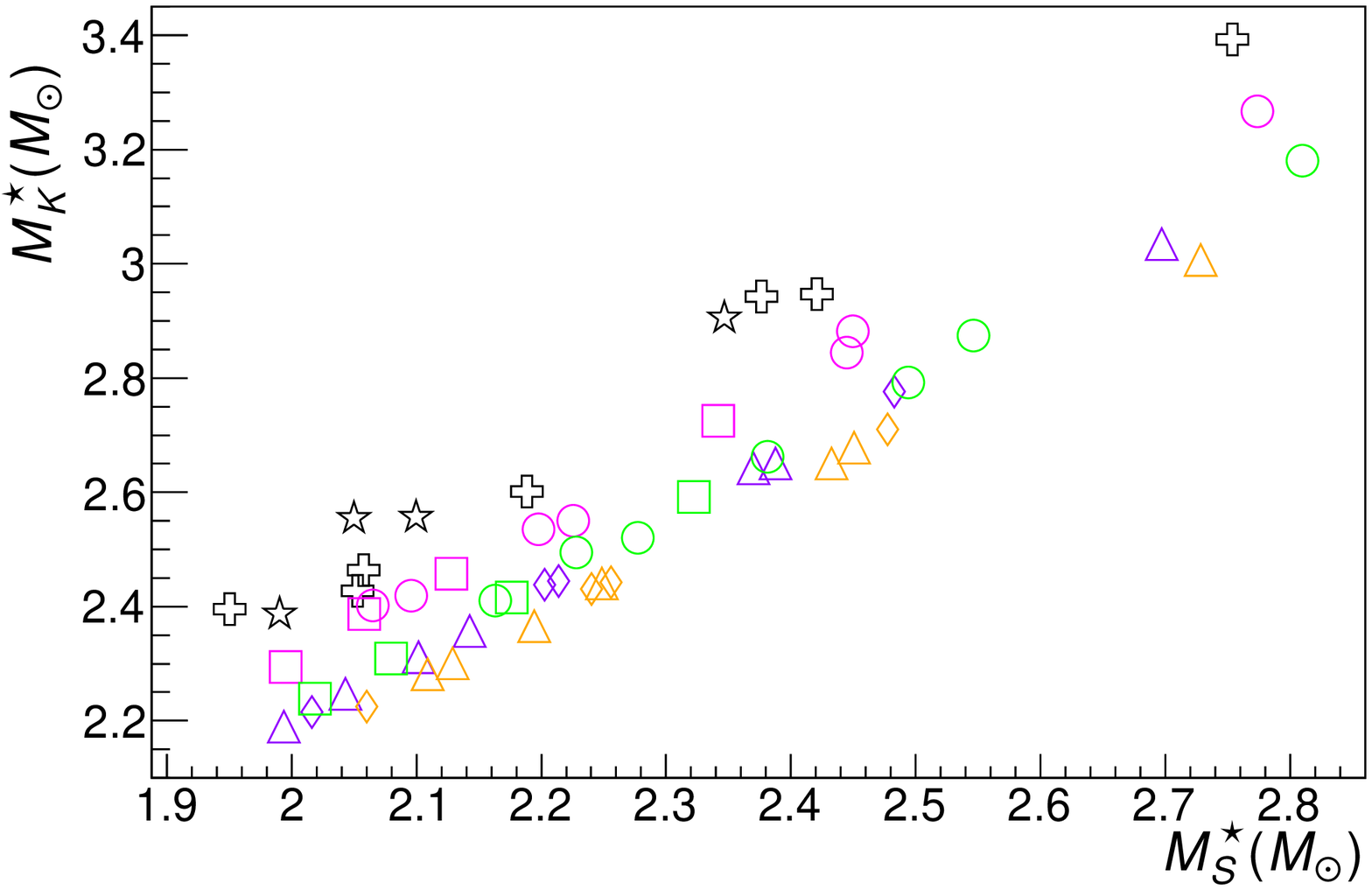}
    \includegraphics[angle=0, width=0.9\columnwidth]{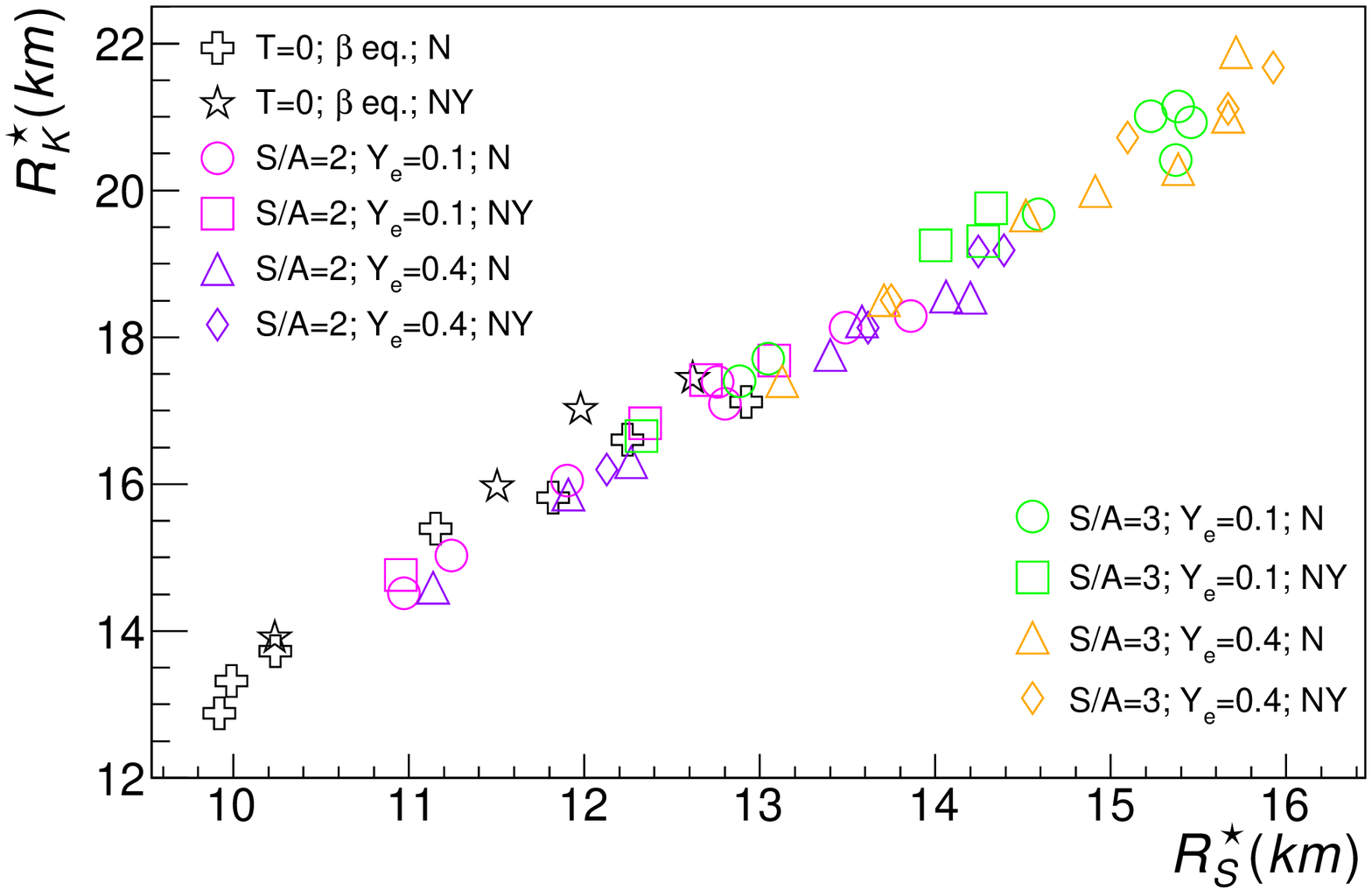}
    \includegraphics[angle=0, width=0.9\columnwidth]{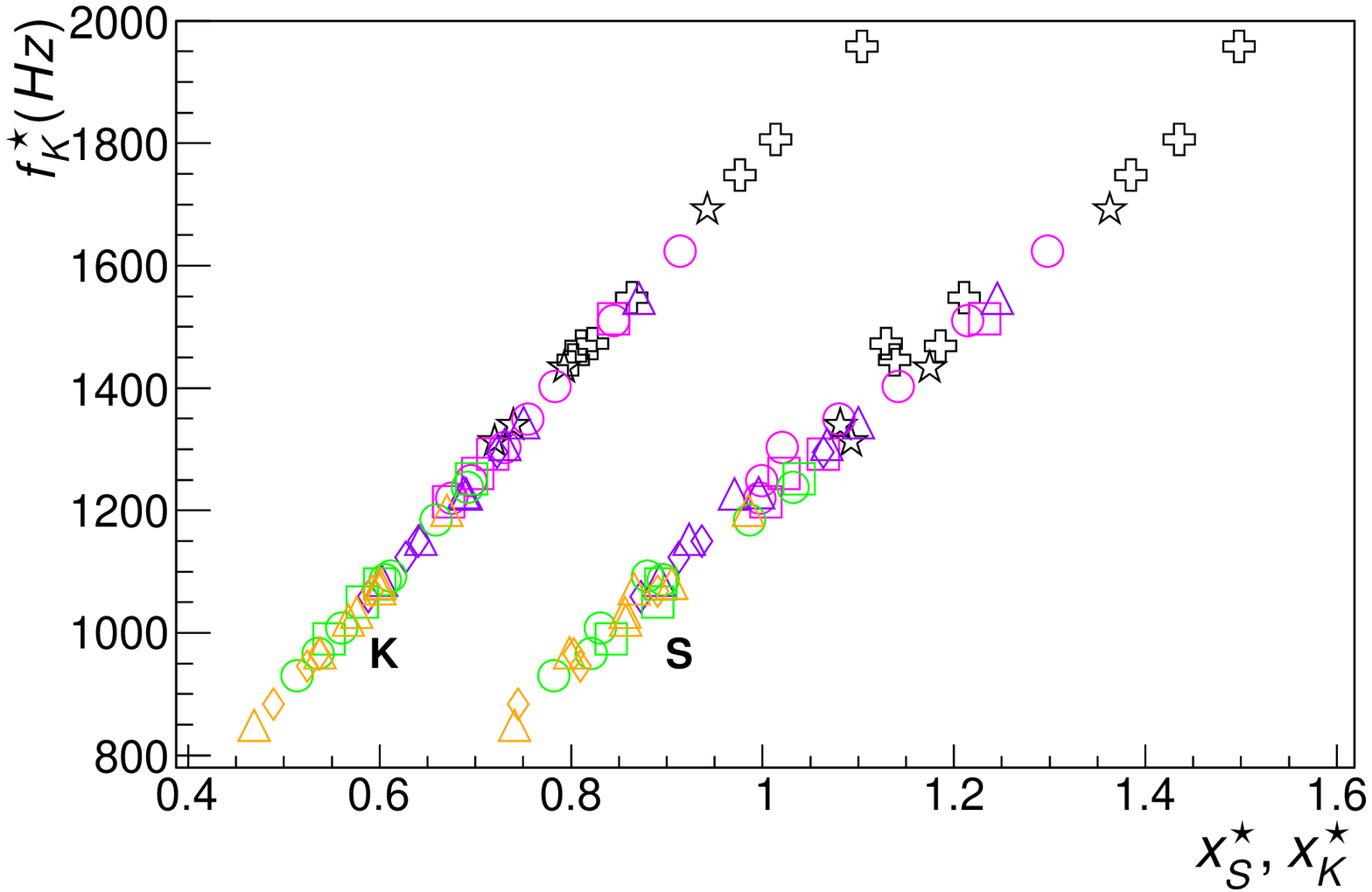}
    \caption{Top panel: maximum gravitational mass at the Kepler limit ($\mmax{K}$) vs. maximum gravitational
      mass of a static star ($\mmax{S}$),
      Eq.~\eqref{eq:MKsbye}; middle panel: equatorial circumferential
      radius of the maximum mass Keplerian configuration ($\rmax{K}$)
      vs. circumferential radius of the maximum mass static configuration
      ($\rmax{S}$), Eq.~\eqref{eq:RKmax}; bottom panel: rotation
      frequency of the maximum mass configuration at the Kepler limit
      $\fmax{K}$ as function of $\xmax{S}$ or $\xmax{K}$, i.e., for the
      maximum mass static ($S$) and Keplerian ($K$) configurations,
      see Eq. \eqref{eq:fKmax}.
      The results correspond
      to eleven EoS models and different thermodynamic conditions
      expressed in terms of $S/A$ and $Y_e$.  Results for cold stars
      are shown for comparison. 
}
  \label{fig:KmaxSmax}
\end{figure}

\begin{figure}
  \includegraphics[angle=0, width=0.8\columnwidth]{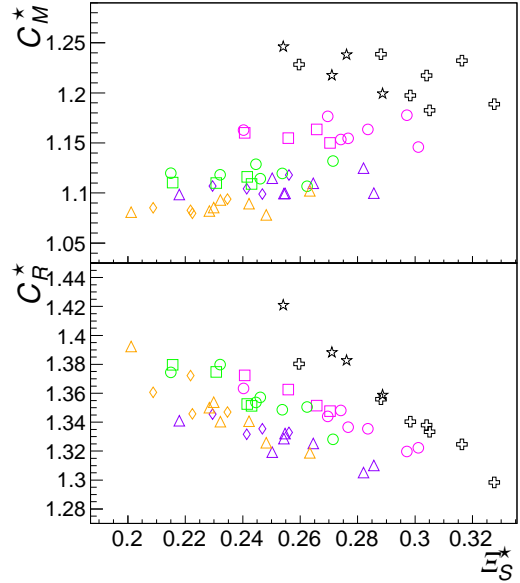}
  \caption{The dependence of $\cmax{M}$ (top) and $\cmax{R}$ (middle)
    on the compactness of the maximum mass static configuration.
    The same thermodynamic conditions and EoS models as in Fig. \ref{fig:KmaxSmax}
    are considered.}
  \label{fig:CvsXi} 
\end{figure}

\begin{table*}
  \begin{tabular}{|l|c|c|c|c|}
    \hline 
    Thermo. cond. &  $C^{\star}_M$ & $C^{\star}_R$ & $C^{\star}_f$ &  $C'^{\star}_f$  \\ \hline
    $T=0$, $\beta$-eq. & 1.2187 (0.0064) & 1.3587 (0.0104)  & 1259.63 (9.72) & 1795.30 (4.35) \\
    $S/A=2$, $Y_e=0.1$ & 1.1617 (0.0032) & 1.3459 (0.0051)  & 1237.68 (5.47) & 1791.62 (2.69) \\
    $S/A=2$, $Y_e=0.4$ & 1.1084 (0.0029) & 1.3282 (0.0038)  & 1231.58 (4.46) & 1789.23 (2.62) \\
    $S/A=3$, $Y_e=0.1$ & 1.1181 (0.0026) & 1.3593 (0.0051)  & 1201.11 (5.60) & 1798.54 (2.17)  \\
    $S/A=3$, $Y_e=0.4$ & 1.0877 (0.0023) & 1.3506 (0.0063)  & 1199.01 (7.10) & 1798.92 (2.17)  \\
   \hline   
  \end{tabular}
  \caption{Fitting parameters entering Eqs. \eqref{eq:MKsbye},
    \eqref{eq:RKmax}, \eqref{eq:fKmax} and their standard errors (in
    parenthesis), under different thermodynamic conditions specified
    in the first column.  }
  \label{tab:fitparams}
\end{table*}


In this subsection, we are interested in a particular class of
universal relations, among the parameters of non-rotating and
maximally rotating (at the mass-shedding limit) stars. The original
motivation for studying these relations was to constrain on the
stellar radii using the measurements of masses and frequencies of
sub-millisecond
pulsars~\cite{HZ_Nature_1989,Friedman_PRL_1989,Shapiro_Nature_1989}.
The non-observation of a rapidly rotating pulsar in the remnant of
SN1987A led to a declining interest in these relations, although
searches of sub-millisecond pulsars continued~\cite{Cordes2004}.  The
fastest rotating pulsar observed to date~\cite{Hessels_2006} rotates
at 716 Hz, which is still far from Kepler frequencies predicted by the
various EoS of dense matter. The gravitational wave event GW170817
and the attempt to deduce a maximum mass constraint for a non-rotating
cold neutron star stimulated several recent studies of
rigidly~\cite{Koliogiannis_PRC_2020} and differentially rotating
stars~\cite{Bozzola_17,Bozzola_19}. Furthermore, the GW190814 event
rekindled the interest in the subject within the scenario in which the
light component of this merger event is a maximally rotating compact
star~\cite{Zhang2020,Sedrakian2020,Dexheimer2020,Bombaci2020,Tan2020,Tsokaros2020,Biswas2020,Fattoyev2020}.

Equation~\eqref{eq:KTOV} which expresses the maximum gravitational mass of the Keplerian configuration as a function of the maximum mass of a non-rotating star is an example of such relations. It was initially
proposed in \cite{Cook1994,Lasota_ApJ_1996} and later on confirmed by extensive computations in \cite{Breu_2016}.  Other examples are a relation
between the circumferential equatorial radius of the maximum mass
configuration at the Kepler limit and the circumferential radius of
the maximum mass static configuration~\cite{Cook1994,Lasota_ApJ_1996},
\begin{equation}
  \rmax{K}=\cmax{R} \rmax{\mathrm{TOV}},
  \label{eq:RKRS_max_cold}
  \end{equation}
  and the dependence of the rotation frequency of this maximum mass
  configuration at the Kepler limit on mass and radius of the
  non-rotating maximum mass configuration~\cite{HZ_Nature_1989,Friedman_PRL_1989,Shapiro_Nature_1989,Haensel_AA_1995},
\begin{equation}
  \fmax{K} = \cmax{f} \xmax{\mathrm{TOV}},
  \label{eq:fKxS_max_cold}
\end{equation}
where
$\xmax{\mathrm{TOV}}=\left(\mmax{\mathrm{TOV}}/M_{\odot}\right)^{1/2}
\cdot \left(10~{\rm km}/\rmax{\mathrm{TOV}}\right)^{3/2}$. This
functional form is actually identical to the Newtonian expression for
the mass shedding frequency of a rotating sphere, see also the
discussion in \cite{Shapiro_Nature_1989} about its justification in
the relativistic case.

Motivated by the findings of Ref.~\cite{Raduta_MNRAS_2020} we
reinterpret Eqs. (\ref{eq:KTOV}), (\ref{eq:RKRS_max_cold}) and
(\ref{eq:fKxS_max_cold}) as relations between properties of maximum
mass Keplerian and static configurations with identical thermodynamic
conditions
\begin{equation}
  \mmax{K}(S/A,Y_e) = \cmax{M}(S/A,Y_e) \mmax{S}(S/A,Y_e)~,
\label{eq:MKsbye}
\end{equation}
\begin{equation}
\rmax{K}(S/A,Y_e) = \cmax{R}(S/A,Y_e) \rmax{S}(S/A,Y_e)~,
  \label{eq:RKmax} 
\end{equation}
and
\begin{equation}
  \fmax{K}(S/A,Y_e) = \cmax{f}(S/A,Y_e) x^{\star}_{S}(S/A,Y_e)~,
  \label{eq:fKmax}
\end{equation}
which implies that the coefficients $\cmax{i}$, {$i\in M,R, f$} depend
on two additional thermodynamic parameters, which are chosen here to
be $S/A$ and $Y_e$.  The subscript $S$ refers to static,
hot configurations and the subscript ``TOV'' refers to cold
static stars in $\beta$-equilibrium.

The relations \eqref{eq:MKsbye}, \eqref{eq:RKmax}, \eqref{eq:fKmax}
are shown in Fig. \ref{fig:KmaxSmax} for various combinations of
$S/A=2,3$ and $Y_e=0.1,0.4$ and eleven different EoS
models. Nature does of course not supply us with hot stars under
  these idealized conditions with constant $S/A$ and $Y_e$. For the
  sake of the argument, we have chosen these values from the typical
  range of values we encounter in the central part of hot stars,
  i.e. proto-neutron stars or the binary merger remnants. For
completeness, we show also the results corresponding to cold
stars. The values of $\cmax{i}$ obtained by a fit to these results are
provided in Table \ref{tab:fitparams} for each considered
thermodynamic condition.  In the bottom panel of
Fig.~\ref{fig:KmaxSmax} the dependence of $\fmax{K}$ on $\xmax{K}$ is
considered, too. As a trivial consequence of the linear dependencies
in Eqs. \eqref{eq:MKsbye}, \eqref{eq:RKmax}, \eqref{eq:fKmax} one
finds again a linear relation $\fmax{K} = C'^{\star}_f \xmax{K}$
\cite{Koliogiannis_PRC_2020}.  Our results show that universality
holds reasonably well for hot rapidly rotating stars as well if the
same constant $S/A$- and $Y_e$-values are considered. Similar results
were obtained and discussed for non-rotating in
Ref.~\cite{Raduta_MNRAS_2020}. Moreover, since our set of EoS models
contains purely nucleonic models as well as models with hyperons, we
conclude that these relations are insensitive to the baryonic
composition of matter, be it purely nucleonic or with an admixture of
hyperons. As mentioned above, the proportionality coefficients depend,
however, on the thermodynamic conditions. A small residual dependence
of $\cmax{R}$ on the EoS remains.  It arises, as previously discussed
for cold stars~\cite{Lasota_ApJ_1996}, from a weak dependence of the
maximum mass static configuration on the compactness $\Xi^\star_S
=\mmax{S}/\rmax{S}$.
The $\Xi^\star_S$-dependence of $\cmax{M}$ and $\cmax{R}$ is depicted in
Fig.~\ref{fig:CvsXi} for the same EoS models and thermodynamic
conditions as in Fig. \ref{fig:KmaxSmax}.

Refs.~\cite{Shapiro_Nature_1989,Lattimer_Science_2004,Haensel_AA_2009}
suggested that relations analogous to Eqs.~\eqref{eq:RKRS_max_cold}
and \eqref{eq:fKxS_max_cold} hold for stars with the same
gravitational mass (and not only at the maximum of a sequence).  These
can again be generalized to configurations with the fixed $S/A$ and
$Y_e$ to find
\begin{equation}
  R_K (M) = C_R R_S(M)~,
  \label{eq:RK}
\end{equation}
\begin{equation}
  f_K (M) = C_f x_{S}(M)~,
    \label{eq:fK}
\end{equation}
where $x_{S} = \left[\left(M/M_{\odot} \right)\cdot
  \left(10~{\mathrm{km}}/R_S (M) \right)^{3}\right]^{1/2}$.

\begin{figure*}
  \begin{center}
    \includegraphics[angle=0, width=0.45\linewidth]{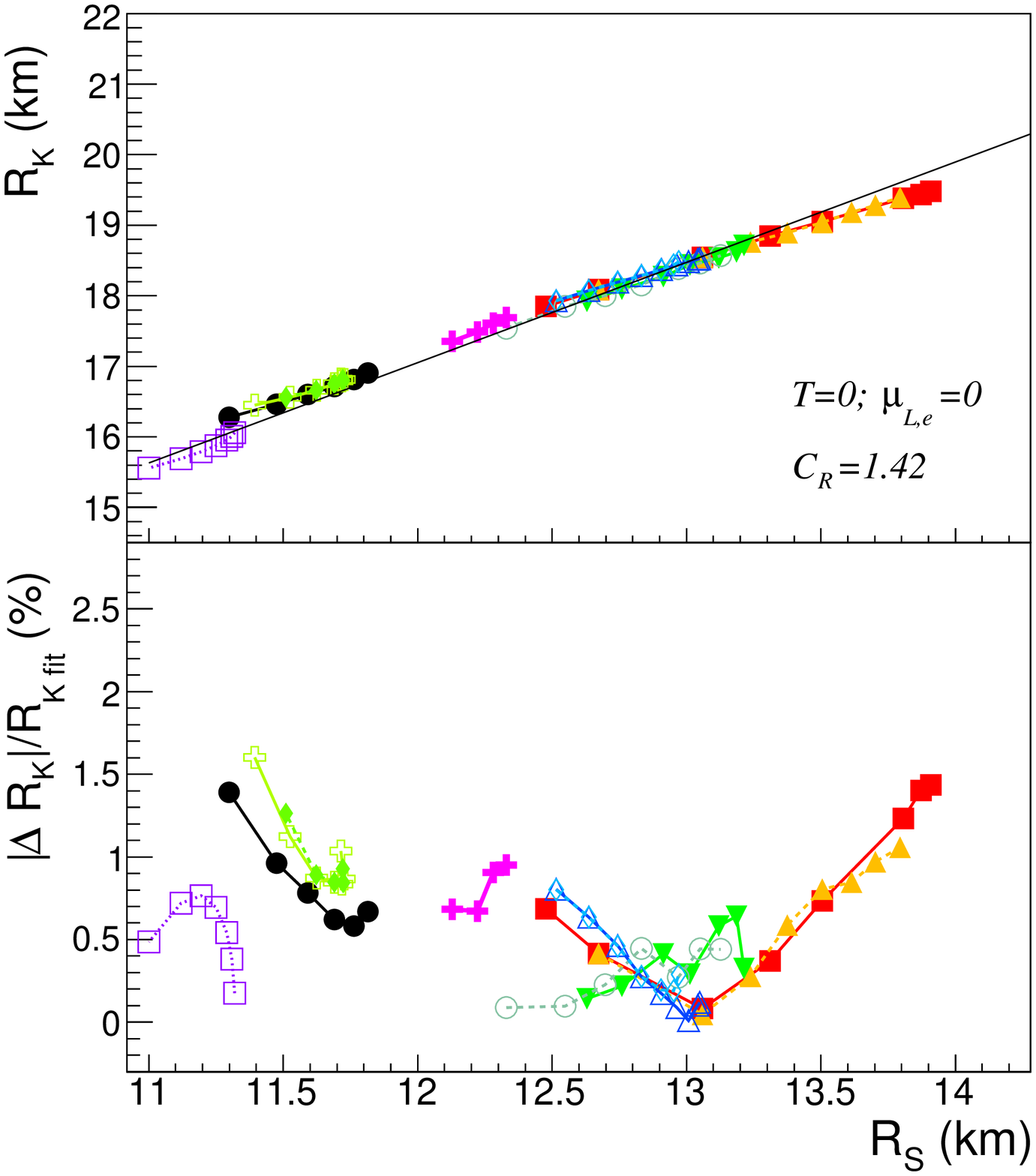}
    \includegraphics[angle=0, width=0.45\linewidth]{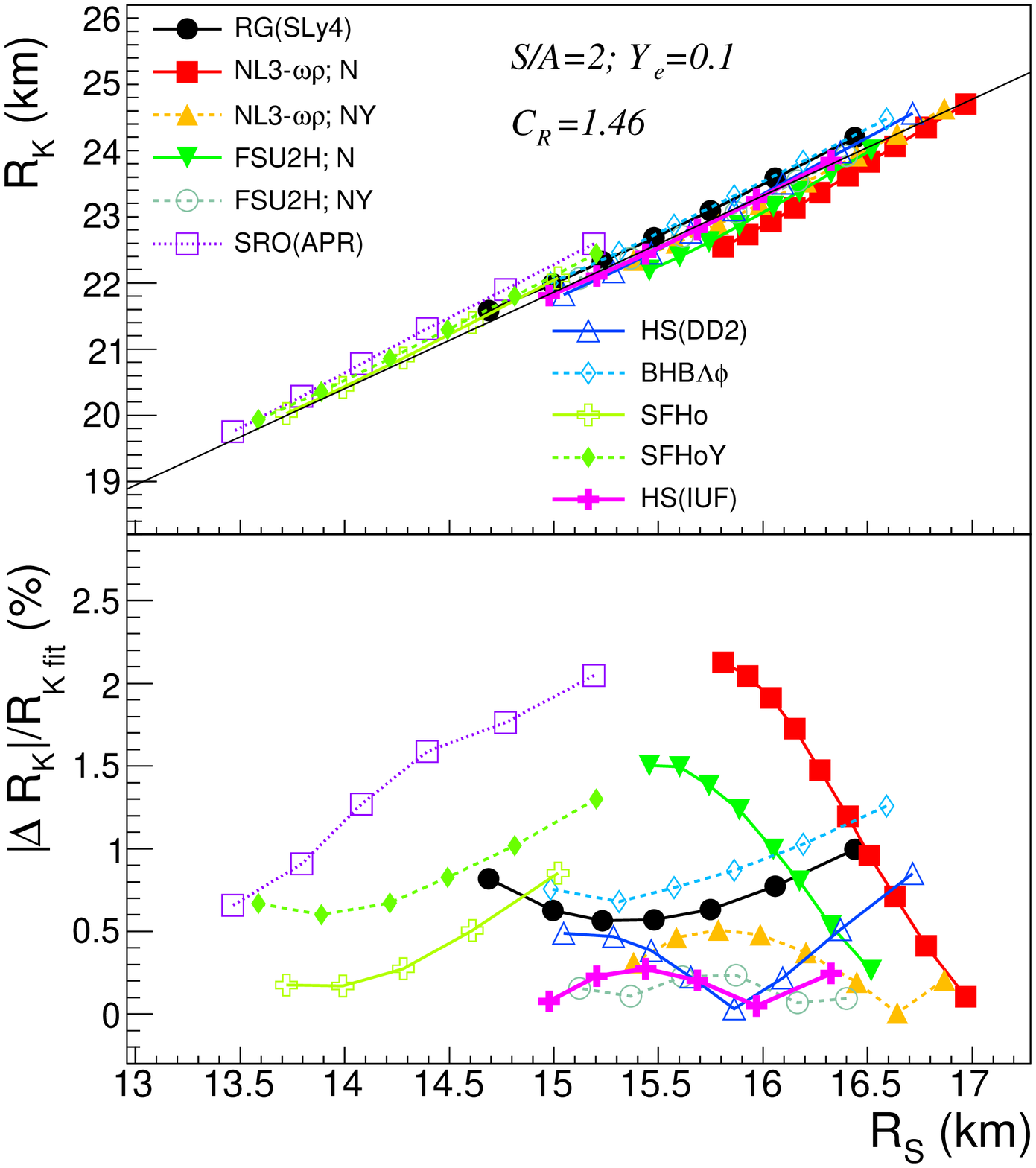}
 \end{center}
  \caption{Equatorial circumferential radius of the Keplerian
    configuration $R_K$ vs. radius of the non-rotating configuration
    $R_S$ for the same mass (top panels) and relative residual errors
    with respect to the fit employing Eq.~\eqref{eq:RK} (bottom panels).  The
    value of the fit parameter $C_R$ is mentioned in the top
    panels and result of Eq.~\eqref{eq:RK} using this value is shown by a 
    solid line.  Left and right panels corresponds to cold stars
    and, respectively, hot stars with $S/A=2$ and $Y_e=0.1$.  The
    results are shown for a set of EoS models as indicated by the
    labels.}
  \label{fig:RK-RNS}
\end{figure*}
\begin{figure*}
  \begin{center}
    \includegraphics[angle=0, width=0.45\linewidth]{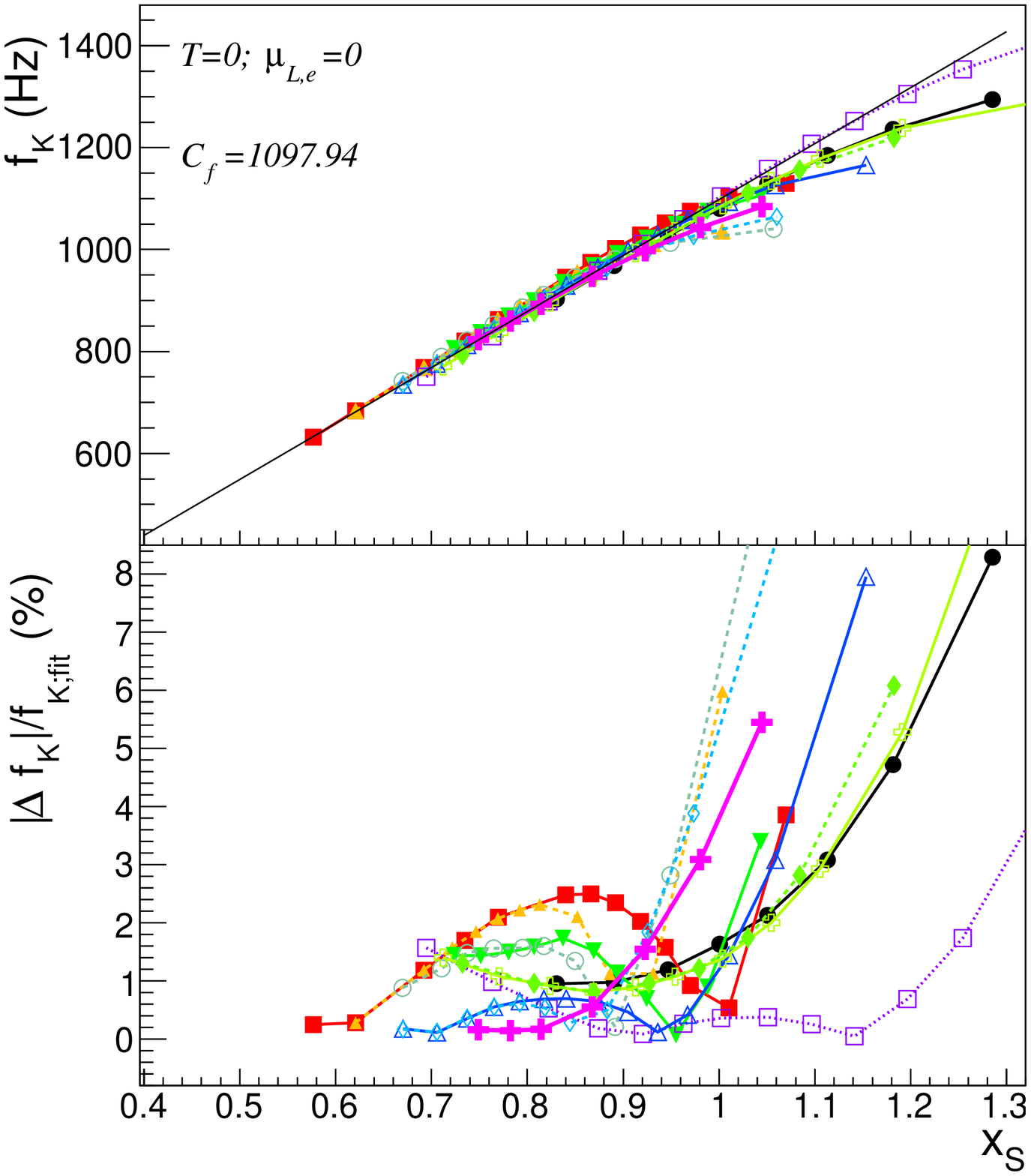}
    \includegraphics[angle=0, width=0.45\linewidth]{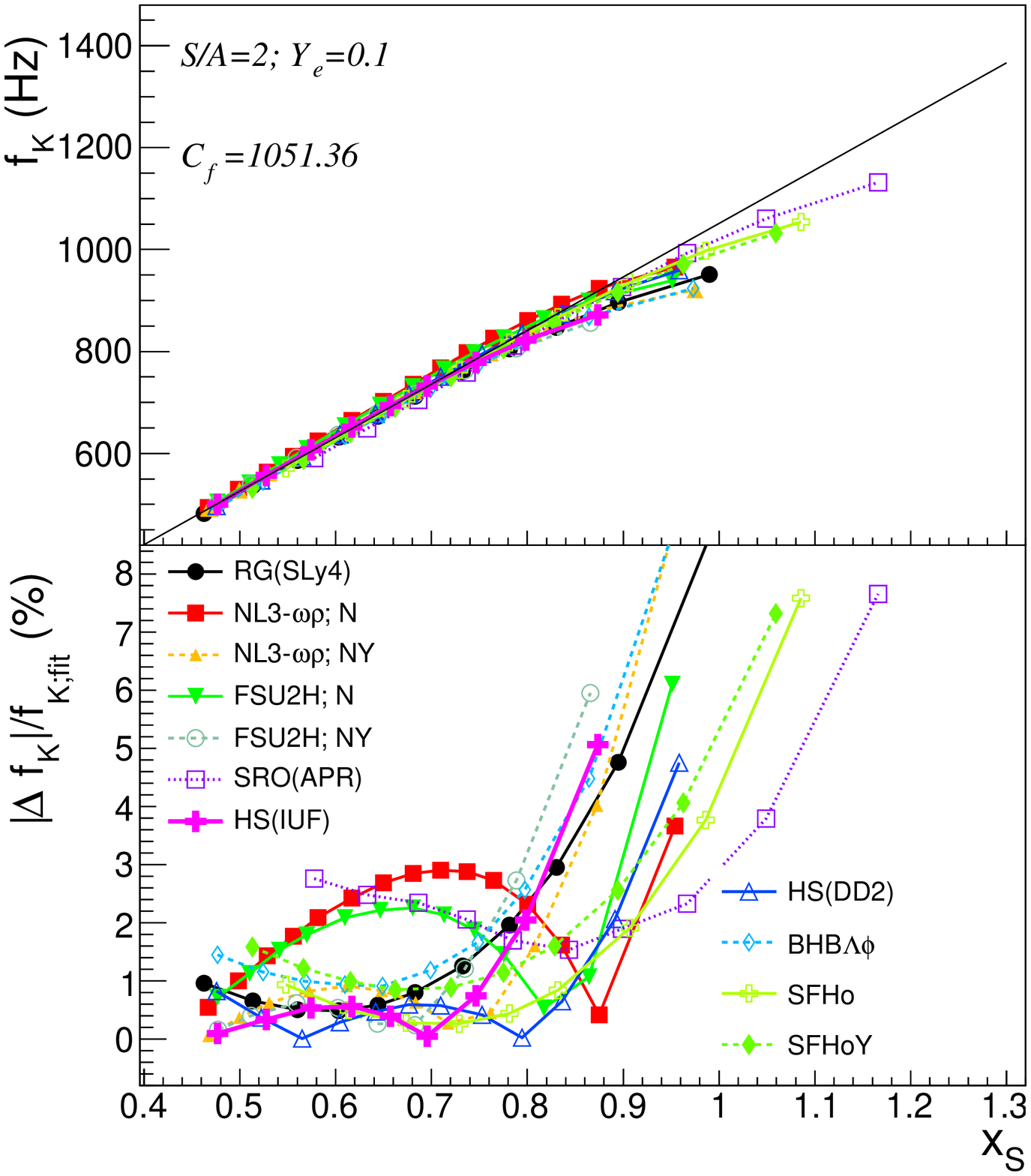}
 \end{center}
 \caption{Rotation frequency at the Kepler limit $f_{K}$ as a
     function of the parameter $x$ corresponding to a static
     configuration with the same mass (top panels) and relative
     residual errors with respect to the fit employing
     Eq.~\eqref{eq:fK} (bottom panels).  The value of the fit
     parameter $C_f$ is mentioned in the top panels, Eq.~\eqref{eq:fK}
     using this value is indicated by a solid line.  Left and right
     panels corresponds to cold stars and, respectively, hot stars
     with $S/A=2$ and $Y_e=0.1$.  The results are shown for a set of
     EoS models as indicated by the labels.}
 \label{fig:fK-xS}
\end{figure*}
Eqs. \eqref{eq:RK} and \eqref{eq:fK} are plotted in
Figs. \ref{fig:RK-RNS} and \ref{fig:fK-xS} for our
collection of eleven EoS. The same for a cold star as well as for
stars with ($S/A=2$, $Y_e=0.1$) are also plotted.  It can be seen that
the relation \eqref{eq:RK} holds, but the proportionality constant
$C_R$ slightly depends on the EoS for finite $S/A$.
The relation (\ref{eq:fK}) is confirmed too. The observed deviations
occur only for $M_S \gtrsim 0.7$-$0.8 \mmax{S}$, in agreement with previous
findings~\cite{Haensel_AA_2009}. We thus find again that the different
thermodynamic conditions lead to different values for the
proportionality coefficients in Eqs.~\eqref{eq:RK} and \eqref{eq:fK},
but the linear relationships remain well fulfilled.

\subsection{Relations between global parameters of the maximum mass
  configuration at the Kepler limit}
\label{ssec:KKmax}

\begin{figure}
  \begin{center}
    \includegraphics[angle=0, width=0.99\columnwidth]{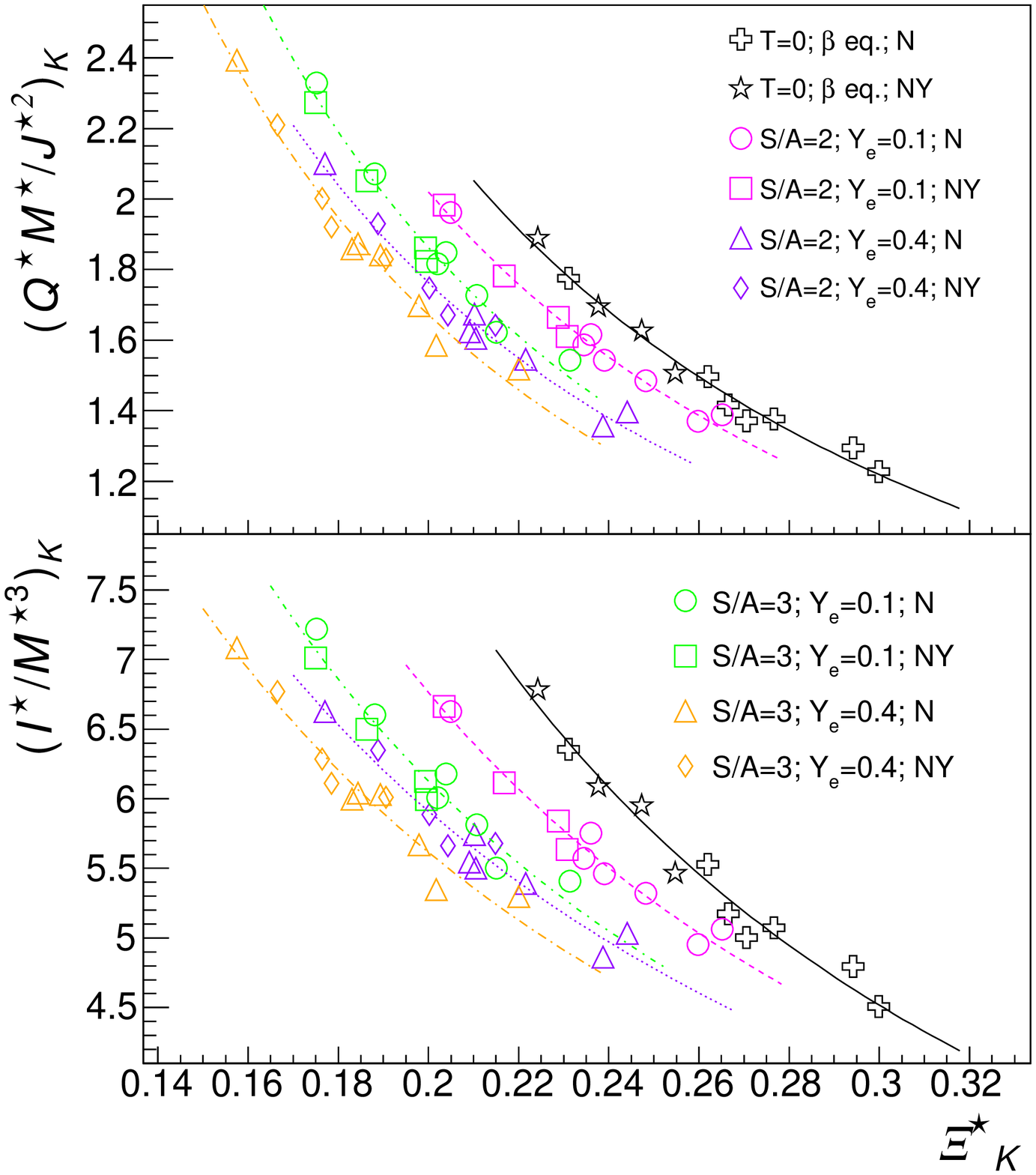}
  \end{center}
  \caption{Relations between global properties of maximum mass
    configurations at the Kepler limit: normalized moment of inertia
    $\bar I$ as function of the star's compactness (bottom) and
    normalized quadrupole moment $\bar Q$ as function of compactness
    (top).  Compactness is here defined with the equatorial radius.
    The results correspond to eleven different EoS models and various
    thermodynamic conditions as indicated in the legend. 
    The lines correspond to Eqs. \eqref{eq:IbarC} and \eqref{eq:QbarC},
    respectively; the values of the fitting parameters are provided in
      Tables \ref{tab:fitparams_IC} and \ref{tab:fitparams_QC}. 
  }
  \label{fig:IQC_K}
\end{figure}
\begin{table}
  \begin{tabular}{|l|c|c|}
    \hline 
    Thermo. cond. &  $a_1$ & $a_2$ \\ \hline
    $T=0$, $\beta$-eq. & 0.9398 (0.1093) & 0.1246 (0.0277)  \\
    $S/A=2$, $Y_e=0.1$ & 1.1632 (0.0788) & 0.0379 (0.0180) \\
    $S/A=2$, $Y_e=0.4$ & 1.2474 (0.0867) & -0.0130 (0.0178) \\
    $S/A=3$, $Y_e=0.1$ & 1.1458 (0.0976) & 0.0159 (0.0190) \\
    $S/A=3$, $Y_e=0.4$ & 1.1777 (0.0844) & -0.0109 (0.0153)  \\
   \hline   
  \end{tabular}
  \caption{Fitting parameters entering Eq. \eqref{eq:IbarC},
    and their standard errors (in parenthesis),
    under different thermodynamic conditions specified
    in the first column.  }
  \label{tab:fitparams_IC}
\end{table}

\begin{table}
  \begin{tabular}{|l|c|c|}
    \hline 
    Thermo. cond. &  $b_1$ & $b_2$ \\ \hline
    $T=0$, $\beta$-eq. & 0.2129 (0.0248) & 0.0458 (0.0063)  \\
    $S/A=2$, $Y_e=0.1$ & 0.2148 (0.0192) & 0.0378 (0.0044) \\
    $S/A=2$, $Y_e=0.4$ & 0.2228 (0.0251) & 0.0259 (0.0052) \\
    $S/A=3$, $Y_e=0.1$ & 0.1749 (0.0272) & 0.0395 (0.0053) \\
    $S/A=3$, $Y_e=0.4$ & 0.1877 (0.0270) & 0.0293 (0.0049) \\
   \hline   
  \end{tabular}
  \caption{Fitting parameters entering Eq. \eqref{eq:QbarC},
    and their standard errors (in parenthesis),
    under different thermodynamic conditions specified
    in the first column.  }
  \label{tab:fitparams_QC}
\end{table}
For cold compact stars in $\beta$-equilibrium numerous other universal
relations between global properties have been found, notably the 
so-called ``I-Love-Q'' relations~\cite{Yagi_Science_2013,Yagi:2014qua}
between the moment of inertia ($I$), the tidal deformability
($\lambda$), and the quadrupole moment ($Q$). In this context,
different relations expressing global properties in terms of the
star's compactness $\Xi$ have received much attention,
too~\cite{Ravenhall_ApJ_1994,Lattimer_ApJ_2005,Maselli_PRD_2013,Breu_2016}.

Here, we will consider as an example two such relations and investigate
whether they hold for rapidly rotating hot compact stars. These are $\bar I =
I/M^3$ and $\bar Q = Q M/J^2$, with $J$ standing for the angular
momentum, expressed as polynomials of $\Xi^{-1}$
\begin{eqnarray}
  \bar I &=& a_1 \Xi^{-1} + a_2 \Xi^{-2}~,  
  \label{eq:IbarC} \\
  \bar Q &=& b_1 \Xi^{-1} + b_2 \Xi^{-2}~.
  \label{eq:QbarC}   
\end{eqnarray}
Slightly different polynomial expressions of $\bar I$ and $\bar Q$
  in terms of $\Xi^{-1}$ have been previously proposed in
  ~\cite{Breu_2016}, who have also shown that they are universal for
  rigidly and slowly rotating cold, $\beta$-equilibrated stars. In
Ref.~\cite{Raduta_MNRAS_2020} these relations were shown to be
universal also for hot stars, as long as the same pair of constant
$S/A$ and $Y_e/Y_L$ is considered.

More specifically, we will investigate the behavior of the different
quantities taken for the maximum mass Keplerian configuration, {\it
  i.e.}, we study $\bar{I}^{\star}_K$ and $\bar{Q}^{\star}_K$ as a
function of $\Xi^\star_K =\mmax{K}/\rmax{K}$. Note that because of
rotational stretching of the star, the equatorial and polar radii are
different; we recall that $\rmax{K}$ refers to the equatorial
circumferential one. Fig. \ref{fig:IQC_K} depicts these
relationships. Each symbol indicates a particular EoS model and the
different colors differentiate different thermodynamic conditions
among $S/A=2,3$ and $Y_e=0.1,0.4$.  Results for cold stars in
  $\beta$-equilibrium are shown by black symbols for comparison.
Results of fits using Eqs. \eqref{eq:IbarC} and \eqref{eq:QbarC} are
illustrated with lines in Fig. \eqref{fig:IQC_K}; values of the
  fitting parameters entering eqs. \eqref{eq:IbarC} and
  \eqref{eq:QbarC} are provided in Tables \ref{tab:fitparams_IC} and
  \ref{tab:fitparams_QC}. 
These fits reproduce the exact results with good accuracy; the
reduced $\chi^2$-values are of the order of 
  $10^{-3}(10^{-2})$ for $\bar Q^{\star}$ vs. $\Xi^{\star}$ ($\bar I$
vs. $\Xi^{\star}$) are and slightly increasing with $S/A$.  Although
some scattering is seen in Fig.~\ref{fig:IQC_K}, a functional form
similar to the one obtained for slowly rotating stars applies
reasonably well to the maximum mass configuration at the Kepler limit,
too, and universality is again reasonably well fulfilled. However,
the relative displacement of points corresponding to a given
combination of entropy and electron fraction indicates that the values
of the parameters $a_i,b_i$ entering Eqs.~\eqref{eq:IbarC}, and
\eqref{eq:QbarC} depend on thermodynamic conditions, as expected.


\section{Maximum mass of rigidly rotating hot stars}
\label{sec:mkepler}

As well-known, for cold compact stars the value of $\mmax{K}$, is
$20\%$ larger than $\mmax{\mathrm{TOV}}$, independent of the
EoS~\cite{Cook1994,Lasota_ApJ_1996,Breu_2016}. As seen in the previous
section, the value of $\cmax{M} \approx 1.2$ is, however, only valid
if both, $\mmax{\mathrm{TOV}}$ and $\mmax{K}$ are computed for cold, $\beta$-equilibrated,
stars.  The assumption of a cold star fails for the merger remnant, as
the EoS obtains significant thermal corrections and a hot star
potentially out of $\beta$-equilibrium should be considered for
$\mmax{K}$, as has been argued in the case of GW170817
event~\cite{Margalit_17,Ruiz2018,Rezzolla_2018,Shibata_2019a}.  The purpose of this section is to investigate the relation between
$\mmax{K}$ for various thermodynamic conditions and the cold
$\mmax{\rm TOV}$ to verify to which extent thermal and out of
equilibrium effects can change the estimated value of
$\mmax{\rm TOV}$.

\begin{figure}
     \includegraphics[width=\columnwidth]{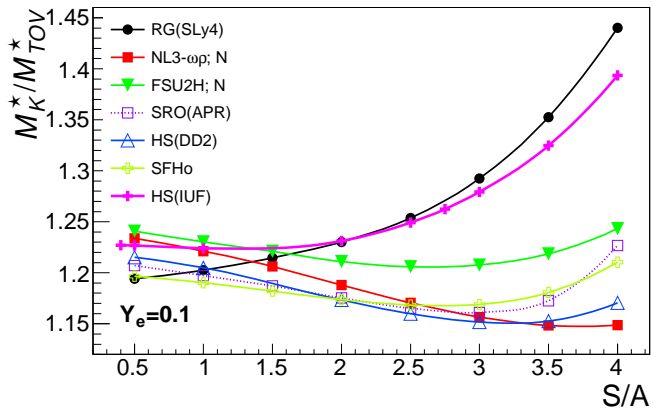}
     \caption{Dependence of $\cmax{M}$ [see Eq.~\eqref{eq:KTOV}] on
       entropy per baryon $S/A$ for fixed $Y_e=0.1$ and for various
       nucleonic EoS as labeled.  }
     \label{fig:4}
\end{figure}

What are the effects of finite-temperature EoS on the maximum masses
of a static and a rapidly rotating star, respectively? First,
compact stars expand due to thermal effects
(e.g.~\cite{Sumiyoshi_AASS_1999,Raduta_MNRAS_2020}), therefore a
same-mass hot star will have a larger radius than its cold
counterpart.  Consequently, the larger centrifugal force acting on
particles on the stellar surface will be larger and, therefore, the
Keplerian limit will be achieved for smaller frequencies, which will
result in smaller masses at the Kepler limit.  Second, the thermal
pressure adds to the degeneracy pressure which means that a larger
mass can be supported against the gravitational pull. Thus we see that
there is an interplay between two competing effects.
Fig.~\ref{fig:4} shows the variation of $\cmax{M}$ with $S/A$
for different purely nucleonic EoS and a constant electron fraction of
$Y_e$ = 0.1. The value of the Keplerian maximum mass $\mmax{K}$ is
normalized to that of the TOV maximum mass $\mmax{\mathrm{TOV}}$
computed for a cold star.  An inspection of Fig.~\ref{fig:4} shows
that one EoS model (RG(SLy4)) manifests a
  monotonic increase of $\cmax{M}$ over the considered $S/A$ range
  while the remaining six models show a non-monotonic behavior; the
  position of the minimum value of $\cmax{M}$ for the latter category
  of models is situated in the domain $1 \leq S/A \leq 3.5$.  This
variety of behaviors is associated with the interplay between the
effects of the increase of the pressure due to the thermal
contribution and expansion of the star with temperature and the
associated reduction of the Keplerian frequency. The first effect
increases $\cmax{M}$, whereas the second one decreases it.
In addition to the two factors described above, $\cmax{M}$ is expected
to depend on the composition of matter as well.  The reason is that
different compositions and electron fractions were shown to influence
the maximum mass and the star's
compactness~\cite{Pons_1999,Marques_PRC_2017,Raduta_MNRAS_2020}, too.

\begin{figure}[t]
    \includegraphics[scale=0.4]{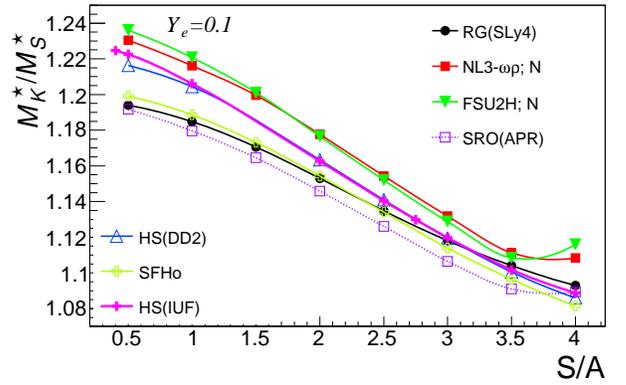}
    \caption{Same as in Fig. \ref{fig:4} except that the
      normalization is done by the maximum gravitational mass of the
      non-rotating star with the same $S/A$ and $Y_e$ values.
}
    \label{fig:norm}
\end{figure}

To disentangle the different effects discussed above,
Fig.~\ref{fig:norm} shows $\mmax{K}$ this time normalized to the
maximum mass of a non-rotating configuration with the same values of
$S/A$ and $Y_e$ (instead of the non-rotating TOV mass of a cold star). In
this way, we eliminate the thermal and $Y_e$-dependence and we observe
the change in $\cmax{M}$ due entirely to the expansion of the
star. Indeed, the masses in Fig.~\ref{fig:norm} are observed to almost
linearly decrease with $S/A$ and increasing radii as expected.  For
completeness, we reproduce in Table~\ref{tab:dd2} as an example the
results in the case of the HS(DD2) EoS. The compactness is given here
for the non-rotating configuration as an indication for the expansion
of the star with increasing entropy.
We thus confirm the earlier expectation born out from the analysis of
Fig.~\ref{fig:4}, namely that for low entropies the variation in the
mass is controlled predominantly by the expansion. As the entropy
increases, however, the thermal effects lead to a substantial increase
in mass and outweigh the effect due to the growth in radius and thus
reduced compactness.
\begin{table}
  \begin{tabular}{|c|c|c|c|c|c|s}
    \hline 
    $S/A$ & $\mmax{K}$ & $\mmax{B}$ & $\rmax{K}$ & $\Xi^{\star}_S$ & $n_B^{(c)}$   \\
     ($k_B$)& ($M_\odot$)& ($M_\odot$) & (km) & & ($\mathrm{fm}^{-3}$) \\
    \hline
    1 & 2.92 & 3.44 & 16.0 & 0.27 & 0.72 \\
    2 & 2.84 & 3.27 & 17.1 & 0.25 & 0.72 \\
    3 & 2.79 & 3.09 & 19.7 & 0.21 & 0.65 \\
    4 & 2.84 & 3.01 & 26.3 & 0.16 & 0.46 \\
   \hline   
  \end{tabular}
  \caption{Dependence on $S/A$ of some global properties of the maximum mass
    configuration of stars at Kepler limit for the HS(DD2)
    EoS~\cite{HS,Typel_PRC_2010} and for fixed $Y_e=0.1$.  Listed are
    gravitational and baryonic masses, equatorial circumferential
    radius, compactness of the non-rotating configuration and central
    baryonic number density.  }
  \label{tab:dd2}
\end{table}

\begin{figure}[t] 
     \includegraphics[scale=0.4]{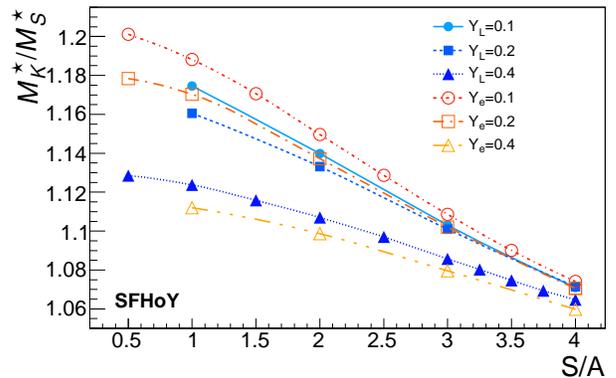}
     \caption{The same dependence as in Fig.~\ref{fig:norm}, for three
       cases of constant electron fraction $Y_e$ and constant lepton
       fraction $Y_L$ and one specific EoS model, 
       SFHoY~\cite{Fortin_PASA_2018}.  }
     \label{fig:yeyl}
\end{figure}

Up to now, we have investigated configurations with a particular value
of constant electron fraction, $Y_e = 0.1$. As discussed above, the
value of the electron fraction influences maximum masses and radii
and thus our results. Also, at the center of the merger
remnant, neutrinos are trapped at least during early
post-merger~\cite{Endrizzi2019} such that a related question is to
which extent choosing constant electron or constant lepton fraction
$Y_L$ changes our findings. To examine the dependence on
$Y_e$ and $Y_L$, we show in Fig.~\ref{fig:yeyl} the maximum masses at
Kepler frequency normalized to the non-rotating maximum mass as
function of $S/A$ for different values of constant $Y_e$ and
$Y_L$. The SFHoY EoS model~\cite{Fortin_PASA_2018} has been chosen for
that purpose, we have checked that other EoS models behave
qualitatively similarly.

First, since neutrinos themselves only contribute weakly to the EoS at
high density and therefore only have a very small impact on maximum
masses, we observe that the main difference between choosing $Y_e$ or
$Y_L$ arises from the fact that the electron fraction is equal to the
hadronic charge fraction $Y_Q$, whereas due to the presence of
neutrinos $Y_L \neq Y_Q$. This shift in $Y_Q$ induces a different
behavior of the hadronic part of the EoS which is well visible in the
maximum masses. This implies, too, that for our study it is sufficient
to vary either $Y_e$ or $Y_L$ if the range is chosen large
enough. Second, since a higher electron/lepton fraction increases the
star's radius, the Kepler frequency is lower
and the supported mass, too. Thus the ratio of the Kepler maximum mass
$\mmax{K}$ and the static one $\mmax{S}$ decreases with increasing
$Y_e/Y_L$ with the most pronounced reduction observed at low entropies
per baryon, where thermal effects are small. A related question is
whether the presence of muons would change our results. It is obvious
that in equilibrium, for the thermodynamic conditions considered here,
charged muons will be abundant. In contrast to core-collapse
supernovae, where there are no muons in the progenitor star and
complete equilibrium has to be reached by dynamical reactions (see
e.g. \cite{Bollig_PRL_2017}), the two neutron stars before merger
contain already muons, such that the merger remnant should indeed
contain a non-negligible fraction of muons. The EoS itself is,
however, still dominated by the hadronic part, such that again the
influence of muons on our results would manifest itself only by a
potential shift in the hadronic charge fraction since in the presence
of charged muons we have $Y_Q = Y_e + Y_\mu$.  In the following
discussion we choose $Y_e = 0.1$, which should be close to the
conditions in the central part of the merger remnant, see
e.g.~Ref.\cite{Perego2019}, keeping in mind that, if the electron
fraction in the merger remnant is higher, then $\cmax{M}$ is reduced.

\subsection{Comparison between nucleonic and hyperonic equations of state}
So far, when selecting the EoS of dense matter, we assumed that
neutron star matter contains nucleons and leptons.  At densities
exceeding several times the nuclear saturation density, non-nucleonic
degrees of freedom, such as hyperons, meson-condensates, and even
quark matter may appear~\cite{Glend2000}. Below we explore the effect
of different compositions on the observables discussed by comparing
the results for purely nucleonic EoS with those obtained in the models
allowing for the presence of hyperons.  In the present context, the
focus will be on the changes in the composition of matter at finite
temperature favoring the onset of hyperons~\cite
{Oertel_PRC_2012,Oertel_EPJA_2016}, which is expected to change the
value of $\cmax{M}$ at high entropies.

\begin{figure*}[t]
    \includegraphics[width=0.8\textwidth]{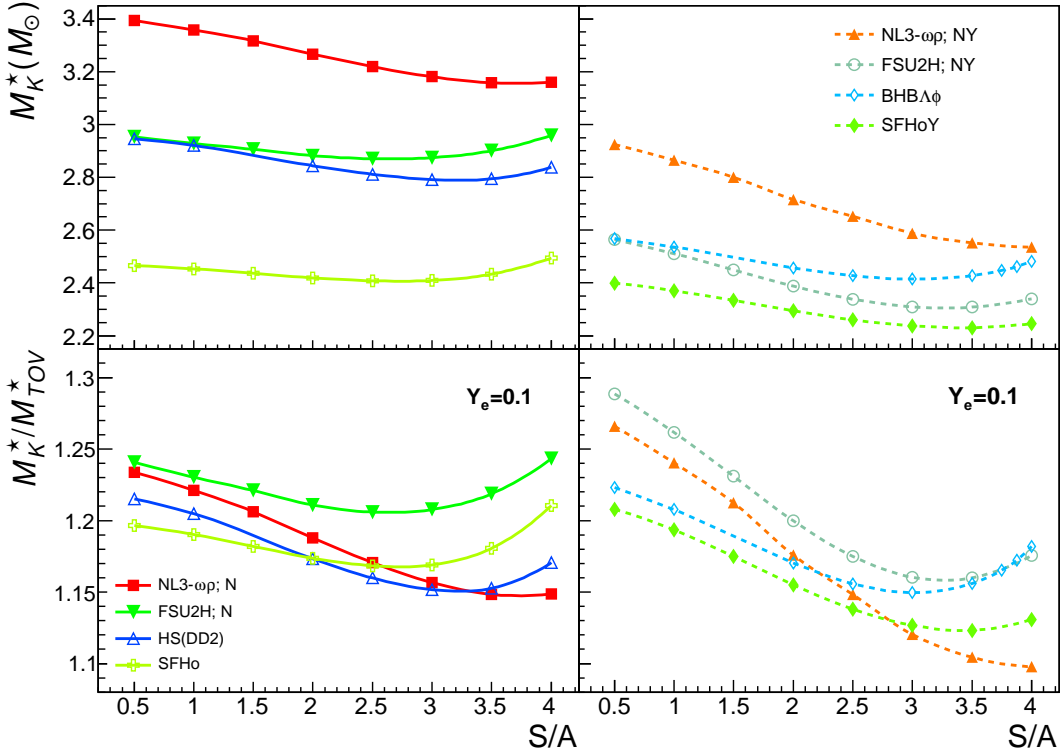}
 \caption{Dependence of $\mmax{K}$ (upper panels) and
   $\mmax{K}/\mmax{\mathrm{TOV}}$ (lower panels) on entropy per baryon
   (see also Fig. \ref{fig:4}). The left two panels correspond to
   nucleonic EoS, the right two panels to EoS which allow for
   hyperons. }
 \label{fig:hyperons}
 \end{figure*}
Fig.~\ref{fig:hyperons} depicts this comparison in detail. The bottom
panels display $\cmax{M}$ vs $S/A$ for four different EoS models and
their hyperonic counterparts. Although qualitatively the behavior for
all the EoS models is the same, a quantitative difference exists
between the purely nucleonic models and those with an admixture of
hyperons.  More precisely, for low $S/A$-values the hyperonic models
start with higher values of the ratio $\cmax{K}/\cmax{\mathrm{TOV}}$
and manifest a much stronger decrease of $\cmax{M}$ with $S/A$ than
the nucleonic models. To understand this, different effects have to be
considered. First, $\mmax{\mathrm{TOV}}$ for hyperonic models is much
smaller than for purely nucleonic models, since the presence of
hyperons softens the EoS. Second, this softening reduces the radius,
thus leading to a comparatively higher rotation frequency and
supported mass at Kepler limit, see the upper panels in
Fig.~\ref{fig:hyperons}. The increasing abundance of hyperons with
increasing $S/A$ leads to a less pronounced increase in the supported
mass due to thermal effects, which explains the more pronounced
decrease in $\mmax{K}/\mmax{\mathrm{TOV}}$ with $S/A$.

\section{Maximum TOV mass from GW170817}
\label{sec:GW170817}

The event GW170817 and its electromagnetic counterpart have been used
by several authors to place an upper limit on the value of the maximum
mass of static cold compact star configurations,
$\mmax{\mathrm{TOV}}$~\cite{Margalit_17,Ruiz2018,Rezzolla_2018,Shibata_2019a}.
In Ref.~\cite{Margalit_17} a selection of microscopic zero-temperature
EoS were approximated by piecewise polytropes and a maximum mass
$\mmax{\mathrm{TOV}}\le 2.17M_{\odot}$ was inferred from conservative
estimates of energy deposited into the short-gamma-ray burst and
kilonova ejecta.  Ref.~\cite{Rezzolla_2018} used the universal
relation between the mass of Keplerian configurations and static ones,
derived for cold compact stars, see Eq.~(\ref{eq:KTOV}), to place a
limit $\mmax{\mathrm{TOV}} \le 2.16^{+0.17}_{-0.15}M_{\odot}$
consistent with the one derived in Ref.~\cite{Margalit_17}.  A weaker
constraint $\mmax{\mathrm{TOV}} \le 2.3M_{\odot}$ was found in
Ref.~\cite{Shibata_2019a}, who used EoS based on ad-hoc piecewise
polytropic parameterization in combination with the angular momentum
conservation and numerical simulation to show that the merger remnant
at the onset of collapse to a black hole needs not to rotate
rapidly. Note that Ref.~\cite{Bauswein2020} derived a lower (instead
of an upper) limit on $\mmax{\mathrm{TOV}}$ from straightforward
numerical simulations, i.e. without use of universal relations, on the
basis of the fact that no prompt BH collapse has been observed.

The physical picture of the GW170817 event that underlies the
argumentation for placing the upper limit on $\mmax{\mathrm{TOV}}$ is
as follows~\cite{Rezzolla_2018,Shibata_2019a,Ruiz2018}. Initially, the merger leaves behind a hypermassive neutron
star (HMNS) which is differentially rotating. The HMNS star spins-down
by losses to gravitational and neutrino radiation, as well as mass ejection,
whereas the internal dissipation leads to
vanishing internal shears and eventually to uniform rotation. (The 
  magneto-dipole radiation due to the star's $B$-field can be
  neglected over the time-scales of 10 ms.) At
this stage, the star is in the region of stability of supramassive
neutron stars, which support themselves against gravitational collapse
due to uniform rotation. Subsequently, the star crosses the stability
line beyond which it is unstable to collapse. While in principle the
star may cross this line (which connects  $M_{\mathrm{TOV}}$ and
$M_{K}$) at any point, it has been argued that the dynamics of the
merger suggest that this crossing occurs in the vicinity of $\mmax{K}$
(see, however, Ref.~\cite{Shibata_2019a}, where this assumption has
been questioned and the resulting corrections to the limits have been
explored.  Since the slower rotation implies a larger maximum mass
limit, one should keep in mind that our estimate below may be relaxed
somewhat.)

The extraction of the upper limit circumvents the full dynamical study
and uses  the baryon mass conservation between the instances of
creation of HMNS in the merger (hereafter $t=0$) and the moment of
collapse to a black-hole ($t = t_c$), which reads
\begin{eqnarray}
  \label{eq:mass_conservation}
M_B (t_c,S/A,Y_e)=  M_B(0) - M_{\rm out } -M_{\rm ej} ,
\end{eqnarray}
where $M_{\rm out }$ refers to the baryon mass of the torus formed
around the black-hole, after the merger and $M_{\rm ej}$ refers to the
baryon mass of the ejecta. The left-hand-side of
\eqref{eq:mass_conservation} refers here to a {\it hot} supramassive
compact star at the instance of collapse, $ M_B(0)$ is the baryonic
mass of the HMNS formed in the merger at the initial time $t=0$.

As already mentioned in the introduction, the previous estimates of
the $\mmax{\mathrm{TOV}}$ were based on EoS of cold baryonic matter,
{\it i.e.} they do not account for the thermal pressure in the BNS
merger remnant and consider in particular the cold mass on the
left-hand side of Eq.~(\ref{eq:mass_conservation}). Numerical
simulations, however, show evidence that the BNS merger remnant is
heated up to temperatures of the order of tens of MeV. Thus, it is
necessary to carry out the analysis of the post-merger remnant taking
into account the finite-temperature EoS of baryonic matter.

In
the left-hand side of Eq.~\eqref{eq:mass_conservation} we now
substitute
\begin{eqnarray}
  M_B (t_c,S/A,Y_e) &=& \eta(S/A,Y_e) M (t_c,S/A,Y_e)\nonumber\\
  &=&  \eta(S/A,Y_e){\mmax{K}} (S/A,Y_e),
\end{eqnarray}
where the second equality assumes that at the instance of collapse the
star is rotating at the maximum of its rotational speed, consistent
with Ref.~\cite{Rezzolla_2018}, but see also
Ref.~\cite{Shibata_2019a}. The coefficient $\eta(S/A,Y_e) $ relates
the baryonic and gravitational masses of the hot compact star at the
instance of collapse and is an EoS-dependent quantity. On the
right-hand side of Eq.~(\ref{eq:mass_conservation}) we introduce the
same quantity for the newly formed object via $M_B(0) = \eta(0) M(0)$,
where $M(0) = 2.73M_{\odot}$~\cite{LIGO_Virgo2018b} is the
gravitational mass of the merger as measured during inspiral for the
GW170817 event, {\it i.e.} for cold stars.
Thus, the mass conservation
equation \eqref{eq:mass_conservation} can be rewritten as
\begin{eqnarray}
  \label{eq:mass_conservation2}
  {\mmax{K}} (S/A,Y_e)=  \frac{1}{\eta(S/A,Y_e)}\left[ \eta(0) M(0)-
  M_{\rm out } -M_{\rm ej} \right].
  \nonumber\\
\end{eqnarray}
It has been estimated from the analysis of GW170817 that
$M_{\rm ej} \simeq 0.03-0.05M_{\odot}$~\cite{Kasen_2017} and
$0.02\le M_{\rm out}\le 0.1 M_{\odot}$~\cite{Shibata_2019a}.  Taking
$M_{\rm out} = 0.06\pm 0.04M_{\odot}$ and
$M_{\rm ej}= 0.04\pm 0.01M_{\odot}$ we have
$M_{\rm out}+M_{\rm ej} = 0.1\pm 0.041$.  Thus, the masses on the
right-hand side are fixed within the given limits and the knowledge of
the two $\eta$-coefficients allows one to estimate the Keplerian
maximum mass of a hot supramassive compact star on the left-hand side
of Eq.~\eqref{eq:mass_conservation2}.

\begin{figure}[t]
  \begin{center}
    \includegraphics[width=0.45\textwidth]{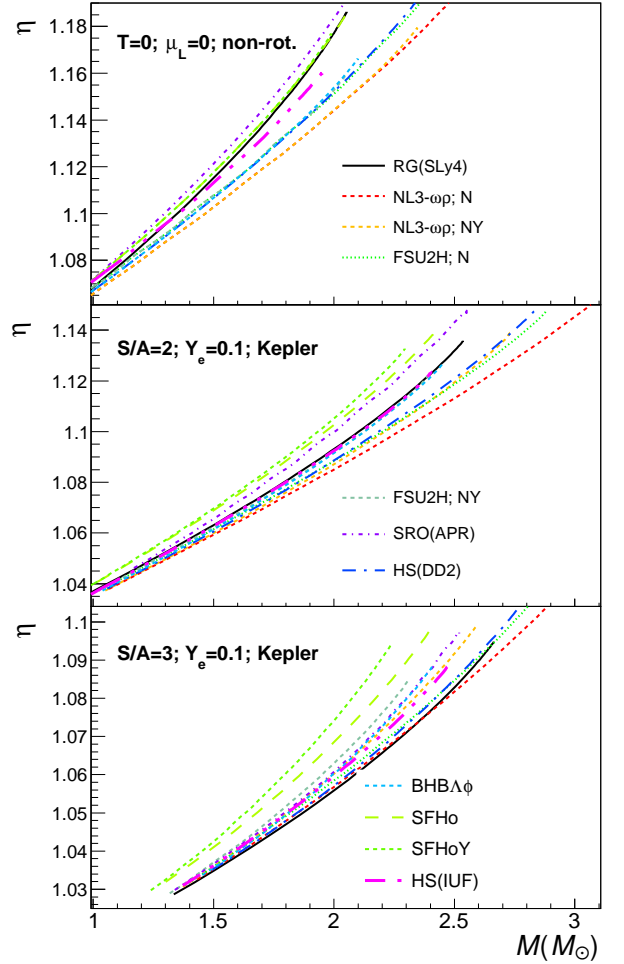}
 \end{center}
 \caption{Dependence of the $\eta$ parameter on the gravitational mass
   for spherically symmetric (non-rotating) stars at $T = 0$ and in
   $\beta$-equilibrium (top), for hot stars rotating at the Kepler
   limit for $S/A = 2$ (middle) and $S/A = 3$ (bottom panel) for fixed
   $Y_e = 0.1$.
}
 \label{fig:eta}
 \end{figure}

As illustrated in Fig.~\ref{fig:eta}, for cold compact stars based on
our collection of EoS we have $\eta(0) \simeq 1.120 \pm 0.002$ for
$M = 1.6 M_{\odot}$ and $\eta(0) \simeq 1.085 \pm 0.001$ for
$M = 1.2 M_{\odot}$. The chosen  values of gravitational masses bracket the
range $1.2\le M_{\odot} \le 1.6 $ from which the masses of two stars
are drawn to  add up to the gravitational mass
$2.73^{+0.04}_{-0.01} M_{\odot}$ of the merger remnant at
$t = 0$~\cite{LIGO_Virgo2018b}. For our estimates we adopt the value
$\eta(0) \simeq 1.1004^{+0.0014}_{-0.0003}$ leading to
$M_B(0) = 3.00^{+0.05}_{-0.01}M_\odot$.  We extract values of
$\eta(S/A,Y_e)$ for two values of entropy as given in
Fig.~\ref{fig:eta} assuming that the star is rotating at the Keplerian
frequency. We then find that $\eta(2,0.1)\simeq 1.139\pm 0.004$ and
$\eta(3,0.1)\simeq 1.099\pm 0.003$. For the quantity
$(M_{\rm out} + M_{\rm ej})/\eta(S/A, Y_e)$ we obtain $0.087\pm 0.036$
and $0.091 \pm 0.037$ for $S/A = 2$ and 3 and $Y_e = 0.1$,
respectively.
Substituting the numerical values we find 
\begin{eqnarray}
  \label{eq:limit_hot_K}
  {\mmax{K}} (2,0.1)=2.55^{+0.06}_{-0.04}, \quad
  {\mmax{K}} (3,0.1)= 2.64^{+0.06}_{-0.04}. \nonumber\\
\end{eqnarray}
It was shown recently that several universal relations hold for hot,
{\it isentropic stars out of $\beta$-equilibrium}
~\cite{Raduta_MNRAS_2020}, if thermodynamic conditions in terms of
entropy per baryon and electron/lepton fraction are fixed. In
Section~\ref{sec:universal} we have extended these findings to relations
between stars rotating at Kepler frequency and non-rotating ones.  The
above limits can thus be used to set a limit on the {\it maximum mass
  of non-rotating hot compact stars}, using Eq.~\eqref{eq:MKsbye} and
fitting parameters in Table~\ref{tab:fitparams}. We find
\begin{eqnarray}
  \label{eq:limit_hot_L}
  {\mmax{S}} (2,0.1)=2.19^{+0.05}_{-0.03}, \quad
  {\mmax{S}} (3,0.1)= 2.36^{+0.05}_{-0.04}.\nonumber\\
\end{eqnarray}
We can also use the limits \eqref{eq:limit_hot_K} in
combination with the results shown in Fig.~\eqref{fig:4} to deduce an
upper limit on {\it the maximum mass of cold compact stars}. Let us
stress that in this case universality is lost, and $\cmax{M}$ assumes
values in a range $1.15 < \cmax{M} < 1.23$ ($S/A = 2$) and $1.10 <
\cmax{M} < 1.29$ ($S/A = 3$) for the eleven EoS models considered
here.
The average values $\cmax{M} = 1.19\pm 0.04$
for $S = 2$ and $\cmax{M} = 1.18\pm 0.11$ for $S = 3$ can now be
used to obtain, respectively,
\begin{eqnarray}
  \label{eq:limit_hot_TOV}
  \mmax{\mathrm{TOV}}  =2.15^{+0.09 + 0.16}_{-0.07 - 0.16}, \,
\mmax{\mathrm{TOV}} = 2.24^{+0.10+0.44}_{-0.07-0.44}.
\end{eqnarray}
In this last relation, the errors correspond to 2$\sigma$ standard
deviation.  Here and in the formulas for the masses above the error
propagation for the upper and lower limits was computed
independently. The first uncertainty thereby stems from the
  propagation of errors from $\mmax{K}$, whereas the second part
  indicates the EoS dependence in $\cmax{M}$. When comparing the
limits \eqref{eq:limit_hot_TOV} with those of previous
works~\cite{Rezzolla_2018,Shibata_2019a}, one should keep in mind that
we used a (recent) value for the mass $M(0)$, which is slightly lower
than the value of $2.74 M_\odot$~\cite{LIGO_Virgo2017c} used in these
studies. Our limits on $\mmax{\mathrm{TOV}}$ would have been higher
had we adopted the larger value of $M(0)$. It is seen that, if just
before collapse the supermassive neutron star has average entropy per
baryon $S/A = 3$, then the estimate of the TOV mass is significantly
relaxed compared to the bound placed in
Refs.~\cite{Margalit_17,Rezzolla_2018,Ruiz2018}. According to the
discussion in Sec.~\ref{sec:mkepler}, a higher electron fraction in
the merger remnant would further relax the bound on the TOV mass.
Please note that we have considered stars at constant entropy per
  baryon and constant electron fraction, whereas a realistic merger
  remnant shows in particular strong entropy
  gradients~\cite{Perego2019,Camelio2020}. The values of $S/A = 2$ and
  $3$ can be roughly taken as typical average values for the inner
  part of the merger remnant, thus most relevant for the mass. As is
  obvious from the difference in the results for $S/A = 2$ and $S/A =
  3$, the detailed entropy profile influences the final limit for
  $\mmax{\mathrm{TOV}}$. These profiles cannot be measured and the
  exact entropy distribution in the remnant depends on many
  parameters, among others the EoS. Including the uncertainty on the
  exact entropy profiles would considerably increase the global
  uncertainty and further relax the limits.

The limit we found is similar to the one in Ref.~\cite{Shibata_2019a}
but for a physically different reason. The last fact indicates that
lifting the assumption that the star rotates at the Keplerian
frequency would further loosen the bound on the TOV mass. Let us,
however, stress the fact that universality is lost when extracting the
cold TOV mass limits \eqref{eq:limit_hot_TOV} from the information on
the hot merger remnant, independent of the assumption about rotation
at collapse, {\it i.e.}, these final limits become EoS dependent.

\section{Summary and conclusions}
\label{sec:conclusions}

In this work, we have addressed two interrelated topics that rely on
the knowledge of finite temperature EoS of dense matter. First, we
have extended the universal relations, previously found for hot slowly
rotating compact stars, to rapidly rotating stars. In particular we
considered in detail the mass-shedding (Keplerian) limit.  Secondly,
we discussed an improvement of the previous maximum mass limits for
non-rotating compact stars obtained from the GW170817 event in the
scenario where the merger remnant is a hypermassive compact star that
collapses to a black hole upon crossing the neutral stability line as
a supramassive (uniformly rotating) compact star.

Our analysis was carried out using a variety of finite-temperature
EoS. The collection used includes relativistic density functional
theory based EoS with nucleonic degrees of freedom as well as EoS
models allowing for the presence of hyperons. These EoS satisfy the
astrophysical constraints on neutron stars and nuclear data (nuclear
binding energies, rms radii, etc).  As an alternative to the covariant
description, we used a non-relativistic model based on a Skyrme-type
functional and a parameterization of a microscopic model. In this
way, we were able to bracket the range of possible predictions for the
observables stemming from various EoS with different underlying
methods of modeling.

When considering universal relations, we followed the strategy of
Ref.~\cite{Raduta_MNRAS_2020} to search universality under the {\it
  same thermodynamical conditions,} meaning that we compare
observables of the same star or various rotating and non-rotating
configurations at the same fixed entropy per baryon $S/A$ and electron
fraction $Y_e$.  Specifically, we considered a class of relations
which connect the Keplerian configurations with their non-rotating
counterparts given by Eqs.~\eqref{eq:MKsbye}-\eqref{eq:fKmax}
generalizing the earlier zero-temperature studies to the
finite-temperature case. We find that these relations are universal
(in the sense of independence on the EoS) to good accuracy. Similarly,
finite-temperature universality propagates beyond zero-temperature
results for the relations connecting radii and frequencies of the same
mass Keplerian and non-rotating stars, see Eqs.~~\eqref{eq:RK} and
\eqref{eq:fK}.  Finally, we have verified (partially) the validity of
the $I$-Love-$Q$ relations by computing the first and the last
quantity of the triple, specifically, $\bar I = I/M^3$ and $\bar Q = Q
M/J^2$ for maximum-mass Keplerian configurations.  We find that
  the functional dependence of these quantities for the maximum mass
  configurations at the Kepler limit on the compactness of the star is
  similar to the one obtained for slowly rotating stars.

The relation between the maximum masses of non-rotating and Keplerian
sequences is an important link needed for placing limits on the
maximum mass of a cold, non-rotating star from studies of the
millisecond pulsars or gravitational wave analysis of binary neutron
star mergers.  We have explored this relation for finite-temperature
stars finding that there are two competing effects: one is the thermal
expansion of the star, which reduces the Kepler frequency and,
implicitly, the star's mass at this limit and the additional thermal
pressure which makes a star of a given mass more stable against
collapse. If the static and maximally rotating configurations are
taken at the same values of $S/A$ and $Y_e$, then we find universality
of the coefficient relating their masses, see Fig.~\ref{fig:norm}.

The second important application of our analysis concerns the upper
limit on the maximum mass of a non-rotating cold compact
star. Several works, using various methods and scenarios, claimed
that this maximum mass can be tightly constrained using the GW170817
event~\cite{Margalit_17,Rezzolla_2018,Ruiz2018,Shibata_2019a} to the
range $\mmax{\mathrm{TOV}} \le 2.17-2.3M_{\odot}$, where the upper
range in this limit arises when considering below-Keplerian rotations,
instead of Keplerian ones. We have improved on the previous analysis
by extracting the ratio of the baryonic to gravitational masses for
hot compact stars of given $S/A$ and $Y_e$ and applying this to the same
scenario.  Our central finding is that  the upper
limit on the maximum mass of static, cold neutron stars is
$$ 2.15^{+0.09+0.16}_{-0.07-0.16}\le \mmax{\mathrm{TOV}} \le
2.24^{+0.10+0.44}_{-0.07-0.44} $$ for a typical parameter range $2\le S/A\le 3$
and $Y_e = 0.1$ of the hot merger remnant.  Note that the large
  error in the case of $S=3$ is dominated by the non-universal
  behavior $C_M^\star$ as displayed in Fig.~\ref{fig:4}. We thus conclude that
accounting for the finite temperature of the merger remnant relaxes
the derived constraints on the maximum mass of the cold, static
compact star, obtained in
Refs.~\cite{Margalit_17,Rezzolla_2018,Ruiz2018,Shibata_2019a}. In
particular, universality is lost and the final number becomes EoS
dependent due to the EoS dependence of $\cmax{M}$. In case the
collapse to a black hole does not occur at the maximum possible mass
of supramassive compact stars~\cite{Shibata_2019a}, as we assumed
here, the upper limit will become less stringent.

\section*{ACKNOWLEDGMENTS}
We thank N. Stergioulas for useful comments on the manuscript. 
This work has been partially funded by the European COST Action
CA16214 PHAROS “The multi-messenger physics and astrophysics of
neutron stars”.  A. R. R. acknowledges support from UEFISCDI (Grant
No. PN-III-P4-ID-PCE-2020-0293).  The work of M.~O. has been supported
by the Observatoire de Paris through the action f\'ed\'eratrice
``PhyFog''.  A.~S. acknowledges the support by the Deutsche
Forschungsgemeinschaft (Grant No. SE 1836/5-1).  The authors
gratefully acknowledge the Italian Istituto Nazionale de Fisica
Nucleare (INFN), the French Centre National de la Recherche
Scientifique (CNRS) and the Netherlands Organization for Scientific
Research for the construction and operation of the Virgo detector and
the creation and support of the EGO consortium.

\begin{appendix}
  \section{Influence of the surface definition on results}
\label{sec:app}

In this appendix, we discuss the sensitivity of our results on the
density at which the surface is located. This is essential for
establishing the validity of our results and conclusions.  The
available data for most finite-temperature EoS models are limited to
temperatures above $T = 0.1$ MeV, such that for a range of entropy per
baryon, no solution for the EoS at very low densities can be found. In
practice, for the values of $S/A$ considered, many of the EoS models
used did not have solutions for densities below roughly
($~10^{-6}$-$10^{-7}$ $\mathrm{fm}^{-3}$). This calls for an
extrapolation of the required thermodynamic quantities from the
densities where solutions were available to lower
densities. Extrapolation of thermodynamic quantities
introduces an error in the EoS. To avoid the above-stated
extrapolation we define the surface of the star at $n_B = 10^{-5}$
$\mathrm{fm}^{-3}$ uniformly in our modelling. This surface definition
allows us to use the data provided for every EoS model in the parameter range
used in our calculations.
  
To gauge the amount by which the value of the maximum mass changes
with a variation of the location of the surface, we refer to the results
for $\mmax{K}$ in Section~\ref{sec:mkepler}. We verified that changing
the surface density from $n_B = 10^{-7}$ $\mathrm{fm}^{-3}$ to
$10^{-5}$ $\mathrm{fm}^{-3}$ resulted in a change of the value of
$\mmax{K}$ only in the third decimal. The extrapolation has thereby
been performed assuming linear dependencies of $\log \varepsilon$ and
$\log p$ on $\log n_B$ with parameters calculated over the densities
covering the lowest available data, $10^{-5} \leq n_B \leq
10^{-4}$ fm$^{-3}$. The small change in $\mmax{K}$ can be understood from
the fact that the maximum mass is sensitive only to the high-density
physics.

To quantify the uncertainties on the results in
Section~\ref{sec:universal}, we consider again two different values of
the density at which we define the surface of the star. This time, in
addition to the value of $n_B=10^{-5}$ fm$^{-3}$ for the surface
density, we take a surface at $n_B=10^{-8}$ fm$^{-3}$, implying again
an extrapolation of EoS data over the domain for which data are not
available. We find that the extension of the surface by locating it at
a lower density diminishes the maximum rotation frequency and
that the higher the entropy per baryon the larger the induced
differences in all studied quantities. However, neither the Kepler
frequency, nor the quadrupole moment, the moment of inertia or the
values of the gravitational mass in the ranges discussed in
Section~\ref{sec:universal} are modified by more than a few {\it per mille}
upon varying the location of the surface.  We, therefore, conclude that
we can safely define the surface at $n_B = 10^{-5}$
$\mathrm{fm}^{-3}$.
\end{appendix}

\bibliography{FastRot.bib}

\begin{thebibliography}{118}%
\makeatletter
\providecommand \@ifxundefined [1]{%
 \@ifx{#1\undefined}
}%
\providecommand \@ifnum [1]{%
 \ifnum #1\expandafter \@firstoftwo
 \else \expandafter \@secondoftwo
 \fi
}%
\providecommand \@ifx [1]{%
 \ifx #1\expandafter \@firstoftwo
 \else \expandafter \@secondoftwo
 \fi
}%
\providecommand \natexlab [1]{#1}%
\providecommand \enquote  [1]{``#1''}%
\providecommand \bibnamefont  [1]{#1}%
\providecommand \bibfnamefont [1]{#1}%
\providecommand \citenamefont [1]{#1}%
\providecommand \href@noop [0]{\@secondoftwo}%
\providecommand \href [0]{\begingroup \@sanitize@url \@href}%
\providecommand \@href[1]{\@@startlink{#1}\@@href}%
\providecommand \@@href[1]{\endgroup#1\@@endlink}%
\providecommand \@sanitize@url [0]{\catcode `\\12\catcode `\$12\catcode
  `\&12\catcode `\#12\catcode `\^12\catcode `\_12\catcode `\%12\relax}%
\providecommand \@@startlink[1]{}%
\providecommand \@@endlink[0]{}%
\providecommand \url  [0]{\begingroup\@sanitize@url \@url }%
\providecommand \@url [1]{\endgroup\@href {#1}{\urlprefix }}%
\providecommand \urlprefix  [0]{URL }%
\providecommand \Eprint [0]{\href }%
\providecommand \doibase [0]{http://dx.doi.org/}%
\providecommand \selectlanguage [0]{\@gobble}%
\providecommand \bibinfo  [0]{\@secondoftwo}%
\providecommand \bibfield  [0]{\@secondoftwo}%
\providecommand \translation [1]{[#1]}%
\providecommand \BibitemOpen [0]{}%
\providecommand \bibitemStop [0]{}%
\providecommand \bibitemNoStop [0]{.\EOS\space}%
\providecommand \EOS [0]{\spacefactor3000\relax}%
\providecommand \BibitemShut  [1]{\csname bibitem#1\endcsname}%
\let\auto@bib@innerbib\@empty
\bibitem [{\citenamefont {Demorest}\ \emph {et~al.}(2010)\citenamefont
  {Demorest}, \citenamefont {Pennucci}, \citenamefont {Ransom}, \citenamefont
  {Roberts},\ and\ \citenamefont {Hessels}}]{Demorest:2010bx}%
  \BibitemOpen
  \bibfield  {author} {\bibinfo {author} {\bibfnamefont {P.}~\bibnamefont
  {Demorest}}, \bibinfo {author} {\bibfnamefont {T.}~\bibnamefont {Pennucci}},
  \bibinfo {author} {\bibfnamefont {S.}~\bibnamefont {Ransom}}, \bibinfo
  {author} {\bibfnamefont {M.}~\bibnamefont {Roberts}}, \ and\ \bibinfo
  {author} {\bibfnamefont {J.}~\bibnamefont {Hessels}},\ }\href {\doibase
  10.1038/nature09466} {\bibfield  {journal} {\bibinfo  {journal} {Nature}\
  }\textbf {\bibinfo {volume} {467}},\ \bibinfo {pages} {1081} (\bibinfo {year}
  {2010})},\ \Eprint {http://arxiv.org/abs/1010.5788} {arXiv:1010.5788
  [astro-ph.HE]} \BibitemShut {NoStop}%
\bibitem [{\citenamefont {Antoniadis}\ \emph {et~al.}(2013)\citenamefont
  {Antoniadis} \emph {et~al.}}]{Antoniadis:2013pzd}%
  \BibitemOpen
  \bibfield  {author} {\bibinfo {author} {\bibfnamefont {J.}~\bibnamefont
  {Antoniadis}} \emph {et~al.},\ }\href {\doibase 10.1126/science.1233232}
  {\bibfield  {journal} {\bibinfo  {journal} {Science}\ }\textbf {\bibinfo
  {volume} {340}},\ \bibinfo {pages} {6131} (\bibinfo {year} {2013})},\ \Eprint
  {http://arxiv.org/abs/1304.6875} {arXiv:1304.6875 [astro-ph.HE]} \BibitemShut
  {NoStop}%
\bibitem [{\citenamefont {Cromartie}\ \emph {et~al.}(2019)\citenamefont
  {Cromartie} \emph {et~al.}}]{Cromartie:2019kug}%
  \BibitemOpen
  \bibfield  {author} {\bibinfo {author} {\bibfnamefont {H.}~\bibnamefont
  {Cromartie}} \emph {et~al.} (\bibinfo {collaboration} {NANOGrav}),\ }\href
  {\doibase 10.1038/s41550-019-0880-2} {\bibfield  {journal} {\bibinfo
  {journal} {Nature Astron.}\ }\textbf {\bibinfo {volume} {4}},\ \bibinfo
  {pages} {72} (\bibinfo {year} {2019})},\ \Eprint
  {http://arxiv.org/abs/1904.06759} {arXiv:1904.06759 [astro-ph.HE]}
  \BibitemShut {NoStop}%
\bibitem [{\citenamefont {\"Ozel}\ and\ \citenamefont
  {Freire}(2016)}]{Ozel:2016oaf}%
  \BibitemOpen
  \bibfield  {author} {\bibinfo {author} {\bibfnamefont {F.}~\bibnamefont
  {\"Ozel}}\ and\ \bibinfo {author} {\bibfnamefont {P.}~\bibnamefont
  {Freire}},\ }\href {\doibase 10.1146/annurev-astro-081915-023322} {\bibfield
  {journal} {\bibinfo  {journal} {Ann. Rev. Astron. Astrophys.}\ }\textbf
  {\bibinfo {volume} {54}},\ \bibinfo {pages} {401} (\bibinfo {year} {2016})},\
  \Eprint {http://arxiv.org/abs/1603.02698} {arXiv:1603.02698 [astro-ph.HE]}
  \BibitemShut {NoStop}%
\bibitem [{\citenamefont {Watts}\ \emph {et~al.}(2015)\citenamefont {Watts}
  \emph {et~al.}}]{Watts:2014tja}%
  \BibitemOpen
  \bibfield  {author} {\bibinfo {author} {\bibfnamefont {A.}~\bibnamefont
  {Watts}} \emph {et~al.},\ }\bibfield  {booktitle} {\emph {\bibinfo
  {booktitle} {{Proceedings, Advancing Astrophysics with the Square Kilometre
  Array (AASKA14): Giardini Naxos, Italy, June 9-13, 2014}}},\ }\href {\doibase
  10.22323/1.215.0043} {\bibfield  {journal} {\bibinfo  {journal} {PoS}\
  }\textbf {\bibinfo {volume} {AASKA14}},\ \bibinfo {pages} {043} (\bibinfo
  {year} {2015})},\ \Eprint {http://arxiv.org/abs/1501.00042} {arXiv:1501.00042
  [astro-ph.SR]} \BibitemShut {NoStop}%
\bibitem [{\citenamefont {Watts}\ \emph {et~al.}(2016)\citenamefont {Watts}
  \emph {et~al.}}]{Watts:2016uzu}%
  \BibitemOpen
  \bibfield  {author} {\bibinfo {author} {\bibfnamefont {A.~L.}\ \bibnamefont
  {Watts}} \emph {et~al.},\ }\href {\doibase 10.1103/RevModPhys.88.021001}
  {\bibfield  {journal} {\bibinfo  {journal} {Rev. Mod. Phys.}\ }\textbf
  {\bibinfo {volume} {88}},\ \bibinfo {pages} {021001} (\bibinfo {year}
  {2016})},\ \Eprint {http://arxiv.org/abs/1602.01081} {arXiv:1602.01081
  [astro-ph.HE]} \BibitemShut {NoStop}%
\bibitem [{\citenamefont {Watts}\ \emph {et~al.}(2019)\citenamefont {Watts}
  \emph {et~al.}}]{Watts:2018iom}%
  \BibitemOpen
  \bibfield  {author} {\bibinfo {author} {\bibfnamefont {A.~L.}\ \bibnamefont
  {Watts}} \emph {et~al.},\ }\href {\doibase 10.1007/s11433-017-9188-4}
  {\bibfield  {journal} {\bibinfo  {journal} {Sci. China Phys. Mech. Astron.}\
  }\textbf {\bibinfo {volume} {62}},\ \bibinfo {pages} {29503} (\bibinfo {year}
  {2019})}\BibitemShut {NoStop}%
\bibitem [{\citenamefont {Riley}\ \emph {et~al.}(2019)\citenamefont {Riley}
  \emph {et~al.}}]{Riley_2019}%
  \BibitemOpen
  \bibfield  {author} {\bibinfo {author} {\bibfnamefont {T.~E.}\ \bibnamefont
  {Riley}} \emph {et~al.},\ }\href {\doibase 10.3847/2041-8213/ab481c}
  {\bibfield  {journal} {\bibinfo  {journal} {ApJ Lett.}\ }\textbf {\bibinfo
  {volume} {887}},\ \bibinfo {pages} {L21} (\bibinfo {year} {2019})},\ \Eprint
  {http://arxiv.org/abs/1912.05702} {arXiv:1912.05702 [astro-ph.HE]}
  \BibitemShut {NoStop}%
\bibitem [{\citenamefont {Miller}\ \emph {et~al.}(2019)\citenamefont {Miller}
  \emph {et~al.}}]{Miller_2019}%
  \BibitemOpen
  \bibfield  {author} {\bibinfo {author} {\bibfnamefont {M.~C.}\ \bibnamefont
  {Miller}} \emph {et~al.},\ }\href {\doibase 10.3847/2041-8213/ab50c5}
  {\bibfield  {journal} {\bibinfo  {journal} {ApJLett.}\ }\textbf {\bibinfo
  {volume} {887}},\ \bibinfo {pages} {L24} (\bibinfo {year} {2019})},\ \Eprint
  {http://arxiv.org/abs/1912.05705} {arXiv:1912.05705 [astro-ph.HE]}
  \BibitemShut {NoStop}%
\bibitem [{\citenamefont {Abbott}\ \emph
  {et~al.}(2017{\natexlab{a}})\citenamefont {Abbott} \emph
  {et~al.}}]{Abbott_2017}%
  \BibitemOpen
  \bibfield  {author} {\bibinfo {author} {\bibfnamefont {B.~P.}\ \bibnamefont
  {Abbott}} \emph {et~al.} (\bibinfo {collaboration} {LIGO Scientific
  Collaboration and Virgo Collaboration}),\ }\href {\doibase
  10.1103/PhysRevLett.119.161101} {\bibfield  {journal} {\bibinfo  {journal}
  {Phys. Rev. Lett.}\ }\textbf {\bibinfo {volume} {119}},\ \bibinfo {pages}
  {161101} (\bibinfo {year} {2017}{\natexlab{a}})}\BibitemShut {NoStop}%
\bibitem [{\citenamefont {Abbott}\ \emph
  {et~al.}(2020{\natexlab{a}})\citenamefont {Abbott} \emph
  {et~al.}}]{Abbott:2020uma}%
  \BibitemOpen
  \bibfield  {author} {\bibinfo {author} {\bibfnamefont {B.}~\bibnamefont
  {Abbott}} \emph {et~al.} (\bibinfo {collaboration} {LIGO Scientific,
  Virgo}),\ }\href {\doibase 10.3847/2041-8213/ab75f5} {\bibfield  {journal}
  {\bibinfo  {journal} {Astrophys. J. Lett.}\ }\textbf {\bibinfo {volume}
  {892}},\ \bibinfo {pages} {L3} (\bibinfo {year} {2020}{\natexlab{a}})},\
  \Eprint {http://arxiv.org/abs/2001.01761} {arXiv:2001.01761 [astro-ph.HE]}
  \BibitemShut {NoStop}%
\bibitem [{\citenamefont {Abbott}\ \emph
  {et~al.}(2017{\natexlab{b}})\citenamefont {Abbott}, \citenamefont {Abbott},
  \citenamefont {Abbott}, \citenamefont {Acernese}, \citenamefont {Ackley}
  \emph {et~al.}}]{LIGO_Virgo2017b}%
  \BibitemOpen
  \bibfield  {author} {\bibinfo {author} {\bibfnamefont {B.~P.}\ \bibnamefont
  {Abbott}}, \bibinfo {author} {\bibfnamefont {R.}~\bibnamefont {Abbott}},
  \bibinfo {author} {\bibfnamefont {T.~D.}\ \bibnamefont {Abbott}}, \bibinfo
  {author} {\bibfnamefont {F.}~\bibnamefont {Acernese}}, \bibinfo {author}
  {\bibfnamefont {K.}~\bibnamefont {Ackley}},  \emph {et~al.},\ }\href@noop {}
  {\bibfield  {journal} {\bibinfo  {journal} {ApJL}\ }\textbf {\bibinfo
  {volume} {848}},\ \bibinfo {pages} {L13} (\bibinfo {year}
  {2017}{\natexlab{b}})}\BibitemShut {NoStop}%
\bibitem [{\citenamefont {Abbott}\ \emph {et~al.}(2018)\citenamefont {Abbott},
  \citenamefont {Abbott}, \citenamefont {Abbott}, \citenamefont {Acernese},
  \citenamefont {Ackley} \emph {et~al.}}]{LIGO_Virgo2018a}%
  \BibitemOpen
  \bibfield  {author} {\bibinfo {author} {\bibfnamefont {B.~P.}\ \bibnamefont
  {Abbott}}, \bibinfo {author} {\bibfnamefont {R.}~\bibnamefont {Abbott}},
  \bibinfo {author} {\bibfnamefont {T.~D.}\ \bibnamefont {Abbott}}, \bibinfo
  {author} {\bibfnamefont {F.}~\bibnamefont {Acernese}}, \bibinfo {author}
  {\bibfnamefont {K.}~\bibnamefont {Ackley}},  \emph {et~al.} (\bibinfo
  {collaboration} {The LIGO Scientific Collaboration and the Virgo
  Collaboration}),\ }\href@noop {} {\bibfield  {journal} {\bibinfo  {journal}
  {PhRvL}\ }\textbf {\bibinfo {volume} {121}},\ \bibinfo {pages} {161101}
  (\bibinfo {year} {2018})}\BibitemShut {NoStop}%
\bibitem [{\citenamefont {Abbott}\ \emph
  {et~al.}(2019{\natexlab{a}})\citenamefont {Abbott}, \citenamefont {Abbott},
  \citenamefont {Abbott}, \citenamefont {Acernese}, \citenamefont {Ackley}
  \emph {et~al.}}]{LIGO_Virgo2018b}%
  \BibitemOpen
  \bibfield  {author} {\bibinfo {author} {\bibfnamefont {B.~P.}\ \bibnamefont
  {Abbott}}, \bibinfo {author} {\bibfnamefont {R.}~\bibnamefont {Abbott}},
  \bibinfo {author} {\bibfnamefont {T.~D.}\ \bibnamefont {Abbott}}, \bibinfo
  {author} {\bibfnamefont {F.}~\bibnamefont {Acernese}}, \bibinfo {author}
  {\bibfnamefont {K.}~\bibnamefont {Ackley}},  \emph {et~al.} (\bibinfo
  {collaboration} {LIGO Scientific Collaboration and Virgo Collaboration}),\
  }\href@noop {} {\bibfield  {journal} {\bibinfo  {journal} {PhRvX}\ }\textbf
  {\bibinfo {volume} {9}},\ \bibinfo {pages} {011001} (\bibinfo {year}
  {2019}{\natexlab{a}})}\BibitemShut {NoStop}%
\bibitem [{\citenamefont {Malik}\ \emph {et~al.}(2018)\citenamefont {Malik},
  \citenamefont {Alam}, \citenamefont {Fortin}, \citenamefont {Provid\^encia},
  \citenamefont {Agrawal}, \citenamefont {Jha}, \citenamefont {Kumar},\ and\
  \citenamefont {Patra}}]{Malik_18}%
  \BibitemOpen
  \bibfield  {author} {\bibinfo {author} {\bibfnamefont {T.}~\bibnamefont
  {Malik}}, \bibinfo {author} {\bibfnamefont {N.}~\bibnamefont {Alam}},
  \bibinfo {author} {\bibfnamefont {M.}~\bibnamefont {Fortin}}, \bibinfo
  {author} {\bibfnamefont {C.}~\bibnamefont {Provid\^encia}}, \bibinfo {author}
  {\bibfnamefont {B.}~\bibnamefont {Agrawal}}, \bibinfo {author} {\bibfnamefont
  {T.}~\bibnamefont {Jha}}, \bibinfo {author} {\bibfnamefont {B.}~\bibnamefont
  {Kumar}}, \ and\ \bibinfo {author} {\bibfnamefont {S.}~\bibnamefont
  {Patra}},\ }\href {\doibase 10.1103/PhysRevC.98.035804} {\bibfield  {journal}
  {\bibinfo  {journal} {Phys. Rev. C}\ }\textbf {\bibinfo {volume} {98}},\
  \bibinfo {pages} {035804} (\bibinfo {year} {2018})},\ \Eprint
  {http://arxiv.org/abs/1805.11963} {arXiv:1805.11963 [nucl-th]} \BibitemShut
  {NoStop}%
\bibitem [{\citenamefont {Paschalidis}\ \emph {et~al.}(2018)\citenamefont
  {Paschalidis}, \citenamefont {Yagi}, \citenamefont {Alvarez-Castillo},\ and\
  \citenamefont {Sedrakian}}]{Paschalidis2018}%
  \BibitemOpen
  \bibfield  {author} {\bibinfo {author} {\bibfnamefont {V.}~\bibnamefont
  {Paschalidis}}, \bibinfo {author} {\bibfnamefont {K.}~\bibnamefont {Yagi}},
  \bibinfo {author} {\bibfnamefont {D.~B.}\ \bibnamefont {Alvarez-Castillo},
  \bibfnamefont {David~Blaschke}}, \ and\ \bibinfo {author} {\bibfnamefont
  {A.}~\bibnamefont {Sedrakian}},\ }\href@noop {} {\bibfield  {journal}
  {\bibinfo  {journal} {PhRvD}\ }\textbf {\bibinfo {volume} {97}},\ \bibinfo
  {pages} {084038} (\bibinfo {year} {2018})}\BibitemShut {NoStop}%
\bibitem [{\citenamefont {Dexheimer}\ \emph {et~al.}(2019)\citenamefont
  {Dexheimer}, \citenamefont {de~Oliveira~Gomes}, \citenamefont {Schramm},\
  and\ \citenamefont {Pais}}]{Dexheimer_18}%
  \BibitemOpen
  \bibfield  {author} {\bibinfo {author} {\bibfnamefont {V.}~\bibnamefont
  {Dexheimer}}, \bibinfo {author} {\bibfnamefont {R.}~\bibnamefont
  {de~Oliveira~Gomes}}, \bibinfo {author} {\bibfnamefont {S.}~\bibnamefont
  {Schramm}}, \ and\ \bibinfo {author} {\bibfnamefont {H.}~\bibnamefont
  {Pais}},\ }\href {\doibase 10.1088/1361-6471/ab01f0} {\bibfield  {journal}
  {\bibinfo  {journal} {J. Phys. G}\ }\textbf {\bibinfo {volume} {46}},\
  \bibinfo {pages} {034002} (\bibinfo {year} {2019})},\ \Eprint
  {http://arxiv.org/abs/1810.06109} {arXiv:1810.06109 [nucl-th]} \BibitemShut
  {NoStop}%
\bibitem [{\citenamefont {Capano}\ \emph {et~al.}(2020)\citenamefont {Capano},
  \citenamefont {Tews}, \citenamefont {Brown}, \citenamefont {Margalit},
  \citenamefont {De}, \citenamefont {Kumar}, \citenamefont {Brown},
  \citenamefont {Krishnan},\ and\ \citenamefont {Reddy}}]{Capano_19}%
  \BibitemOpen
  \bibfield  {author} {\bibinfo {author} {\bibfnamefont {C.~D.}\ \bibnamefont
  {Capano}}, \bibinfo {author} {\bibfnamefont {I.}~\bibnamefont {Tews}},
  \bibinfo {author} {\bibfnamefont {S.~M.}\ \bibnamefont {Brown}}, \bibinfo
  {author} {\bibfnamefont {B.}~\bibnamefont {Margalit}}, \bibinfo {author}
  {\bibfnamefont {S.}~\bibnamefont {De}}, \bibinfo {author} {\bibfnamefont
  {S.}~\bibnamefont {Kumar}}, \bibinfo {author} {\bibfnamefont {D.~A.}\
  \bibnamefont {Brown}}, \bibinfo {author} {\bibfnamefont {B.}~\bibnamefont
  {Krishnan}}, \ and\ \bibinfo {author} {\bibfnamefont {S.}~\bibnamefont
  {Reddy}},\ }\href {\doibase 10.1038/s41550-020-1014-6} {\bibfield  {journal}
  {\bibinfo  {journal} {Nature Astron.}\ }\textbf {\bibinfo {volume} {4}},\
  \bibinfo {pages} {625} (\bibinfo {year} {2020})},\ \Eprint
  {http://arxiv.org/abs/1908.10352} {arXiv:1908.10352 [astro-ph.HE]}
  \BibitemShut {NoStop}%
\bibitem [{\citenamefont {Li}\ and\ \citenamefont
  {Sedrakian}(2019)}]{Li:2019tjx}%
  \BibitemOpen
  \bibfield  {author} {\bibinfo {author} {\bibfnamefont {J.~J.}\ \bibnamefont
  {Li}}\ and\ \bibinfo {author} {\bibfnamefont {A.}~\bibnamefont {Sedrakian}},\
  }\href {\doibase 10.3847/2041-8213/ab1090} {\bibfield  {journal} {\bibinfo
  {journal} {Astrophys. J. Lett.}\ }\textbf {\bibinfo {volume} {874}},\
  \bibinfo {pages} {L22} (\bibinfo {year} {2019})},\ \Eprint
  {http://arxiv.org/abs/1904.02006} {arXiv:1904.02006 [nucl-th]} \BibitemShut
  {NoStop}%
\bibitem [{\citenamefont {G\"uven}\ \emph {et~al.}(2020)\citenamefont
  {G\"uven}, \citenamefont {Bozkurt}, \citenamefont {Khan},\ and\ \citenamefont
  {Margueron}}]{Guven_20}%
  \BibitemOpen
  \bibfield  {author} {\bibinfo {author} {\bibfnamefont {H.}~\bibnamefont
  {G\"uven}}, \bibinfo {author} {\bibfnamefont {K.}~\bibnamefont {Bozkurt}},
  \bibinfo {author} {\bibfnamefont {E.}~\bibnamefont {Khan}}, \ and\ \bibinfo
  {author} {\bibfnamefont {J.}~\bibnamefont {Margueron}},\ }\href {\doibase
  10.1103/PhysRevC.102.015805} {\bibfield  {journal} {\bibinfo  {journal}
  {Phys. Rev. C}\ }\textbf {\bibinfo {volume} {102}},\ \bibinfo {pages}
  {015805} (\bibinfo {year} {2020})},\ \Eprint
  {http://arxiv.org/abs/2001.10259} {arXiv:2001.10259 [nucl-th]} \BibitemShut
  {NoStop}%
\bibitem [{\citenamefont {{Li}}\ \emph {et~al.}(2020)\citenamefont {{Li}},
  \citenamefont {{Sedrakian}},\ and\ \citenamefont {{Alford}}}]{Li2020PhRvD}%
  \BibitemOpen
  \bibfield  {author} {\bibinfo {author} {\bibfnamefont {J.~J.}\ \bibnamefont
  {{Li}}}, \bibinfo {author} {\bibfnamefont {A.}~\bibnamefont {{Sedrakian}}}, \
  and\ \bibinfo {author} {\bibfnamefont {M.}~\bibnamefont {{Alford}}},\ }\href
  {\doibase 10.1103/PhysRevD.101.063022} {\bibfield  {journal} {\bibinfo
  {journal} {\prd}\ }\textbf {\bibinfo {volume} {101}},\ \bibinfo {eid}
  {063022} (\bibinfo {year} {2020})}\BibitemShut {NoStop}%
\bibitem [{\citenamefont {{Marczenko}}\ \emph {et~al.}(2020)\citenamefont
  {{Marczenko}}, \citenamefont {{Blaschke}}, \citenamefont {{Redlich}},\ and\
  \citenamefont {{Sasaki}}}]{Marczenko2020AA}%
  \BibitemOpen
  \bibfield  {author} {\bibinfo {author} {\bibfnamefont {M.}~\bibnamefont
  {{Marczenko}}}, \bibinfo {author} {\bibfnamefont {D.}~\bibnamefont
  {{Blaschke}}}, \bibinfo {author} {\bibfnamefont {K.}~\bibnamefont
  {{Redlich}}}, \ and\ \bibinfo {author} {\bibfnamefont {C.}~\bibnamefont
  {{Sasaki}}},\ }\href {\doibase 10.1051/0004-6361/202038211} {\bibfield
  {journal} {\bibinfo  {journal} {Astron. Astr.}\ }\textbf {\bibinfo {volume}
  {643}},\ \bibinfo {eid} {A82} (\bibinfo {year} {2020})},\ \Eprint
  {http://arxiv.org/abs/2004.09566} {arXiv:2004.09566 [astro-ph.HE]}
  \BibitemShut {NoStop}%
\bibitem [{\citenamefont {Radice}\ \emph {et~al.}(2018)\citenamefont {Radice},
  \citenamefont {Perego}, \citenamefont {Zappa},\ and\ \citenamefont
  {Bernuzzi}}]{Radice_17}%
  \BibitemOpen
  \bibfield  {author} {\bibinfo {author} {\bibfnamefont {D.}~\bibnamefont
  {Radice}}, \bibinfo {author} {\bibfnamefont {A.}~\bibnamefont {Perego}},
  \bibinfo {author} {\bibfnamefont {F.}~\bibnamefont {Zappa}}, \ and\ \bibinfo
  {author} {\bibfnamefont {S.}~\bibnamefont {Bernuzzi}},\ }\href {\doibase
  10.3847/2041-8213/aaa402} {\bibfield  {journal} {\bibinfo  {journal}
  {Astrophys. J. Lett.}\ }\textbf {\bibinfo {volume} {852}},\ \bibinfo {pages}
  {L29} (\bibinfo {year} {2018})},\ \Eprint {http://arxiv.org/abs/1711.03647}
  {arXiv:1711.03647 [astro-ph.HE]} \BibitemShut {NoStop}%
\bibitem [{\citenamefont {Margalit}\ and\ \citenamefont
  {Metzger}(2017)}]{Margalit_17}%
  \BibitemOpen
  \bibfield  {author} {\bibinfo {author} {\bibfnamefont {B.}~\bibnamefont
  {Margalit}}\ and\ \bibinfo {author} {\bibfnamefont {B.~D.}\ \bibnamefont
  {Metzger}},\ }\href {\doibase 10.3847/2041-8213/aa991c} {\bibfield  {journal}
  {\bibinfo  {journal} {Astrophys. J. Lett.}\ }\textbf {\bibinfo {volume}
  {850}},\ \bibinfo {pages} {L19} (\bibinfo {year} {2017})},\ \Eprint
  {http://arxiv.org/abs/1710.05938} {arXiv:1710.05938 [astro-ph.HE]}
  \BibitemShut {NoStop}%
\bibitem [{\citenamefont {Shibata}\ \emph {et~al.}(2017)\citenamefont
  {Shibata}, \citenamefont {Fujibayashi}, \citenamefont {Hotokezaka},
  \citenamefont {Kiuchi}, \citenamefont {Kyutoku}, \citenamefont {Sekiguchi},\
  and\ \citenamefont {Tanaka}}]{Shibata_17}%
  \BibitemOpen
  \bibfield  {author} {\bibinfo {author} {\bibfnamefont {M.}~\bibnamefont
  {Shibata}}, \bibinfo {author} {\bibfnamefont {S.}~\bibnamefont
  {Fujibayashi}}, \bibinfo {author} {\bibfnamefont {K.}~\bibnamefont
  {Hotokezaka}}, \bibinfo {author} {\bibfnamefont {K.}~\bibnamefont {Kiuchi}},
  \bibinfo {author} {\bibfnamefont {K.}~\bibnamefont {Kyutoku}}, \bibinfo
  {author} {\bibfnamefont {Y.}~\bibnamefont {Sekiguchi}}, \ and\ \bibinfo
  {author} {\bibfnamefont {M.}~\bibnamefont {Tanaka}},\ }\href {\doibase
  10.1103/PhysRevD.96.123012} {\bibfield  {journal} {\bibinfo  {journal} {Phys.
  Rev. D}\ }\textbf {\bibinfo {volume} {96}},\ \bibinfo {pages} {123012}
  (\bibinfo {year} {2017})},\ \Eprint {http://arxiv.org/abs/1710.07579}
  {arXiv:1710.07579 [astro-ph.HE]} \BibitemShut {NoStop}%
\bibitem [{\citenamefont {Bauswein}\ \emph {et~al.}(2017)\citenamefont
  {Bauswein}, \citenamefont {Just}, \citenamefont {Janka},\ and\ \citenamefont
  {Stergioulas}}]{Bauswein_17}%
  \BibitemOpen
  \bibfield  {author} {\bibinfo {author} {\bibfnamefont {A.}~\bibnamefont
  {Bauswein}}, \bibinfo {author} {\bibfnamefont {O.}~\bibnamefont {Just}},
  \bibinfo {author} {\bibfnamefont {H.-T.}\ \bibnamefont {Janka}}, \ and\
  \bibinfo {author} {\bibfnamefont {N.}~\bibnamefont {Stergioulas}},\ }\href
  {\doibase 10.3847/2041-8213/aa9994} {\bibfield  {journal} {\bibinfo
  {journal} {Astrophys. J. Lett.}\ }\textbf {\bibinfo {volume} {850}},\
  \bibinfo {pages} {L34} (\bibinfo {year} {2017})},\ \Eprint
  {http://arxiv.org/abs/1710.06843} {arXiv:1710.06843 [astro-ph.HE]}
  \BibitemShut {NoStop}%
\bibitem [{\citenamefont {Coughlin}\ \emph {et~al.}(2019)\citenamefont
  {Coughlin}, \citenamefont {Dietrich}, \citenamefont {Margalit},\ and\
  \citenamefont {Metzger}}]{Coughlin_18}%
  \BibitemOpen
  \bibfield  {author} {\bibinfo {author} {\bibfnamefont {M.~W.}\ \bibnamefont
  {Coughlin}}, \bibinfo {author} {\bibfnamefont {T.}~\bibnamefont {Dietrich}},
  \bibinfo {author} {\bibfnamefont {B.}~\bibnamefont {Margalit}}, \ and\
  \bibinfo {author} {\bibfnamefont {B.~D.}\ \bibnamefont {Metzger}},\ }\href
  {\doibase 10.1093/mnrasl/slz133} {\bibfield  {journal} {\bibinfo  {journal}
  {Mon. Not. Roy. Astron. Soc.}\ }\textbf {\bibinfo {volume} {489}},\ \bibinfo
  {pages} {L91} (\bibinfo {year} {2019})},\ \Eprint
  {http://arxiv.org/abs/1812.04803} {arXiv:1812.04803 [astro-ph.HE]}
  \BibitemShut {NoStop}%
\bibitem [{\citenamefont {Rezzolla}\ \emph {et~al.}(2018)\citenamefont
  {Rezzolla}, \citenamefont {Most},\ and\ \citenamefont
  {Weih}}]{Rezzolla_2018}%
  \BibitemOpen
  \bibfield  {author} {\bibinfo {author} {\bibfnamefont {L.}~\bibnamefont
  {Rezzolla}}, \bibinfo {author} {\bibfnamefont {E.~R.}\ \bibnamefont {Most}},
  \ and\ \bibinfo {author} {\bibfnamefont {L.~R.}\ \bibnamefont {Weih}},\
  }\href {\doibase 10.3847/2041-8213/aaa401} {\bibfield  {journal} {\bibinfo
  {journal} {The Astrophysical Journal}\ }\textbf {\bibinfo {volume} {852}},\
  \bibinfo {pages} {L25} (\bibinfo {year} {2018})}\BibitemShut {NoStop}%
\bibitem [{\citenamefont {{Shibata}}\ \emph {et~al.}(2019)\citenamefont
  {{Shibata}}, \citenamefont {{Zhou}}, \citenamefont {{Kiuchi}},\ and\
  \citenamefont {{Fujibayashi}}}]{Shibata_2019a}%
  \BibitemOpen
  \bibfield  {author} {\bibinfo {author} {\bibfnamefont {M.}~\bibnamefont
  {{Shibata}}}, \bibinfo {author} {\bibfnamefont {E.}~\bibnamefont {{Zhou}}},
  \bibinfo {author} {\bibfnamefont {K.}~\bibnamefont {{Kiuchi}}}, \ and\
  \bibinfo {author} {\bibfnamefont {S.}~\bibnamefont {{Fujibayashi}}},\ }\href
  {\doibase 10.1103/PhysRevD.100.023015} {\bibfield  {journal} {\bibinfo
  {journal} {\prd}\ }\textbf {\bibinfo {volume} {100}},\ \bibinfo {eid}
  {023015} (\bibinfo {year} {2019})},\ \Eprint
  {http://arxiv.org/abs/1905.03656} {arXiv:1905.03656 [astro-ph.HE]}
  \BibitemShut {NoStop}%
\bibitem [{\citenamefont {{Ruiz}}\ \emph {et~al.}(2018)\citenamefont {{Ruiz}},
  \citenamefont {{Shapiro}},\ and\ \citenamefont {{Tsokaros}}}]{Ruiz2018}%
  \BibitemOpen
  \bibfield  {author} {\bibinfo {author} {\bibfnamefont {M.}~\bibnamefont
  {{Ruiz}}}, \bibinfo {author} {\bibfnamefont {S.~L.}\ \bibnamefont
  {{Shapiro}}}, \ and\ \bibinfo {author} {\bibfnamefont {A.}~\bibnamefont
  {{Tsokaros}}},\ }\href {\doibase 10.1103/PhysRevD.97.021501} {\bibfield
  {journal} {\bibinfo  {journal} {\prd}\ }\textbf {\bibinfo {volume} {97}},\
  \bibinfo {eid} {021501} (\bibinfo {year} {2018})},\ \Eprint
  {http://arxiv.org/abs/1711.00473} {arXiv:1711.00473 [astro-ph.HE]}
  \BibitemShut {NoStop}%
\bibitem [{\citenamefont {Abbott}\ \emph
  {et~al.}(2020{\natexlab{b}})\citenamefont {Abbott} \emph
  {et~al.}}]{Abbott_2020}%
  \BibitemOpen
  \bibfield  {author} {\bibinfo {author} {\bibfnamefont {R.}~\bibnamefont
  {Abbott}} \emph {et~al.} (\bibinfo {collaboration} {LIGO Scientific,
  Virgo}),\ }\href {\doibase 10.3847/2041-8213/ab960f} {\bibfield  {journal}
  {\bibinfo  {journal} {Astrophys. J. Lett.}\ }\textbf {\bibinfo {volume}
  {896}},\ \bibinfo {pages} {L44} (\bibinfo {year} {2020}{\natexlab{b}})},\
  \Eprint {http://arxiv.org/abs/2006.12611} {arXiv:2006.12611 [astro-ph.HE]}
  \BibitemShut {NoStop}%
\bibitem [{\citenamefont {{Zhang}}\ and\ \citenamefont
  {{Li}}(2020)}]{Zhang2020}%
  \BibitemOpen
  \bibfield  {author} {\bibinfo {author} {\bibfnamefont {N.-B.}\ \bibnamefont
  {{Zhang}}}\ and\ \bibinfo {author} {\bibfnamefont {B.-A.}\ \bibnamefont
  {{Li}}},\ }\href {\doibase 10.3847/1538-4357/abb470} {\bibfield  {journal}
  {\bibinfo  {journal} {\apj}\ }\textbf {\bibinfo {volume} {902}},\ \bibinfo
  {eid} {38} (\bibinfo {year} {2020})},\ \Eprint
  {http://arxiv.org/abs/2007.02513} {arXiv:2007.02513 [astro-ph.HE]}
  \BibitemShut {NoStop}%
\bibitem [{\citenamefont {Sedrakian}\ \emph {et~al.}(2020)\citenamefont
  {Sedrakian}, \citenamefont {Weber},\ and\ \citenamefont
  {Li}}]{Sedrakian2020}%
  \BibitemOpen
  \bibfield  {author} {\bibinfo {author} {\bibfnamefont {A.}~\bibnamefont
  {Sedrakian}}, \bibinfo {author} {\bibfnamefont {F.}~\bibnamefont {Weber}}, \
  and\ \bibinfo {author} {\bibfnamefont {J.~J.}\ \bibnamefont {Li}},\ }\href
  {\doibase 10.1103/PhysRevD.102.041301} {\bibfield  {journal} {\bibinfo
  {journal} {Phys. Rev. D}\ }\textbf {\bibinfo {volume} {102}},\ \bibinfo
  {pages} {041301} (\bibinfo {year} {2020})},\ \Eprint
  {http://arxiv.org/abs/2007.09683} {arXiv:2007.09683 [astro-ph.HE]}
  \BibitemShut {NoStop}%
\bibitem [{\citenamefont {Dexheimer}\ \emph {et~al.}(2020)\citenamefont
  {Dexheimer}, \citenamefont {Gomes}, \citenamefont {Kl\"ahn}, \citenamefont
  {Han},\ and\ \citenamefont {Salinas}}]{Dexheimer2020}%
  \BibitemOpen
  \bibfield  {author} {\bibinfo {author} {\bibfnamefont {V.}~\bibnamefont
  {Dexheimer}}, \bibinfo {author} {\bibfnamefont {R.}~\bibnamefont {Gomes}},
  \bibinfo {author} {\bibfnamefont {T.}~\bibnamefont {Kl\"ahn}}, \bibinfo
  {author} {\bibfnamefont {S.}~\bibnamefont {Han}}, \ and\ \bibinfo {author}
  {\bibfnamefont {M.}~\bibnamefont {Salinas}},\ }\href@noop {} {\  (\bibinfo
  {year} {2020})},\ \Eprint {http://arxiv.org/abs/2007.08493} {arXiv:2007.08493
  [astro-ph.HE]} \BibitemShut {NoStop}%
\bibitem [{\citenamefont {Tan}\ \emph {et~al.}(2020)\citenamefont {Tan},
  \citenamefont {Noronha-Hostler},\ and\ \citenamefont {Yunes}}]{Tan2020}%
  \BibitemOpen
  \bibfield  {author} {\bibinfo {author} {\bibfnamefont {H.}~\bibnamefont
  {Tan}}, \bibinfo {author} {\bibfnamefont {J.}~\bibnamefont
  {Noronha-Hostler}}, \ and\ \bibinfo {author} {\bibfnamefont {N.}~\bibnamefont
  {Yunes}},\ }\href@noop {} {\  (\bibinfo {year} {2020})},\ \Eprint
  {http://arxiv.org/abs/2006.16296} {arXiv:2006.16296 [astro-ph.HE]}
  \BibitemShut {NoStop}%
\bibitem [{\citenamefont {Tsokaros}\ \emph {et~al.}(2020)\citenamefont
  {Tsokaros}, \citenamefont {Ruiz},\ and\ \citenamefont
  {Shapiro}}]{Tsokaros2020}%
  \BibitemOpen
  \bibfield  {author} {\bibinfo {author} {\bibfnamefont {A.}~\bibnamefont
  {Tsokaros}}, \bibinfo {author} {\bibfnamefont {M.}~\bibnamefont {Ruiz}}, \
  and\ \bibinfo {author} {\bibfnamefont {S.~L.}\ \bibnamefont {Shapiro}},\
  }\href {\doibase 10.3847/1538-4357/abc421} {\  (\bibinfo {year} {2020}),\
  10.3847/1538-4357/abc421},\ \Eprint {http://arxiv.org/abs/2007.05526}
  {arXiv:2007.05526 [astro-ph.HE]} \BibitemShut {NoStop}%
\bibitem [{\citenamefont {Biswas}\ \emph {et~al.}(2020)\citenamefont {Biswas},
  \citenamefont {Nandi}, \citenamefont {Char}, \citenamefont {Bose},\ and\
  \citenamefont {Stergioulas}}]{Biswas2020}%
  \BibitemOpen
  \bibfield  {author} {\bibinfo {author} {\bibfnamefont {B.}~\bibnamefont
  {Biswas}}, \bibinfo {author} {\bibfnamefont {R.}~\bibnamefont {Nandi}},
  \bibinfo {author} {\bibfnamefont {P.}~\bibnamefont {Char}}, \bibinfo {author}
  {\bibfnamefont {S.}~\bibnamefont {Bose}}, \ and\ \bibinfo {author}
  {\bibfnamefont {N.}~\bibnamefont {Stergioulas}},\ }\href@noop {} {\
  (\bibinfo {year} {2020})},\ \Eprint {http://arxiv.org/abs/2010.02090}
  {arXiv:2010.02090 [astro-ph.HE]} \BibitemShut {NoStop}%
\bibitem [{\citenamefont {{Fattoyev}}\ \emph {et~al.}(2020)\citenamefont
  {{Fattoyev}}, \citenamefont {{Horowitz}}, \citenamefont {{Piekarewicz}},\
  and\ \citenamefont {{Reed}}}]{Fattoyev2020}%
  \BibitemOpen
  \bibfield  {author} {\bibinfo {author} {\bibfnamefont {F.~J.}\ \bibnamefont
  {{Fattoyev}}}, \bibinfo {author} {\bibfnamefont {C.~J.}\ \bibnamefont
  {{Horowitz}}}, \bibinfo {author} {\bibfnamefont {J.}~\bibnamefont
  {{Piekarewicz}}}, \ and\ \bibinfo {author} {\bibfnamefont {B.}~\bibnamefont
  {{Reed}}},\ }\href@noop {} {\bibfield  {journal} {\bibinfo  {journal} {arXiv
  e-prints}\ ,\ \bibinfo {eid} {arXiv:2007.03799}} (\bibinfo {year} {2020})},\
  \Eprint {http://arxiv.org/abs/2007.03799} {arXiv:2007.03799 [nucl-th]}
  \BibitemShut {NoStop}%
\bibitem [{\citenamefont {{Nathanail}}\ \emph {et~al.}(2021)\citenamefont
  {{Nathanail}}, \citenamefont {{Most}},\ and\ \citenamefont
  {{Rezzolla}}}]{Nathanail2021}%
  \BibitemOpen
  \bibfield  {author} {\bibinfo {author} {\bibfnamefont {A.}~\bibnamefont
  {{Nathanail}}}, \bibinfo {author} {\bibfnamefont {E.~R.}\ \bibnamefont
  {{Most}}}, \ and\ \bibinfo {author} {\bibfnamefont {L.}~\bibnamefont
  {{Rezzolla}}},\ }\href@noop {} {\bibfield  {journal} {\bibinfo  {journal}
  {arXiv e-prints}\ ,\ \bibinfo {eid} {arXiv:2101.01735}} (\bibinfo {year}
  {2021})},\ \Eprint {http://arxiv.org/abs/2101.01735} {arXiv:2101.01735
  [astro-ph.HE]} \BibitemShut {NoStop}%
\bibitem [{\citenamefont {{Horvath}}\ and\ \citenamefont
  {{Moraes}}(2020)}]{Horvath2020}%
  \BibitemOpen
  \bibfield  {author} {\bibinfo {author} {\bibfnamefont {J.~E.}\ \bibnamefont
  {{Horvath}}}\ and\ \bibinfo {author} {\bibfnamefont {P.~H.~R.~S.}\
  \bibnamefont {{Moraes}}},\ }\href@noop {} {\bibfield  {journal} {\bibinfo
  {journal} {arXiv e-prints}\ ,\ \bibinfo {eid} {arXiv:2012.00917}} (\bibinfo
  {year} {2020})},\ \Eprint {http://arxiv.org/abs/2012.00917} {arXiv:2012.00917
  [astro-ph.HE]} \BibitemShut {NoStop}%
\bibitem [{\citenamefont {Bombaci}\ \emph {et~al.}(2020)\citenamefont
  {Bombaci}, \citenamefont {Drago}, \citenamefont {Logoteta}, \citenamefont
  {Pagliara},\ and\ \citenamefont {Vidana}}]{Bombaci2020}%
  \BibitemOpen
  \bibfield  {author} {\bibinfo {author} {\bibfnamefont {I.}~\bibnamefont
  {Bombaci}}, \bibinfo {author} {\bibfnamefont {A.}~\bibnamefont {Drago}},
  \bibinfo {author} {\bibfnamefont {D.}~\bibnamefont {Logoteta}}, \bibinfo
  {author} {\bibfnamefont {G.}~\bibnamefont {Pagliara}}, \ and\ \bibinfo
  {author} {\bibfnamefont {I.}~\bibnamefont {Vidana}},\ }\href@noop {} {\
  (\bibinfo {year} {2020})},\ \Eprint {http://arxiv.org/abs/2010.01509}
  {arXiv:2010.01509 [nucl-th]} \BibitemShut {NoStop}%
\bibitem [{\citenamefont {{Astashenok}}\ \emph {et~al.}(2020)\citenamefont
  {{Astashenok}}, \citenamefont {{Capozziello}}, \citenamefont {{Odintsov}},\
  and\ \citenamefont {{Oikonomou}}}]{Astashenok2020}%
  \BibitemOpen
  \bibfield  {author} {\bibinfo {author} {\bibfnamefont {A.~V.}\ \bibnamefont
  {{Astashenok}}}, \bibinfo {author} {\bibfnamefont {S.}~\bibnamefont
  {{Capozziello}}}, \bibinfo {author} {\bibfnamefont {S.~D.}\ \bibnamefont
  {{Odintsov}}}, \ and\ \bibinfo {author} {\bibfnamefont {V.~K.}\ \bibnamefont
  {{Oikonomou}}},\ }\href {\doibase 10.1016/j.physletb.2020.135910} {\bibfield
  {journal} {\bibinfo  {journal} {Physics Letters B}\ }\textbf {\bibinfo
  {volume} {811}},\ \bibinfo {eid} {135910} (\bibinfo {year} {2020})},\ \Eprint
  {http://arxiv.org/abs/2008.10884} {arXiv:2008.10884 [gr-qc]} \BibitemShut
  {NoStop}%
\bibitem [{\citenamefont {Shibata}\ and\ \citenamefont
  {Taniguchi}(2011)}]{Shibata_11}%
  \BibitemOpen
  \bibfield  {author} {\bibinfo {author} {\bibfnamefont {M.}~\bibnamefont
  {Shibata}}\ and\ \bibinfo {author} {\bibfnamefont {K.}~\bibnamefont
  {Taniguchi}},\ }\href {\doibase 10.12942/lrr-2011-6} {\bibfield  {journal}
  {\bibinfo  {journal} {Living Rev.Rel.}\ }\textbf {\bibinfo {volume} {14}},\
  \bibinfo {pages} {6} (\bibinfo {year} {2011})}\BibitemShut {NoStop}%
\bibitem [{\citenamefont {{Rosswog}}(2015)}]{Rosswog_15}%
  \BibitemOpen
  \bibfield  {author} {\bibinfo {author} {\bibfnamefont {S.}~\bibnamefont
  {{Rosswog}}},\ }\href {\doibase 10.1142/S0218271815300128} {\bibfield
  {journal} {\bibinfo  {journal} {Int. J. Mod. Phys. D}\ }\textbf {\bibinfo
  {volume} {24}},\ \bibinfo {eid} {1530012} (\bibinfo {year} {2015})},\ \Eprint
  {http://arxiv.org/abs/1501.02081} {arXiv:1501.02081 [astro-ph.HE]}
  \BibitemShut {NoStop}%
\bibitem [{\citenamefont {{Baiotti}}\ and\ \citenamefont
  {{Rezzolla}}(2017)}]{Baiotti_2017}%
  \BibitemOpen
  \bibfield  {author} {\bibinfo {author} {\bibfnamefont {L.}~\bibnamefont
  {{Baiotti}}}\ and\ \bibinfo {author} {\bibfnamefont {L.}~\bibnamefont
  {{Rezzolla}}},\ }\href {\doibase 10.1088/1361-6633/aa67bb} {\bibfield
  {journal} {\bibinfo  {journal} {Rep.Prog. Phys.}\ }\textbf {\bibinfo {volume}
  {80}},\ \bibinfo {eid} {096901} (\bibinfo {year} {2017})},\ \Eprint
  {http://arxiv.org/abs/1607.03540} {arXiv:1607.03540 [gr-qc]} \BibitemShut
  {NoStop}%
\bibitem [{\citenamefont {{Chatziioannou}}(2020)}]{Chatziioannou2020GReGr}%
  \BibitemOpen
  \bibfield  {author} {\bibinfo {author} {\bibfnamefont {K.}~\bibnamefont
  {{Chatziioannou}}},\ }\href {\doibase 10.1007/s10714-020-02754-3} {\bibfield
  {journal} {\bibinfo  {journal} {General Relativity and Gravitation}\ }\textbf
  {\bibinfo {volume} {52}},\ \bibinfo {eid} {109} (\bibinfo {year} {2020})},\
  \Eprint {http://arxiv.org/abs/2006.03168} {arXiv:2006.03168 [gr-qc]}
  \BibitemShut {NoStop}%
\bibitem [{\citenamefont {Oertel}\ \emph {et~al.}(2017)\citenamefont {Oertel},
  \citenamefont {Hempel}, \citenamefont {Klahn},\ and\ \citenamefont
  {Typel}}]{Oertel_RMP_2017}%
  \BibitemOpen
  \bibfield  {author} {\bibinfo {author} {\bibfnamefont {M.}~\bibnamefont
  {Oertel}}, \bibinfo {author} {\bibfnamefont {M.}~\bibnamefont {Hempel}},
  \bibinfo {author} {\bibfnamefont {T.}~\bibnamefont {Klahn}}, \ and\ \bibinfo
  {author} {\bibfnamefont {S.}~\bibnamefont {Typel}},\ }\href {\doibase
  10.1103/RevModPhys.89.015007} {\bibfield  {journal} {\bibinfo  {journal}
  {Rev. Mod. Phys.}\ }\textbf {\bibinfo {volume} {89}},\ \bibinfo {pages}
  {015007} (\bibinfo {year} {2017})}\BibitemShut {NoStop}%
\bibitem [{\citenamefont {{Cook}}\ \emph {et~al.}(1994)\citenamefont {{Cook}},
  \citenamefont {{Shapiro}},\ and\ \citenamefont {{Teukolsky}}}]{Cook1994}%
  \BibitemOpen
  \bibfield  {author} {\bibinfo {author} {\bibfnamefont {G.~B.}\ \bibnamefont
  {{Cook}}}, \bibinfo {author} {\bibfnamefont {S.~L.}\ \bibnamefont
  {{Shapiro}}}, \ and\ \bibinfo {author} {\bibfnamefont {S.~A.}\ \bibnamefont
  {{Teukolsky}}},\ }\href {\doibase 10.1086/173934} {\bibfield  {journal}
  {\bibinfo  {journal} {\apj}\ }\textbf {\bibinfo {volume} {424}},\ \bibinfo
  {pages} {823} (\bibinfo {year} {1994})}\BibitemShut {NoStop}%
\bibitem [{\citenamefont {Doneva}\ \emph {et~al.}(2013)\citenamefont {Doneva},
  \citenamefont {Yazadjiev}, \citenamefont {Stergioulas},\ and\ \citenamefont
  {Kokkotas}}]{Doneva:2013rha}%
  \BibitemOpen
  \bibfield  {author} {\bibinfo {author} {\bibfnamefont {D.~D.}\ \bibnamefont
  {Doneva}}, \bibinfo {author} {\bibfnamefont {S.~S.}\ \bibnamefont
  {Yazadjiev}}, \bibinfo {author} {\bibfnamefont {N.}~\bibnamefont
  {Stergioulas}}, \ and\ \bibinfo {author} {\bibfnamefont {K.~D.}\ \bibnamefont
  {Kokkotas}},\ }\href {\doibase 10.1088/2041-8205/781/1/L6} {\bibfield
  {journal} {\bibinfo  {journal} {Astrophys. J.}\ }\textbf {\bibinfo {volume}
  {781}},\ \bibinfo {pages} {L6} (\bibinfo {year} {2013})}\BibitemShut
  {NoStop}%
\bibitem [{\citenamefont {Maselli}\ \emph {et~al.}(2013)\citenamefont
  {Maselli}, \citenamefont {Cardoso}, \citenamefont {Ferrari}, \citenamefont
  {Gualtieri},\ and\ \citenamefont {Pani}}]{Maselli_PRD_2013}%
  \BibitemOpen
  \bibfield  {author} {\bibinfo {author} {\bibfnamefont {A.}~\bibnamefont
  {Maselli}}, \bibinfo {author} {\bibfnamefont {V.}~\bibnamefont {Cardoso}},
  \bibinfo {author} {\bibfnamefont {V.}~\bibnamefont {Ferrari}}, \bibinfo
  {author} {\bibfnamefont {L.}~\bibnamefont {Gualtieri}}, \ and\ \bibinfo
  {author} {\bibfnamefont {P.}~\bibnamefont {Pani}},\ }\href {\doibase
  10.1103/PhysRevD.88.023007} {\bibfield  {journal} {\bibinfo  {journal} {Phys.
  Rev. D}\ }\textbf {\bibinfo {volume} {88}},\ \bibinfo {pages} {023007}
  (\bibinfo {year} {2013})}\BibitemShut {NoStop}%
\bibitem [{\citenamefont {Breu}\ and\ \citenamefont
  {Rezzolla}(2016)}]{Breu_2016}%
  \BibitemOpen
  \bibfield  {author} {\bibinfo {author} {\bibfnamefont {C.}~\bibnamefont
  {Breu}}\ and\ \bibinfo {author} {\bibfnamefont {L.}~\bibnamefont
  {Rezzolla}},\ }\href {\doibase 10.1093/mnras/stw575} {\bibfield  {journal}
  {\bibinfo  {journal} {Mon. Not. Roy. Astron. Soc.}\ }\textbf {\bibinfo
  {volume} {459}},\ \bibinfo {pages} {646} (\bibinfo {year} {2016})},\ \Eprint
  {http://arxiv.org/abs/1601.06083} {arXiv:1601.06083 [gr-qc]} \BibitemShut
  {NoStop}%
\bibitem [{\citenamefont {Bozzola}\ \emph {et~al.}(2018)\citenamefont
  {Bozzola}, \citenamefont {Stergioulas},\ and\ \citenamefont
  {Bauswein}}]{Bozzola_17}%
  \BibitemOpen
  \bibfield  {author} {\bibinfo {author} {\bibfnamefont {G.}~\bibnamefont
  {Bozzola}}, \bibinfo {author} {\bibfnamefont {N.}~\bibnamefont
  {Stergioulas}}, \ and\ \bibinfo {author} {\bibfnamefont {A.}~\bibnamefont
  {Bauswein}},\ }\href {\doibase 10.1093/mnras/stx3002} {\bibfield  {journal}
  {\bibinfo  {journal} {Mon. Not. Roy. Astron. Soc.}\ }\textbf {\bibinfo
  {volume} {474}},\ \bibinfo {pages} {3557} (\bibinfo {year} {2018})},\ \Eprint
  {http://arxiv.org/abs/1709.02787} {arXiv:1709.02787 [gr-qc]} \BibitemShut
  {NoStop}%
\bibitem [{\citenamefont {Bozzola}\ \emph {et~al.}(2019)\citenamefont
  {Bozzola}, \citenamefont {Espino}, \citenamefont {Lewin},\ and\ \citenamefont
  {Paschalidis}}]{Bozzola_19}%
  \BibitemOpen
  \bibfield  {author} {\bibinfo {author} {\bibfnamefont {G.}~\bibnamefont
  {Bozzola}}, \bibinfo {author} {\bibfnamefont {P.~L.}\ \bibnamefont {Espino}},
  \bibinfo {author} {\bibfnamefont {C.~D.}\ \bibnamefont {Lewin}}, \ and\
  \bibinfo {author} {\bibfnamefont {V.}~\bibnamefont {Paschalidis}},\ }\href
  {\doibase 10.1140/epja/i2019-12831-2} {\bibfield  {journal} {\bibinfo
  {journal} {Eur. Phys. J. A}\ }\textbf {\bibinfo {volume} {55}},\ \bibinfo
  {pages} {149} (\bibinfo {year} {2019})},\ \Eprint
  {http://arxiv.org/abs/1905.00028} {arXiv:1905.00028 [astro-ph.HE]}
  \BibitemShut {NoStop}%
\bibitem [{\citenamefont {{Weih}}\ \emph {et~al.}(2018)\citenamefont {{Weih}},
  \citenamefont {{Most}},\ and\ \citenamefont {{Rezzolla}}}]{Weih2018MNRAS}%
  \BibitemOpen
  \bibfield  {author} {\bibinfo {author} {\bibfnamefont {L.~R.}\ \bibnamefont
  {{Weih}}}, \bibinfo {author} {\bibfnamefont {E.~R.}\ \bibnamefont {{Most}}},
  \ and\ \bibinfo {author} {\bibfnamefont {L.}~\bibnamefont {{Rezzolla}}},\
  }\href {\doibase 10.1093/mnrasl/slx178} {\bibfield  {journal} {\bibinfo
  {journal} {\mnras}\ }\textbf {\bibinfo {volume} {473}},\ \bibinfo {pages}
  {L126} (\bibinfo {year} {2018})},\ \Eprint {http://arxiv.org/abs/1709.06058}
  {arXiv:1709.06058 [gr-qc]} \BibitemShut {NoStop}%
\bibitem [{\citenamefont {Marques}\ \emph {et~al.}(2017)\citenamefont
  {Marques}, \citenamefont {Oertel}, \citenamefont {Hempel},\ and\
  \citenamefont {Novak}}]{Marques_PRC_2017}%
  \BibitemOpen
  \bibfield  {author} {\bibinfo {author} {\bibfnamefont {M.}~\bibnamefont
  {Marques}}, \bibinfo {author} {\bibfnamefont {M.}~\bibnamefont {Oertel}},
  \bibinfo {author} {\bibfnamefont {M.}~\bibnamefont {Hempel}}, \ and\ \bibinfo
  {author} {\bibfnamefont {J.}~\bibnamefont {Novak}},\ }\href {\doibase
  10.1103/PhysRevC.96.045806} {\bibfield  {journal} {\bibinfo  {journal} {Phys.
  Rev. C}\ }\textbf {\bibinfo {volume} {96}},\ \bibinfo {pages} {045806}
  (\bibinfo {year} {2017})}\BibitemShut {NoStop}%
\bibitem [{\citenamefont {Nunna}\ \emph {et~al.}(2020)\citenamefont {Nunna},
  \citenamefont {Banik},\ and\ \citenamefont {Chatterjee}}]{Nunna_ApJ_2020}%
  \BibitemOpen
  \bibfield  {author} {\bibinfo {author} {\bibfnamefont {K.~P.}\ \bibnamefont
  {Nunna}}, \bibinfo {author} {\bibfnamefont {S.}~\bibnamefont {Banik}}, \ and\
  \bibinfo {author} {\bibfnamefont {D.}~\bibnamefont {Chatterjee}},\ }\href
  {\doibase 10.3847/1538-4357/ab8f2c} {\bibfield  {journal} {\bibinfo
  {journal} {Astrophys. J.}\ }\textbf {\bibinfo {volume} {896}},\ \bibinfo
  {pages} {109} (\bibinfo {year} {2020})},\ \Eprint
  {http://arxiv.org/abs/2002.07538} {arXiv:2002.07538 [astro-ph.HE]}
  \BibitemShut {NoStop}%
\bibitem [{\citenamefont {Martinon}\ \emph {et~al.}(2014)\citenamefont
  {Martinon}, \citenamefont {Maselli}, \citenamefont {Gualtieri},\ and\
  \citenamefont {Ferrari}}]{Martinon_PRD_2014}%
  \BibitemOpen
  \bibfield  {author} {\bibinfo {author} {\bibfnamefont {G.}~\bibnamefont
  {Martinon}}, \bibinfo {author} {\bibfnamefont {A.}~\bibnamefont {Maselli}},
  \bibinfo {author} {\bibfnamefont {L.}~\bibnamefont {Gualtieri}}, \ and\
  \bibinfo {author} {\bibfnamefont {V.}~\bibnamefont {Ferrari}},\ }\href
  {\doibase 10.1103/PhysRevD.90.064026} {\bibfield  {journal} {\bibinfo
  {journal} {Phys. Rev.}\ }\textbf {\bibinfo {volume} {D90}},\ \bibinfo {pages}
  {064026} (\bibinfo {year} {2014})}\BibitemShut {NoStop}%
\bibitem [{\citenamefont {Lenka}\ \emph {et~al.}(2019)\citenamefont {Lenka},
  \citenamefont {Char},\ and\ \citenamefont {Banik}}]{Lenka_JPG_2019}%
  \BibitemOpen
  \bibfield  {author} {\bibinfo {author} {\bibfnamefont {S.~S.}\ \bibnamefont
  {Lenka}}, \bibinfo {author} {\bibfnamefont {P.}~\bibnamefont {Char}}, \ and\
  \bibinfo {author} {\bibfnamefont {S.}~\bibnamefont {Banik}},\ }\href
  {\doibase 10.1088/1361-6471/ab36a2} {\bibfield  {journal} {\bibinfo
  {journal} {Journal of Physics G: Nuclear and Particle Physics}\ }\textbf
  {\bibinfo {volume} {46}},\ \bibinfo {pages} {105201} (\bibinfo {year}
  {2019})}\BibitemShut {NoStop}%
\bibitem [{\citenamefont {Raduta}\ \emph {et~al.}(2020)\citenamefont {Raduta},
  \citenamefont {Oertel},\ and\ \citenamefont {Sedrakian}}]{Raduta_MNRAS_2020}%
  \BibitemOpen
  \bibfield  {author} {\bibinfo {author} {\bibfnamefont {A.~R.}\ \bibnamefont
  {Raduta}}, \bibinfo {author} {\bibfnamefont {M.}~\bibnamefont {Oertel}}, \
  and\ \bibinfo {author} {\bibfnamefont {A.}~\bibnamefont {Sedrakian}},\ }\href
  {\doibase 10.1093/mnras/staa2491} {\bibfield  {journal} {\bibinfo  {journal}
  {Mon. Not. Roy. Astron. Soc.}\ }\textbf {\bibinfo {volume} {499}},\ \bibinfo
  {pages} {914} (\bibinfo {year} {2020})},\ \Eprint
  {http://arxiv.org/abs/2008.00213} {arXiv:2008.00213 [nucl-th]} \BibitemShut
  {NoStop}%
\bibitem [{\citenamefont {Lasota}\ \emph {et~al.}(1996)\citenamefont {Lasota},
  \citenamefont {Haensel},\ and\ \citenamefont {Abramowicz}}]{Lasota_ApJ_1996}%
  \BibitemOpen
  \bibfield  {author} {\bibinfo {author} {\bibfnamefont {J.-P.}\ \bibnamefont
  {Lasota}}, \bibinfo {author} {\bibfnamefont {P.}~\bibnamefont {Haensel}}, \
  and\ \bibinfo {author} {\bibfnamefont {M.~A.}\ \bibnamefont {Abramowicz}},\
  }\href {\doibase 10.1086/176650} {\bibfield  {journal} {\bibinfo  {journal}
  {Astrophys. J.}\ }\textbf {\bibinfo {volume} {456}},\ \bibinfo {pages} {300}
  (\bibinfo {year} {1996})},\ \Eprint {http://arxiv.org/abs/astro-ph/9508118}
  {arXiv:astro-ph/9508118} \BibitemShut {NoStop}%
\bibitem [{\citenamefont {Goussard}\ \emph {et~al.}(1998)\citenamefont
  {Goussard}, \citenamefont {Haensel},\ and\ \citenamefont {Zdunik}}]{Gouss1}%
  \BibitemOpen
  \bibfield  {author} {\bibinfo {author} {\bibfnamefont {J.~O.}\ \bibnamefont
  {Goussard}}, \bibinfo {author} {\bibfnamefont {P.}~\bibnamefont {Haensel}}, \
  and\ \bibinfo {author} {\bibfnamefont {J.~L.}\ \bibnamefont {Zdunik}},\
  }\href@noop {} {\bibfield  {journal} {\bibinfo  {journal} {Astron.
  Astrophys.}\ }\textbf {\bibinfo {volume} {330}},\ \bibinfo {pages} {1005}
  (\bibinfo {year} {1998})}\BibitemShut {NoStop}%
\bibitem [{\citenamefont {Villain}\ \emph {et~al.}(2004)\citenamefont
  {Villain}, \citenamefont {Pons}, \citenamefont {Cerda-Duran},\ and\
  \citenamefont {Gourgoulhon}}]{Villain2004}%
  \BibitemOpen
  \bibfield  {author} {\bibinfo {author} {\bibfnamefont {L.}~\bibnamefont
  {Villain}}, \bibinfo {author} {\bibfnamefont {J.~A.}\ \bibnamefont {Pons}},
  \bibinfo {author} {\bibfnamefont {P.}~\bibnamefont {Cerda-Duran}}, \ and\
  \bibinfo {author} {\bibfnamefont {E.}~\bibnamefont {Gourgoulhon}},\ }\href
  {\doibase 10.1051/0004-6361:20035619} {\bibfield  {journal} {\bibinfo
  {journal} {Astron. Astrophys.}\ }\textbf {\bibinfo {volume} {418}},\ \bibinfo
  {pages} {283} (\bibinfo {year} {2004})}\BibitemShut {NoStop}%
\bibitem [{\citenamefont {Goussard}\ \emph {et~al.}(1997)\citenamefont
  {Goussard}, \citenamefont {Haensel},\ and\ \citenamefont {Zdunik}}]{Gouss2}%
  \BibitemOpen
  \bibfield  {author} {\bibinfo {author} {\bibfnamefont {J.-O.}\ \bibnamefont
  {Goussard}}, \bibinfo {author} {\bibfnamefont {P.}~\bibnamefont {Haensel}}, \
  and\ \bibinfo {author} {\bibfnamefont {J.~L.}\ \bibnamefont {Zdunik}},\
  }\href@noop {} {\bibfield  {journal} {\bibinfo  {journal} {Astron.
  Astrophys.}\ }\textbf {\bibinfo {volume} {321}},\ \bibinfo {pages} {822}
  (\bibinfo {year} {1997})}\BibitemShut {NoStop}%
\bibitem [{\citenamefont {Camelio}\ \emph {et~al.}(2019)\citenamefont
  {Camelio}, \citenamefont {Dietrich}, \citenamefont {Marques},\ and\
  \citenamefont {Rosswog}}]{Camelio2019}%
  \BibitemOpen
  \bibfield  {author} {\bibinfo {author} {\bibfnamefont {G.}~\bibnamefont
  {Camelio}}, \bibinfo {author} {\bibfnamefont {T.}~\bibnamefont {Dietrich}},
  \bibinfo {author} {\bibfnamefont {M.}~\bibnamefont {Marques}}, \ and\
  \bibinfo {author} {\bibfnamefont {S.}~\bibnamefont {Rosswog}},\ }\href
  {\doibase 10.1103/PhysRevD.100.123001} {\bibfield  {journal} {\bibinfo
  {journal} {Phys. Rev. D}\ }\textbf {\bibinfo {volume} {100}},\ \bibinfo
  {pages} {123001} (\bibinfo {year} {2019})},\ \Eprint
  {http://arxiv.org/abs/1908.11258} {arXiv:1908.11258 [gr-qc]} \BibitemShut
  {NoStop}%
\bibitem [{\citenamefont {Camelio}\ \emph {et~al.}(2020)\citenamefont
  {Camelio}, \citenamefont {Dietrich}, \citenamefont {Rosswog},\ and\
  \citenamefont {Haskell}}]{Camelio2020}%
  \BibitemOpen
  \bibfield  {author} {\bibinfo {author} {\bibfnamefont {G.}~\bibnamefont
  {Camelio}}, \bibinfo {author} {\bibfnamefont {T.}~\bibnamefont {Dietrich}},
  \bibinfo {author} {\bibfnamefont {S.}~\bibnamefont {Rosswog}}, \ and\
  \bibinfo {author} {\bibfnamefont {B.}~\bibnamefont {Haskell}},\ }\href@noop
  {} {\  (\bibinfo {year} {2020})},\ \Eprint {http://arxiv.org/abs/2011.10557}
  {arXiv:2011.10557 [astro-ph.HE]} \BibitemShut {NoStop}%
\bibitem [{\citenamefont {Gourgoulhon}\ \emph {et~al.}(2016)\citenamefont
  {Gourgoulhon}, \citenamefont {Grandclement}, \citenamefont {Marck},
  \citenamefont {Novak},\ and\ \citenamefont {Taniguchi}}]{Lorene}%
  \BibitemOpen
  \bibfield  {author} {\bibinfo {author} {\bibfnamefont {E.}~\bibnamefont
  {Gourgoulhon}}, \bibinfo {author} {\bibfnamefont {P.}~\bibnamefont
  {Grandclement}}, \bibinfo {author} {\bibfnamefont {J.-A.}\ \bibnamefont
  {Marck}}, \bibinfo {author} {\bibfnamefont {J.}~\bibnamefont {Novak}}, \ and\
  \bibinfo {author} {\bibfnamefont {K.}~\bibnamefont {Taniguchi}},\ }\href@noop
  {} {\bibfield  {journal} {\bibinfo  {journal} {Astrophysics Source Code
  Library}\ } (\bibinfo {year} {2016})}\BibitemShut {NoStop}%
\bibitem [{\citenamefont {Bonazzola}\ \emph {et~al.}(1993)\citenamefont
  {Bonazzola}, \citenamefont {Gourgoulhon}, \citenamefont {Salgado},\ and\
  \citenamefont {Marck}}]{BGSM}%
  \BibitemOpen
  \bibfield  {author} {\bibinfo {author} {\bibfnamefont {S.}~\bibnamefont
  {Bonazzola}}, \bibinfo {author} {\bibfnamefont {E.}~\bibnamefont
  {Gourgoulhon}}, \bibinfo {author} {\bibfnamefont {M.}~\bibnamefont
  {Salgado}}, \ and\ \bibinfo {author} {\bibfnamefont {J.}~\bibnamefont
  {Marck}},\ }\href@noop {} {\bibfield  {journal} {\bibinfo  {journal} {A\&A}\
  }\textbf {\bibinfo {volume} {278}},\ \bibinfo {pages} {421} (\bibinfo {year}
  {1993})}\BibitemShut {NoStop}%
\bibitem [{\citenamefont {Stone}\ \emph {et~al.}(2019)\citenamefont {Stone},
  \citenamefont {Dexheimer}, \citenamefont {Guichon},\ and\ \citenamefont
  {Thomas}}]{Stone2019}%
  \BibitemOpen
  \bibfield  {author} {\bibinfo {author} {\bibfnamefont {J.}~\bibnamefont
  {Stone}}, \bibinfo {author} {\bibfnamefont {V.}~\bibnamefont {Dexheimer}},
  \bibinfo {author} {\bibfnamefont {P.}~\bibnamefont {Guichon}}, \ and\
  \bibinfo {author} {\bibfnamefont {A.}~\bibnamefont {Thomas}},\ }\href@noop {}
  {\  (\bibinfo {year} {2019})},\ \Eprint {http://arxiv.org/abs/1906.11100}
  {arXiv:1906.11100 [nucl-th]} \BibitemShut {NoStop}%
\bibitem [{\citenamefont {Abbott}\ \emph
  {et~al.}(2019{\natexlab{b}})\citenamefont {Abbott} \emph
  {et~al.}}]{Abbott_2019}%
  \BibitemOpen
  \bibfield  {author} {\bibinfo {author} {\bibfnamefont {B.}~\bibnamefont
  {Abbott}} \emph {et~al.} (\bibinfo {collaboration} {LIGO Scientific,
  Virgo}),\ }\href {\doibase 10.1103/PhysRevX.9.011001} {\bibfield  {journal}
  {\bibinfo  {journal} {Phys. Rev. X}\ }\textbf {\bibinfo {volume} {9}},\
  \bibinfo {pages} {011001} (\bibinfo {year} {2019}{\natexlab{b}})},\ \Eprint
  {http://arxiv.org/abs/1805.11579} {arXiv:1805.11579 [gr-qc]} \BibitemShut
  {NoStop}%
\bibitem [{\citenamefont {Margueron}\ \emph {et~al.}(2018)\citenamefont
  {Margueron}, \citenamefont {Hoffmann~Casali},\ and\ \citenamefont
  {Gulminelli}}]{Margueron_PRC_2018}%
  \BibitemOpen
  \bibfield  {author} {\bibinfo {author} {\bibfnamefont {J.}~\bibnamefont
  {Margueron}}, \bibinfo {author} {\bibfnamefont {R.}~\bibnamefont
  {Hoffmann~Casali}}, \ and\ \bibinfo {author} {\bibfnamefont {F.}~\bibnamefont
  {Gulminelli}},\ }\href {\doibase 10.1103/PhysRevC.97.025805} {\bibfield
  {journal} {\bibinfo  {journal} {Phys. Rev. C}\ }\textbf {\bibinfo {volume}
  {97}},\ \bibinfo {pages} {025805} (\bibinfo {year} {2018})},\ \Eprint
  {http://arxiv.org/abs/1708.06894} {arXiv:1708.06894 [nucl-th]} \BibitemShut
  {NoStop}%
\bibitem [{\citenamefont {Khan}\ and\ \citenamefont
  {Margueron}(2012)}]{Khan_PRL_2012}%
  \BibitemOpen
  \bibfield  {author} {\bibinfo {author} {\bibfnamefont {E.}~\bibnamefont
  {Khan}}\ and\ \bibinfo {author} {\bibfnamefont {J.}~\bibnamefont
  {Margueron}},\ }\href {\doibase 10.1103/PhysRevLett.109.092501} {\bibfield
  {journal} {\bibinfo  {journal} {Phys. Rev. Lett.}\ }\textbf {\bibinfo
  {volume} {109}},\ \bibinfo {pages} {092501} (\bibinfo {year} {2012})},\
  \Eprint {http://arxiv.org/abs/1204.0399} {arXiv:1204.0399 [nucl-th]}
  \BibitemShut {NoStop}%
\bibitem [{\citenamefont {{Salgado}}\ \emph {et~al.}(1994)\citenamefont
  {{Salgado}}, \citenamefont {{Bonazzola}}, \citenamefont {{Gourgoulhon}},\
  and\ \citenamefont {{Haensel}}}]{Salgado1994}%
  \BibitemOpen
  \bibfield  {author} {\bibinfo {author} {\bibfnamefont {M.}~\bibnamefont
  {{Salgado}}}, \bibinfo {author} {\bibfnamefont {S.}~\bibnamefont
  {{Bonazzola}}}, \bibinfo {author} {\bibfnamefont {E.}~\bibnamefont
  {{Gourgoulhon}}}, \ and\ \bibinfo {author} {\bibfnamefont {P.}~\bibnamefont
  {{Haensel}}},\ }\href@noop {} {\bibfield  {journal} {\bibinfo  {journal}
  {Astron. Astrophys.}\ }\textbf {\bibinfo {volume} {291}},\ \bibinfo {pages}
  {155} (\bibinfo {year} {1994})}\BibitemShut {NoStop}%
\bibitem [{\citenamefont {Pappas}\ and\ \citenamefont
  {Apostolatos}(2012)}]{Pappas2012}%
  \BibitemOpen
  \bibfield  {author} {\bibinfo {author} {\bibfnamefont {G.}~\bibnamefont
  {Pappas}}\ and\ \bibinfo {author} {\bibfnamefont {T.~A.}\ \bibnamefont
  {Apostolatos}},\ }\href {\doibase 10.1103/PhysRevLett.108.231104} {\bibfield
  {journal} {\bibinfo  {journal} {Phys. Rev. Lett.}\ }\textbf {\bibinfo
  {volume} {108}},\ \bibinfo {pages} {231104} (\bibinfo {year} {2012})},\
  \Eprint {http://arxiv.org/abs/1201.6067} {arXiv:1201.6067 [gr-qc]}
  \BibitemShut {NoStop}%
\bibitem [{\citenamefont {Arzoumanian}\ \emph {et~al.}(2018)\citenamefont
  {Arzoumanian} \emph {et~al.}}]{Arzoumanian2018}%
  \BibitemOpen
  \bibfield  {author} {\bibinfo {author} {\bibfnamefont {Z.}~\bibnamefont
  {Arzoumanian}} \emph {et~al.} (\bibinfo {collaboration} {NANOGrav}),\ }\href
  {\doibase 10.3847/1538-4365/aab5b0} {\bibfield  {journal} {\bibinfo
  {journal} {Astrophys. J. Suppl.}\ }\textbf {\bibinfo {volume} {235}},\
  \bibinfo {pages} {37} (\bibinfo {year} {2018})},\ \Eprint
  {http://arxiv.org/abs/1801.01837} {arXiv:1801.01837 [astro-ph.HE]}
  \BibitemShut {NoStop}%
\bibitem [{\citenamefont {Gulminelli}\ and\ \citenamefont
  {Raduta}(2015)}]{Raduta_PRC_2015}%
  \BibitemOpen
  \bibfield  {author} {\bibinfo {author} {\bibfnamefont {F.}~\bibnamefont
  {Gulminelli}}\ and\ \bibinfo {author} {\bibfnamefont {A.~R.}\ \bibnamefont
  {Raduta}},\ }\href {\doibase 10.1103/PhysRevC.92.055803} {\bibfield
  {journal} {\bibinfo  {journal} {Phys. Rev. C}\ }\textbf {\bibinfo {volume}
  {92}},\ \bibinfo {pages} {055803} (\bibinfo {year} {2015})}\BibitemShut
  {NoStop}%
\bibitem [{\citenamefont {Raduta}\ and\ \citenamefont
  {Gulminelli}(2019)}]{Raduta_NPA_2019}%
  \BibitemOpen
  \bibfield  {author} {\bibinfo {author} {\bibfnamefont {A.}~\bibnamefont
  {Raduta}}\ and\ \bibinfo {author} {\bibfnamefont {F.}~\bibnamefont
  {Gulminelli}},\ }\href {\doibase 10.1016/j.nuclphysa.2018.11.003} {\bibfield
  {journal} {\bibinfo  {journal} {Nucl. Phys. A}\ }\textbf {\bibinfo {volume}
  {983}},\ \bibinfo {pages} {252} (\bibinfo {year} {2019})},\ \Eprint
  {http://arxiv.org/abs/1807.06871} {arXiv:1807.06871 [nucl-th]} \BibitemShut
  {NoStop}%
\bibitem [{\citenamefont {Hempel}\ and\ \citenamefont
  {Schaffner-Bielich}(2010)}]{HS}%
  \BibitemOpen
  \bibfield  {author} {\bibinfo {author} {\bibfnamefont {M.}~\bibnamefont
  {Hempel}}\ and\ \bibinfo {author} {\bibfnamefont {J.}~\bibnamefont
  {Schaffner-Bielich}},\ }\href {\doibase 10.1016/j.nuclphysa.2010.02.010}
  {\bibfield  {journal} {\bibinfo  {journal} {Nucl. Phys. A}\ }\textbf
  {\bibinfo {volume} {837}},\ \bibinfo {pages} {210} (\bibinfo {year}
  {2010})},\ \Eprint {http://arxiv.org/abs/0911.4073} {arXiv:0911.4073
  [nucl-th]} \BibitemShut {NoStop}%
\bibitem [{\citenamefont {Typel}\ \emph {et~al.}(2010)\citenamefont {Typel},
  \citenamefont {R\"opke}, \citenamefont {Kl\"ahn}, \citenamefont {Blaschke},\
  and\ \citenamefont {Wolter}}]{Typel_PRC_2010}%
  \BibitemOpen
  \bibfield  {author} {\bibinfo {author} {\bibfnamefont {S.}~\bibnamefont
  {Typel}}, \bibinfo {author} {\bibfnamefont {G.}~\bibnamefont {R\"opke}},
  \bibinfo {author} {\bibfnamefont {T.}~\bibnamefont {Kl\"ahn}}, \bibinfo
  {author} {\bibfnamefont {D.}~\bibnamefont {Blaschke}}, \ and\ \bibinfo
  {author} {\bibfnamefont {H.~H.}\ \bibnamefont {Wolter}},\ }\href {\doibase
  10.1103/PhysRevC.81.015803} {\bibfield  {journal} {\bibinfo  {journal} {Phys.
  Rev. C}\ }\textbf {\bibinfo {volume} {81}},\ \bibinfo {pages} {015803}
  (\bibinfo {year} {2010})}\BibitemShut {NoStop}%
\bibitem [{\citenamefont {Fischer}\ \emph {et~al.}(2014)\citenamefont
  {Fischer}, \citenamefont {Hempel}, \citenamefont {Sagert}, \citenamefont
  {Suwa},\ and\ \citenamefont {Schaffner-Bielich}}]{Fischer2014}%
  \BibitemOpen
  \bibfield  {author} {\bibinfo {author} {\bibfnamefont {T.}~\bibnamefont
  {Fischer}}, \bibinfo {author} {\bibfnamefont {M.}~\bibnamefont {Hempel}},
  \bibinfo {author} {\bibfnamefont {I.}~\bibnamefont {Sagert}}, \bibinfo
  {author} {\bibfnamefont {Y.}~\bibnamefont {Suwa}}, \ and\ \bibinfo {author}
  {\bibfnamefont {J.}~\bibnamefont {Schaffner-Bielich}},\ }\href {\doibase
  10.1140/epja/i2014-14046-5} {\bibfield  {journal} {\bibinfo  {journal} {Eur.
  Phys. J. A}\ }\textbf {\bibinfo {volume} {50}},\ \bibinfo {pages} {46}
  (\bibinfo {year} {2014})},\ \Eprint {http://arxiv.org/abs/1307.6190}
  {arXiv:1307.6190 [astro-ph.HE]} \BibitemShut {NoStop}%
\bibitem [{\citenamefont {Fattoyev}\ \emph {et~al.}(2010)\citenamefont
  {Fattoyev}, \citenamefont {Horowitz}, \citenamefont {Piekarewicz},\ and\
  \citenamefont {Shen}}]{Fattoyev_PRC_2010}%
  \BibitemOpen
  \bibfield  {author} {\bibinfo {author} {\bibfnamefont {F.~J.}\ \bibnamefont
  {Fattoyev}}, \bibinfo {author} {\bibfnamefont {C.~J.}\ \bibnamefont
  {Horowitz}}, \bibinfo {author} {\bibfnamefont {J.}~\bibnamefont
  {Piekarewicz}}, \ and\ \bibinfo {author} {\bibfnamefont {G.}~\bibnamefont
  {Shen}},\ }\href {\doibase 10.1103/PhysRevC.82.055803} {\bibfield  {journal}
  {\bibinfo  {journal} {Phys. Rev. C}\ }\textbf {\bibinfo {volume} {82}},\
  \bibinfo {pages} {055803} (\bibinfo {year} {2010})}\BibitemShut {NoStop}%
\bibitem [{\citenamefont {Steiner}\ \emph {et~al.}(2013)\citenamefont
  {Steiner}, \citenamefont {Hempel},\ and\ \citenamefont
  {Fischer}}]{Steiner_2013}%
  \BibitemOpen
  \bibfield  {author} {\bibinfo {author} {\bibfnamefont {A.~W.}\ \bibnamefont
  {Steiner}}, \bibinfo {author} {\bibfnamefont {M.}~\bibnamefont {Hempel}}, \
  and\ \bibinfo {author} {\bibfnamefont {T.}~\bibnamefont {Fischer}},\ }\href
  {\doibase 10.1088/0004-637x/774/1/17} {\bibfield  {journal} {\bibinfo
  {journal} {ApJ}\ }\textbf {\bibinfo {volume} {774}},\ \bibinfo {pages} {17}
  (\bibinfo {year} {2013})}\BibitemShut {NoStop}%
\bibitem [{\citenamefont {Pais}\ and\ \citenamefont
  {Provid\^encia}(2016)}]{Pais16}%
  \BibitemOpen
  \bibfield  {author} {\bibinfo {author} {\bibfnamefont {H.}~\bibnamefont
  {Pais}}\ and\ \bibinfo {author} {\bibfnamefont {C.}~\bibnamefont
  {Provid\^encia}},\ }\href {\doibase 10.1103/PhysRevC.94.015808} {\bibfield
  {journal} {\bibinfo  {journal} {Phys. Rev. C}\ }\textbf {\bibinfo {volume}
  {94}},\ \bibinfo {pages} {015808} (\bibinfo {year} {2016})},\ \Eprint
  {http://arxiv.org/abs/1607.05899} {arXiv:1607.05899 [nucl-th]} \BibitemShut
  {NoStop}%
\bibitem [{\citenamefont {Horowitz}\ and\ \citenamefont
  {Piekarewicz}(2001)}]{Horowitz01}%
  \BibitemOpen
  \bibfield  {author} {\bibinfo {author} {\bibfnamefont {C.~J.}\ \bibnamefont
  {Horowitz}}\ and\ \bibinfo {author} {\bibfnamefont {J.}~\bibnamefont
  {Piekarewicz}},\ }\href {\doibase 10.1103/PhysRevLett.86.5647} {\bibfield
  {journal} {\bibinfo  {journal} {Phys. Rev. Lett.}\ }\textbf {\bibinfo
  {volume} {86}},\ \bibinfo {pages} {5647} (\bibinfo {year} {2001})},\ \Eprint
  {http://arxiv.org/abs/astro-ph/0010227} {arXiv:astro-ph/0010227 [astro-ph]}
  \BibitemShut {NoStop}%
\bibitem [{\citenamefont {Tolos}\ \emph
  {et~al.}(2017{\natexlab{a}})\citenamefont {Tolos}, \citenamefont
  {Centelles},\ and\ \citenamefont {Ramos}}]{Tolos17}%
  \BibitemOpen
  \bibfield  {author} {\bibinfo {author} {\bibfnamefont {L.}~\bibnamefont
  {Tolos}}, \bibinfo {author} {\bibfnamefont {M.}~\bibnamefont {Centelles}}, \
  and\ \bibinfo {author} {\bibfnamefont {A.}~\bibnamefont {Ramos}},\ }\href
  {\doibase 10.3847/1538-4357/834/1/3} {\bibfield  {journal} {\bibinfo
  {journal} {Astrophys. J.}\ }\textbf {\bibinfo {volume} {834}},\ \bibinfo
  {pages} {3} (\bibinfo {year} {2017}{\natexlab{a}})},\ \Eprint
  {http://arxiv.org/abs/1610.00919} {arXiv:1610.00919 [astro-ph.HE]}
  \BibitemShut {NoStop}%
\bibitem [{\citenamefont {Tolos}\ \emph
  {et~al.}(2017{\natexlab{b}})\citenamefont {Tolos}, \citenamefont
  {Centelles},\ and\ \citenamefont {Ramos}}]{Tolos17b}%
  \BibitemOpen
  \bibfield  {author} {\bibinfo {author} {\bibfnamefont {L.}~\bibnamefont
  {Tolos}}, \bibinfo {author} {\bibfnamefont {M.}~\bibnamefont {Centelles}}, \
  and\ \bibinfo {author} {\bibfnamefont {A.}~\bibnamefont {Ramos}},\ }\href
  {\doibase 10.1017/pasa.2017.60} {\bibfield  {journal} {\bibinfo  {journal}
  {Publ. Astron. Soc. Austral.}\ }\textbf {\bibinfo {volume} {34}},\ \bibinfo
  {pages} {e065} (\bibinfo {year} {2017}{\natexlab{b}})},\ \Eprint
  {http://arxiv.org/abs/1708.08681} {arXiv:1708.08681 [astro-ph.HE]}
  \BibitemShut {NoStop}%
\bibitem [{\citenamefont {Constantinou}\ \emph {et~al.}(2014)\citenamefont
  {Constantinou}, \citenamefont {Muccioli}, \citenamefont {Prakash},\ and\
  \citenamefont {Lattimer}}]{Constantinou2014}%
  \BibitemOpen
  \bibfield  {author} {\bibinfo {author} {\bibfnamefont {C.}~\bibnamefont
  {Constantinou}}, \bibinfo {author} {\bibfnamefont {B.}~\bibnamefont
  {Muccioli}}, \bibinfo {author} {\bibfnamefont {M.}~\bibnamefont {Prakash}}, \
  and\ \bibinfo {author} {\bibfnamefont {J.~M.}\ \bibnamefont {Lattimer}},\
  }\href {\doibase 10.1103/PhysRevC.89.065802} {\bibfield  {journal} {\bibinfo
  {journal} {Phys. Rev. C}\ }\textbf {\bibinfo {volume} {89}},\ \bibinfo
  {pages} {065802} (\bibinfo {year} {2014})},\ \Eprint
  {http://arxiv.org/abs/1402.6348} {arXiv:1402.6348 [astro-ph.SR]} \BibitemShut
  {NoStop}%
\bibitem [{\citenamefont {Schneider}\ \emph {et~al.}(2019)\citenamefont
  {Schneider}, \citenamefont {Constantinou}, \citenamefont {Muccioli},\ and\
  \citenamefont {Prakash}}]{Schneider2019}%
  \BibitemOpen
  \bibfield  {author} {\bibinfo {author} {\bibfnamefont {A.}~\bibnamefont
  {Schneider}}, \bibinfo {author} {\bibfnamefont {C.}~\bibnamefont
  {Constantinou}}, \bibinfo {author} {\bibfnamefont {B.}~\bibnamefont
  {Muccioli}}, \ and\ \bibinfo {author} {\bibfnamefont {M.}~\bibnamefont
  {Prakash}},\ }\href {\doibase 10.1103/PhysRevC.100.025803} {\bibfield
  {journal} {\bibinfo  {journal} {Phys. Rev. C}\ }\textbf {\bibinfo {volume}
  {100}},\ \bibinfo {pages} {025803} (\bibinfo {year} {2019})},\ \Eprint
  {http://arxiv.org/abs/1901.09652} {arXiv:1901.09652 [nucl-th]} \BibitemShut
  {NoStop}%
\bibitem [{\citenamefont {Akmal}\ \emph {et~al.}(1998)\citenamefont {Akmal},
  \citenamefont {Pandharipande},\ and\ \citenamefont {Ravenhall}}]{APR}%
  \BibitemOpen
  \bibfield  {author} {\bibinfo {author} {\bibfnamefont {A.}~\bibnamefont
  {Akmal}}, \bibinfo {author} {\bibfnamefont {V.}~\bibnamefont
  {Pandharipande}}, \ and\ \bibinfo {author} {\bibfnamefont {D.}~\bibnamefont
  {Ravenhall}},\ }\href {\doibase 10.1103/PhysRevC.58.1804} {\bibfield
  {journal} {\bibinfo  {journal} {Phys. Rev. C}\ }\textbf {\bibinfo {volume}
  {58}},\ \bibinfo {pages} {1804} (\bibinfo {year} {1998})},\ \Eprint
  {http://arxiv.org/abs/nucl-th/9804027} {arXiv:nucl-th/9804027} \BibitemShut
  {NoStop}%
\bibitem [{\citenamefont {Akmal}\ and\ \citenamefont
  {Pandharipande}(1997)}]{Akmal_1997}%
  \BibitemOpen
  \bibfield  {author} {\bibinfo {author} {\bibfnamefont {A.}~\bibnamefont
  {Akmal}}\ and\ \bibinfo {author} {\bibfnamefont {V.}~\bibnamefont
  {Pandharipande}},\ }\href {\doibase 10.1103/PhysRevC.56.2261} {\bibfield
  {journal} {\bibinfo  {journal} {Phys. Rev. C}\ }\textbf {\bibinfo {volume}
  {56}},\ \bibinfo {pages} {2261} (\bibinfo {year} {1997})},\ \Eprint
  {http://arxiv.org/abs/nucl-th/9705013} {arXiv:nucl-th/9705013} \BibitemShut
  {NoStop}%
\bibitem [{\citenamefont {{Banik}}\ \emph {et~al.}(2014)\citenamefont
  {{Banik}}, \citenamefont {{Hempel}},\ and\ \citenamefont
  {{Bandyopadhyay}}}]{Banik_2014}%
  \BibitemOpen
  \bibfield  {author} {\bibinfo {author} {\bibfnamefont {S.}~\bibnamefont
  {{Banik}}}, \bibinfo {author} {\bibfnamefont {M.}~\bibnamefont {{Hempel}}}, \
  and\ \bibinfo {author} {\bibfnamefont {D.}~\bibnamefont {{Bandyopadhyay}}},\
  }\href {\doibase 10.1088/0067-0049/214/2/22} {\bibfield  {journal} {\bibinfo
  {journal} {Astrophys.J.Suppl.}\ }\textbf {\bibinfo {volume} {214}},\ \bibinfo
  {eid} {22} (\bibinfo {year} {2014})}\BibitemShut {NoStop}%
\bibitem [{\citenamefont {Fortin}\ \emph {et~al.}(2018)\citenamefont {Fortin},
  \citenamefont {Oertel},\ and\ \citenamefont
  {Provid{\^e}ncia}}]{Fortin_PASA_2018}%
  \BibitemOpen
  \bibfield  {author} {\bibinfo {author} {\bibfnamefont {M.}~\bibnamefont
  {Fortin}}, \bibinfo {author} {\bibfnamefont {M.}~\bibnamefont {Oertel}}, \
  and\ \bibinfo {author} {\bibfnamefont {C.}~\bibnamefont {Provid{\^e}ncia}},\
  }\href {\doibase 10.1017/pasa.2018.32} {\bibfield  {journal} {\bibinfo
  {journal} {Publ. Astron. Soc. Austral.}\ }\textbf {\bibinfo {volume} {35}},\
  \bibinfo {pages} {44} (\bibinfo {year} {2018})},\ \Eprint
  {http://arxiv.org/abs/1711.09427} {arXiv:1711.09427 [astro-ph.HE]}
  \BibitemShut {NoStop}%
\bibitem [{\citenamefont {Fortin}\ \emph {et~al.}(2020)\citenamefont {Fortin},
  \citenamefont {Raduta}, \citenamefont {Avancini},\ and\ \citenamefont
  {Provid\^encia}}]{Fortin_PRD_2020}%
  \BibitemOpen
  \bibfield  {author} {\bibinfo {author} {\bibfnamefont {M.}~\bibnamefont
  {Fortin}}, \bibinfo {author} {\bibfnamefont {A.~R.}\ \bibnamefont {Raduta}},
  \bibinfo {author} {\bibfnamefont {S.}~\bibnamefont {Avancini}}, \ and\
  \bibinfo {author} {\bibfnamefont {C.}~\bibnamefont {Provid\^encia}},\ }\href
  {\doibase 10.1103/PhysRevD.101.034017} {\bibfield  {journal} {\bibinfo
  {journal} {Phys. Rev. D}\ }\textbf {\bibinfo {volume} {101}},\ \bibinfo
  {pages} {034017} (\bibinfo {year} {2020})},\ \Eprint
  {http://arxiv.org/abs/2001.08036} {arXiv:2001.08036 [hep-ph]} \BibitemShut
  {NoStop}%
\bibitem [{\citenamefont {Typel}\ \emph {et~al.}(2015)\citenamefont {Typel},
  \citenamefont {Oertel},\ and\ \citenamefont {Klähn}}]{Typel_2013}%
  \BibitemOpen
  \bibfield  {author} {\bibinfo {author} {\bibfnamefont {S.}~\bibnamefont
  {Typel}}, \bibinfo {author} {\bibfnamefont {M.}~\bibnamefont {Oertel}}, \
  and\ \bibinfo {author} {\bibfnamefont {T.}~\bibnamefont {Klähn}},\ }\href
  {\doibase 10.1134/S1063779615040061} {\bibfield  {journal} {\bibinfo
  {journal} {Phys. Part. Nucl.}\ }\textbf {\bibinfo {volume} {46}},\ \bibinfo
  {pages} {633} (\bibinfo {year} {2015})}\BibitemShut {NoStop}%
\bibitem [{\citenamefont {{Yagi}}\ and\ \citenamefont
  {{Yunes}}(2017)}]{Yagi2017}%
  \BibitemOpen
  \bibfield  {author} {\bibinfo {author} {\bibfnamefont {K.}~\bibnamefont
  {{Yagi}}}\ and\ \bibinfo {author} {\bibfnamefont {N.}~\bibnamefont
  {{Yunes}}},\ }\href {\doibase 10.1016/j.physrep.2017.03.002} {\bibfield
  {journal} {\bibinfo  {journal} {\physrep}\ }\textbf {\bibinfo {volume}
  {681}},\ \bibinfo {pages} {1} (\bibinfo {year} {2017})},\ \Eprint
  {http://arxiv.org/abs/1608.02582} {arXiv:1608.02582 [gr-qc]} \BibitemShut
  {NoStop}%
\bibitem [{\citenamefont {Haensel}\ and\ \citenamefont
  {Zdunik}(1989)}]{HZ_Nature_1989}%
  \BibitemOpen
  \bibfield  {author} {\bibinfo {author} {\bibfnamefont {P.}~\bibnamefont
  {Haensel}}\ and\ \bibinfo {author} {\bibfnamefont {J.~L.}\ \bibnamefont
  {Zdunik}},\ }\href@noop {} {\bibfield  {journal} {\bibinfo  {journal}
  {Nature}\ }\textbf {\bibinfo {volume} {340}},\ \bibinfo {pages} {617}
  (\bibinfo {year} {1989})}\BibitemShut {NoStop}%
\bibitem [{\citenamefont {Friedman}\ \emph {et~al.}(1989)\citenamefont
  {Friedman}, \citenamefont {Ipser},\ and\ \citenamefont
  {Parker}}]{Friedman_PRL_1989}%
  \BibitemOpen
  \bibfield  {author} {\bibinfo {author} {\bibfnamefont {J.~L.}\ \bibnamefont
  {Friedman}}, \bibinfo {author} {\bibfnamefont {J.~R.}\ \bibnamefont {Ipser}},
  \ and\ \bibinfo {author} {\bibfnamefont {L.}~\bibnamefont {Parker}},\ }\href
  {\doibase 10.1103/PhysRevLett.62.3015} {\bibfield  {journal} {\bibinfo
  {journal} {Phys. Rev. Lett.}\ }\textbf {\bibinfo {volume} {62}},\ \bibinfo
  {pages} {3015} (\bibinfo {year} {1989})}\BibitemShut {NoStop}%
\bibitem [{\citenamefont {{Shapiro}}\ \emph {et~al.}(1989)\citenamefont
  {{Shapiro}}, \citenamefont {{Teukolsky}},\ and\ \citenamefont
  {{Wasserman}}}]{Shapiro_Nature_1989}%
  \BibitemOpen
  \bibfield  {author} {\bibinfo {author} {\bibfnamefont {S.~L.}\ \bibnamefont
  {{Shapiro}}}, \bibinfo {author} {\bibfnamefont {S.~A.}\ \bibnamefont
  {{Teukolsky}}}, \ and\ \bibinfo {author} {\bibfnamefont {I.}~\bibnamefont
  {{Wasserman}}},\ }\href {\doibase 10.1038/340451a0} {\bibfield  {journal}
  {\bibinfo  {journal} {\nat}\ }\textbf {\bibinfo {volume} {340}},\ \bibinfo
  {pages} {451} (\bibinfo {year} {1989})}\BibitemShut {NoStop}%
\bibitem [{\citenamefont {{Haensel}}\ \emph {et~al.}(1995)\citenamefont
  {{Haensel}}, \citenamefont {{Salgado}},\ and\ \citenamefont
  {{Bonazzola}}}]{Haensel_AA_1995}%
  \BibitemOpen
  \bibfield  {author} {\bibinfo {author} {\bibfnamefont {P.}~\bibnamefont
  {{Haensel}}}, \bibinfo {author} {\bibfnamefont {M.}~\bibnamefont
  {{Salgado}}}, \ and\ \bibinfo {author} {\bibfnamefont {S.}~\bibnamefont
  {{Bonazzola}}},\ }\href@noop {} {\bibfield  {journal} {\bibinfo  {journal}
  {Astronomy and Astrophysics}\ }\textbf {\bibinfo {volume} {296}},\ \bibinfo
  {pages} {745} (\bibinfo {year} {1995})}\BibitemShut {NoStop}%
\bibitem [{\citenamefont {{Haensel, P.}}\ \emph {et~al.}(2009)\citenamefont
  {{Haensel, P.}}, \citenamefont {{Zdunik, J. L.}}, \citenamefont {{Bejger,
  M.}},\ and\ \citenamefont {{Lattimer, J. M.}}}]{Haensel_AA_2009}%
  \BibitemOpen
  \bibfield  {author} {\bibinfo {author} {\bibnamefont {{Haensel, P.}}},
  \bibinfo {author} {\bibnamefont {{Zdunik, J. L.}}}, \bibinfo {author}
  {\bibnamefont {{Bejger, M.}}}, \ and\ \bibinfo {author} {\bibnamefont
  {{Lattimer, J. M.}}},\ }\href {\doibase 10.1051/0004-6361/200811605}
  {\bibfield  {journal} {\bibinfo  {journal} {A\&A}\ }\textbf {\bibinfo
  {volume} {502}},\ \bibinfo {pages} {605} (\bibinfo {year}
  {2009})}\BibitemShut {NoStop}%
\bibitem [{\citenamefont {Koliogiannis}\ and\ \citenamefont
  {Moustakidis}(2020)}]{Koliogiannis_PRC_2020}%
  \BibitemOpen
  \bibfield  {author} {\bibinfo {author} {\bibfnamefont {P.~S.}\ \bibnamefont
  {Koliogiannis}}\ and\ \bibinfo {author} {\bibfnamefont {C.~C.}\ \bibnamefont
  {Moustakidis}},\ }\href {\doibase 10.1103/PhysRevC.101.015805} {\bibfield
  {journal} {\bibinfo  {journal} {Phys. Rev. C}\ }\textbf {\bibinfo {volume}
  {101}},\ \bibinfo {pages} {015805} (\bibinfo {year} {2020})}\BibitemShut
  {NoStop}%
\bibitem [{\citenamefont {{Cordes}}\ \emph {et~al.}(2004)\citenamefont
  {{Cordes}}, \citenamefont {{Kramer}}, \citenamefont {{Lazio}}, \citenamefont
  {{Stappers}}, \citenamefont {{Backer}},\ and\ \citenamefont
  {{Johnston}}}]{Cordes2004}%
  \BibitemOpen
  \bibfield  {author} {\bibinfo {author} {\bibfnamefont {J.~M.}\ \bibnamefont
  {{Cordes}}}, \bibinfo {author} {\bibfnamefont {M.}~\bibnamefont {{Kramer}}},
  \bibinfo {author} {\bibfnamefont {T.~J.~W.}\ \bibnamefont {{Lazio}}},
  \bibinfo {author} {\bibfnamefont {B.~W.}\ \bibnamefont {{Stappers}}},
  \bibinfo {author} {\bibfnamefont {D.~C.}\ \bibnamefont {{Backer}}}, \ and\
  \bibinfo {author} {\bibfnamefont {S.}~\bibnamefont {{Johnston}}},\ }\href
  {\doibase 10.1016/j.newar.2004.09.040} {\bibfield  {journal} {\bibinfo
  {journal} {\nar}\ }\textbf {\bibinfo {volume} {48}},\ \bibinfo {pages} {1413}
  (\bibinfo {year} {2004})},\ \Eprint {http://arxiv.org/abs/astro-ph/0505555}
  {arXiv:astro-ph/0505555 [astro-ph]} \BibitemShut {NoStop}%
\bibitem [{\citenamefont {{Hessels}}\ \emph {et~al.}(2006)\citenamefont
  {{Hessels}}, \citenamefont {{Ransom}}, \citenamefont {{Stairs}},
  \citenamefont {{Freire}}, \citenamefont {{Kaspi}},\ and\ \citenamefont
  {{Camilo}}}]{Hessels_2006}%
  \BibitemOpen
  \bibfield  {author} {\bibinfo {author} {\bibfnamefont {J.~W.~T.}\
  \bibnamefont {{Hessels}}}, \bibinfo {author} {\bibfnamefont {S.~M.}\
  \bibnamefont {{Ransom}}}, \bibinfo {author} {\bibfnamefont {I.~H.}\
  \bibnamefont {{Stairs}}}, \bibinfo {author} {\bibfnamefont {P.~C.~C.}\
  \bibnamefont {{Freire}}}, \bibinfo {author} {\bibfnamefont {V.~M.}\
  \bibnamefont {{Kaspi}}}, \ and\ \bibinfo {author} {\bibfnamefont
  {F.}~\bibnamefont {{Camilo}}},\ }\href {\doibase 10.1126/science.1123430}
  {\bibfield  {journal} {\bibinfo  {journal} {Science}\ }\textbf {\bibinfo
  {volume} {311}},\ \bibinfo {pages} {1901} (\bibinfo {year} {2006})},\ \Eprint
  {http://arxiv.org/abs/astro-ph/0601337} {arXiv:astro-ph/0601337 [astro-ph]}
  \BibitemShut {NoStop}%
\bibitem [{\citenamefont {Lattimer}\ and\ \citenamefont
  {Prakash}(2004)}]{Lattimer_Science_2004}%
  \BibitemOpen
  \bibfield  {author} {\bibinfo {author} {\bibfnamefont {J.~M.}\ \bibnamefont
  {Lattimer}}\ and\ \bibinfo {author} {\bibfnamefont {M.}~\bibnamefont
  {Prakash}},\ }\href {\doibase 10.1126/science.1090720} {\bibfield  {journal}
  {\bibinfo  {journal} {Science}\ }\textbf {\bibinfo {volume} {304}},\ \bibinfo
  {pages} {536} (\bibinfo {year} {2004})},\ \Eprint
  {http://arxiv.org/abs/https://science.sciencemag.org/content/304/5670/536.full.pdf}
  {https://science.sciencemag.org/content/304/5670/536.full.pdf} \BibitemShut
  {NoStop}%
\bibitem [{\citenamefont {Yagi}\ and\ \citenamefont
  {Yunes}(2013)}]{Yagi_Science_2013}%
  \BibitemOpen
  \bibfield  {author} {\bibinfo {author} {\bibfnamefont {K.}~\bibnamefont
  {Yagi}}\ and\ \bibinfo {author} {\bibfnamefont {N.}~\bibnamefont {Yunes}},\
  }\href {\doibase 10.1126/science.1236462} {\bibfield  {journal} {\bibinfo
  {journal} {Science}\ }\textbf {\bibinfo {volume} {341}},\ \bibinfo {pages}
  {365} (\bibinfo {year} {2013})},\ \Eprint {http://arxiv.org/abs/1302.4499}
  {arXiv:1302.4499 [gr-qc]} \BibitemShut {NoStop}%
\bibitem [{\citenamefont {Yagi}\ \emph {et~al.}(2014)\citenamefont {Yagi},
  \citenamefont {Stein}, \citenamefont {Pappas}, \citenamefont {Yunes},\ and\
  \citenamefont {Apostolatos}}]{Yagi:2014qua}%
  \BibitemOpen
  \bibfield  {author} {\bibinfo {author} {\bibfnamefont {K.}~\bibnamefont
  {Yagi}}, \bibinfo {author} {\bibfnamefont {L.~C.}\ \bibnamefont {Stein}},
  \bibinfo {author} {\bibfnamefont {G.}~\bibnamefont {Pappas}}, \bibinfo
  {author} {\bibfnamefont {N.}~\bibnamefont {Yunes}}, \ and\ \bibinfo {author}
  {\bibfnamefont {T.~A.}\ \bibnamefont {Apostolatos}},\ }\href {\doibase
  10.1103/PhysRevD.90.063010} {\bibfield  {journal} {\bibinfo  {journal} {Phys.
  Rev. D}\ }\textbf {\bibinfo {volume} {90}},\ \bibinfo {pages} {063010}
  (\bibinfo {year} {2014})},\ \Eprint {http://arxiv.org/abs/1406.7587}
  {arXiv:1406.7587 [gr-qc]} \BibitemShut {NoStop}%
\bibitem [{\citenamefont {{Ravenhall}}\ and\ \citenamefont
  {{Pethick}}(1994)}]{Ravenhall_ApJ_1994}%
  \BibitemOpen
  \bibfield  {author} {\bibinfo {author} {\bibfnamefont {D.~G.}\ \bibnamefont
  {{Ravenhall}}}\ and\ \bibinfo {author} {\bibfnamefont {C.~J.}\ \bibnamefont
  {{Pethick}}},\ }\href {\doibase 10.1086/173935} {\bibfield  {journal}
  {\bibinfo  {journal} {\apj}\ }\textbf {\bibinfo {volume} {424}},\ \bibinfo
  {pages} {846} (\bibinfo {year} {1994})}\BibitemShut {NoStop}%
\bibitem [{\citenamefont {Lattimer}\ and\ \citenamefont
  {Schutz}(2005)}]{Lattimer_ApJ_2005}%
  \BibitemOpen
  \bibfield  {author} {\bibinfo {author} {\bibfnamefont {J.~M.}\ \bibnamefont
  {Lattimer}}\ and\ \bibinfo {author} {\bibfnamefont {B.~F.}\ \bibnamefont
  {Schutz}},\ }\href {\doibase 10.1086/431543} {\bibfield  {journal} {\bibinfo
  {journal} {ApJ}\ }\textbf {\bibinfo {volume} {629}},\ \bibinfo {pages} {979}
  (\bibinfo {year} {2005})}\BibitemShut {NoStop}%
\bibitem [{\citenamefont {{Sumiyoshi}}\ \emph {et~al.}(1999)\citenamefont
  {{Sumiyoshi}}, \citenamefont {{Ib\'a\~nez}},\ and\ \citenamefont
  {{Romero}}}]{Sumiyoshi_AASS_1999}%
  \BibitemOpen
  \bibfield  {author} {\bibinfo {author} {\bibfnamefont {K.}~\bibnamefont
  {{Sumiyoshi}}}, \bibinfo {author} {\bibfnamefont {J.~M.}\ \bibnamefont
  {{Ib\'a\~nez}}}, \ and\ \bibinfo {author} {\bibfnamefont {J.~V.}\
  \bibnamefont {{Romero}}},\ }\href {\doibase 10.1051/aas:1999123} {\bibfield
  {journal} {\bibinfo  {journal} {Astron. Astrophys. Suppl. Ser.}\ }\textbf
  {\bibinfo {volume} {134}},\ \bibinfo {pages} {39} (\bibinfo {year}
  {1999})}\BibitemShut {NoStop}%
\bibitem [{\citenamefont {Pons}\ \emph {et~al.}(1999)\citenamefont {Pons},
  \citenamefont {Reddy}, \citenamefont {Prakash}, \citenamefont {Lattimer},\
  and\ \citenamefont {Miralles}}]{Pons_1999}%
  \BibitemOpen
  \bibfield  {author} {\bibinfo {author} {\bibfnamefont {J.~A.}\ \bibnamefont
  {Pons}}, \bibinfo {author} {\bibfnamefont {S.}~\bibnamefont {Reddy}},
  \bibinfo {author} {\bibfnamefont {M.}~\bibnamefont {Prakash}}, \bibinfo
  {author} {\bibfnamefont {J.~M.}\ \bibnamefont {Lattimer}}, \ and\ \bibinfo
  {author} {\bibfnamefont {J.~A.}\ \bibnamefont {Miralles}},\ }\href {\doibase
  10.1086/306889} {\bibfield  {journal} {\bibinfo  {journal} {The Astrophysical
  Journal}\ }\textbf {\bibinfo {volume} {513}},\ \bibinfo {pages} {780}
  (\bibinfo {year} {1999})}\BibitemShut {NoStop}%
\bibitem [{\citenamefont {Endrizzi}\ \emph {et~al.}(2020)\citenamefont
  {Endrizzi}, \citenamefont {Perego}, \citenamefont {Fabbri}, \citenamefont
  {Branca}, \citenamefont {Radice}, \citenamefont {Bernuzzi}, \citenamefont
  {Giacomazzo}, \citenamefont {Pederiva},\ and\ \citenamefont
  {Lovato}}]{Endrizzi2019}%
  \BibitemOpen
  \bibfield  {author} {\bibinfo {author} {\bibfnamefont {A.}~\bibnamefont
  {Endrizzi}}, \bibinfo {author} {\bibfnamefont {A.}~\bibnamefont {Perego}},
  \bibinfo {author} {\bibfnamefont {F.~M.}\ \bibnamefont {Fabbri}}, \bibinfo
  {author} {\bibfnamefont {L.}~\bibnamefont {Branca}}, \bibinfo {author}
  {\bibfnamefont {D.}~\bibnamefont {Radice}}, \bibinfo {author} {\bibfnamefont
  {S.}~\bibnamefont {Bernuzzi}}, \bibinfo {author} {\bibfnamefont
  {B.}~\bibnamefont {Giacomazzo}}, \bibinfo {author} {\bibfnamefont
  {F.}~\bibnamefont {Pederiva}}, \ and\ \bibinfo {author} {\bibfnamefont
  {A.}~\bibnamefont {Lovato}},\ }\href {\doibase
  10.1140/epja/s10050-019-00018-6} {\bibfield  {journal} {\bibinfo  {journal}
  {Eur. Phys. J. A}\ }\textbf {\bibinfo {volume} {56}},\ \bibinfo {pages} {15}
  (\bibinfo {year} {2020})},\ \Eprint {http://arxiv.org/abs/1908.04952}
  {arXiv:1908.04952 [astro-ph.HE]} \BibitemShut {NoStop}%
\bibitem [{\citenamefont {Bollig}\ \emph {et~al.}(2017)\citenamefont {Bollig},
  \citenamefont {Janka}, \citenamefont {Lohs}, \citenamefont {Martinez-Pinedo},
  \citenamefont {Horowitz},\ and\ \citenamefont {Melson}}]{Bollig_PRL_2017}%
  \BibitemOpen
  \bibfield  {author} {\bibinfo {author} {\bibfnamefont {R.}~\bibnamefont
  {Bollig}}, \bibinfo {author} {\bibfnamefont {H.~T.}\ \bibnamefont {Janka}},
  \bibinfo {author} {\bibfnamefont {A.}~\bibnamefont {Lohs}}, \bibinfo {author}
  {\bibfnamefont {G.}~\bibnamefont {Martinez-Pinedo}}, \bibinfo {author}
  {\bibfnamefont {C.}~\bibnamefont {Horowitz}}, \ and\ \bibinfo {author}
  {\bibfnamefont {T.}~\bibnamefont {Melson}},\ }\href {\doibase
  10.1103/PhysRevLett.119.242702} {\bibfield  {journal} {\bibinfo  {journal}
  {Phys. Rev. Lett.}\ }\textbf {\bibinfo {volume} {119}},\ \bibinfo {pages}
  {242702} (\bibinfo {year} {2017})},\ \Eprint
  {http://arxiv.org/abs/1706.04630} {arXiv:1706.04630 [astro-ph.HE]}
  \BibitemShut {NoStop}%
\bibitem [{\citenamefont {Perego}\ \emph {et~al.}(2019)\citenamefont {Perego},
  \citenamefont {Bernuzzi},\ and\ \citenamefont {Radice}}]{Perego2019}%
  \BibitemOpen
  \bibfield  {author} {\bibinfo {author} {\bibfnamefont {A.}~\bibnamefont
  {Perego}}, \bibinfo {author} {\bibfnamefont {S.}~\bibnamefont {Bernuzzi}}, \
  and\ \bibinfo {author} {\bibfnamefont {D.}~\bibnamefont {Radice}},\ }\href
  {\doibase 10.1140/epja/i2019-12810-7} {\bibfield  {journal} {\bibinfo
  {journal} {Eur. Phys. J. A}\ }\textbf {\bibinfo {volume} {55}},\ \bibinfo
  {pages} {124} (\bibinfo {year} {2019})},\ \Eprint
  {http://arxiv.org/abs/1903.07898} {arXiv:1903.07898 [gr-qc]} \BibitemShut
  {NoStop}%
\bibitem [{\citenamefont {Glendenning}(2000)}]{Glend2000}%
  \BibitemOpen
  \bibfield  {author} {\bibinfo {author} {\bibfnamefont {N.~K.}\ \bibnamefont
  {Glendenning}},\ }\href@noop {} {\emph {\bibinfo {title} {{Compact stars:
  Nuclear physics, particle physics, and general relativity}}}},\ \bibinfo
  {edition} {2nd}\ ed.\ (\bibinfo  {publisher} {Springer-Verlag New York},\
  \bibinfo {year} {2000})\ p.\ \bibinfo {pages} {468}\BibitemShut {NoStop}%
\bibitem [{\citenamefont {Oertel}\ \emph {et~al.}(2012)\citenamefont {Oertel},
  \citenamefont {Fantina},\ and\ \citenamefont {Novak}}]{Oertel_PRC_2012}%
  \BibitemOpen
  \bibfield  {author} {\bibinfo {author} {\bibfnamefont {M.}~\bibnamefont
  {Oertel}}, \bibinfo {author} {\bibfnamefont {A.}~\bibnamefont {Fantina}}, \
  and\ \bibinfo {author} {\bibfnamefont {J.}~\bibnamefont {Novak}},\ }\href
  {\doibase 10.1103/PhysRevC.85.055806} {\bibfield  {journal} {\bibinfo
  {journal} {Phys. Rev. C}\ }\textbf {\bibinfo {volume} {85}},\ \bibinfo
  {pages} {055806} (\bibinfo {year} {2012})},\ \Eprint
  {http://arxiv.org/abs/1202.2679} {arXiv:1202.2679 [nucl-th]} \BibitemShut
  {NoStop}%
\bibitem [{\citenamefont {Oertel}\ \emph {et~al.}(2016)\citenamefont {Oertel},
  \citenamefont {Gulminelli}, \citenamefont {Provid\^encia},\ and\
  \citenamefont {Raduta}}]{Oertel_EPJA_2016}%
  \BibitemOpen
  \bibfield  {author} {\bibinfo {author} {\bibfnamefont {M.}~\bibnamefont
  {Oertel}}, \bibinfo {author} {\bibfnamefont {F.}~\bibnamefont {Gulminelli}},
  \bibinfo {author} {\bibfnamefont {C.}~\bibnamefont {Provid\^encia}}, \ and\
  \bibinfo {author} {\bibfnamefont {A.~R.}\ \bibnamefont {Raduta}},\ }\href
  {\doibase 10.1140/epja/i2016-16050-1} {\bibfield  {journal} {\bibinfo
  {journal} {Eur. Phys. J. A}\ }\textbf {\bibinfo {volume} {52}},\ \bibinfo
  {pages} {50} (\bibinfo {year} {2016})},\ \Eprint
  {http://arxiv.org/abs/1601.00435} {arXiv:1601.00435 [nucl-th]} \BibitemShut
  {NoStop}%
\bibitem [{\citenamefont {Bauswein}\ \emph {et~al.}(2020)\citenamefont
  {Bauswein}, \citenamefont {Blacker}, \citenamefont {Vijayan}, \citenamefont
  {Stergioulas}, \citenamefont {Chatziioannou}, \citenamefont {Clark},
  \citenamefont {Bastian}, \citenamefont {Blaschke}, \citenamefont {Cierniak},\
  and\ \citenamefont {Fischer}}]{Bauswein2020}%
  \BibitemOpen
  \bibfield  {author} {\bibinfo {author} {\bibfnamefont {A.}~\bibnamefont
  {Bauswein}}, \bibinfo {author} {\bibfnamefont {S.}~\bibnamefont {Blacker}},
  \bibinfo {author} {\bibfnamefont {V.}~\bibnamefont {Vijayan}}, \bibinfo
  {author} {\bibfnamefont {N.}~\bibnamefont {Stergioulas}}, \bibinfo {author}
  {\bibfnamefont {K.}~\bibnamefont {Chatziioannou}}, \bibinfo {author}
  {\bibfnamefont {J.~A.}\ \bibnamefont {Clark}}, \bibinfo {author}
  {\bibfnamefont {N.-U.~F.}\ \bibnamefont {Bastian}}, \bibinfo {author}
  {\bibfnamefont {D.~B.}\ \bibnamefont {Blaschke}}, \bibinfo {author}
  {\bibfnamefont {M.}~\bibnamefont {Cierniak}}, \ and\ \bibinfo {author}
  {\bibfnamefont {T.}~\bibnamefont {Fischer}},\ }\href {\doibase
  10.1103/PhysRevLett.125.141103} {\bibfield  {journal} {\bibinfo  {journal}
  {Phys. Rev. Lett.}\ }\textbf {\bibinfo {volume} {125}},\ \bibinfo {pages}
  {141103} (\bibinfo {year} {2020})},\ \Eprint
  {http://arxiv.org/abs/2004.00846} {arXiv:2004.00846 [astro-ph.HE]}
  \BibitemShut {NoStop}%
\bibitem [{\citenamefont {{Kasen}}\ \emph {et~al.}(2017)\citenamefont
  {{Kasen}}, \citenamefont {{Metzger}}, \citenamefont {{Barnes}}, \citenamefont
  {{Quataert}},\ and\ \citenamefont {{Ramirez-Ruiz}}}]{Kasen_2017}%
  \BibitemOpen
  \bibfield  {author} {\bibinfo {author} {\bibfnamefont {D.}~\bibnamefont
  {{Kasen}}}, \bibinfo {author} {\bibfnamefont {B.}~\bibnamefont {{Metzger}}},
  \bibinfo {author} {\bibfnamefont {J.}~\bibnamefont {{Barnes}}}, \bibinfo
  {author} {\bibfnamefont {E.}~\bibnamefont {{Quataert}}}, \ and\ \bibinfo
  {author} {\bibfnamefont {E.}~\bibnamefont {{Ramirez-Ruiz}}},\ }\href
  {\doibase 10.1038/nature24453} {\bibfield  {journal} {\bibinfo  {journal}
  {\nat}\ }\textbf {\bibinfo {volume} {551}},\ \bibinfo {pages} {80} (\bibinfo
  {year} {2017})},\ \Eprint {http://arxiv.org/abs/1710.05463} {arXiv:1710.05463
  [astro-ph.HE]} \BibitemShut {NoStop}%
\bibitem [{\citenamefont {Abbott}\ \emph
  {et~al.}(2017{\natexlab{c}})\citenamefont {Abbott}, \citenamefont {Abbott},
  \citenamefont {Abbott}, \citenamefont {Acernese}, \citenamefont {Ackley}
  \emph {et~al.}}]{LIGO_Virgo2017c}%
  \BibitemOpen
  \bibfield  {author} {\bibinfo {author} {\bibfnamefont {B.~P.}\ \bibnamefont
  {Abbott}}, \bibinfo {author} {\bibfnamefont {R.}~\bibnamefont {Abbott}},
  \bibinfo {author} {\bibfnamefont {T.~D.}\ \bibnamefont {Abbott}}, \bibinfo
  {author} {\bibfnamefont {F.}~\bibnamefont {Acernese}}, \bibinfo {author}
  {\bibfnamefont {K.}~\bibnamefont {Ackley}},  \emph {et~al.} (\bibinfo
  {collaboration} {LIGO Scientific Collaboration and Virgo Collaboration}),\
  }\href@noop {} {\bibfield  {journal} {\bibinfo  {journal} {PhRvL}\ }\textbf
  {\bibinfo {volume} {119}},\ \bibinfo {pages} {161101} (\bibinfo {year}
  {2017}{\natexlab{c}})}\BibitemShut {NoStop}%
\end{thebibliography}%

\end{document}